\documentclass[10pt,a4paper]{iopart}
\usepackage{graphicx,color}
\usepackage{dcolumn}
\usepackage{epsfig}
\usepackage{setstack}
\usepackage{ulem}

\usepackage{iopams}
\usepackage{euscript}
\usepackage{amssymb}
\usepackage{amsthm}
\usepackage{newlfont}
\usepackage{mathrsfs}
\usepackage{subfigure}
\usepackage{todonotes}
\usepackage{hyperref}

\def\bea{\begin{eqnarray}}
\def\eea{\end{eqnarray}}
\def\ba{\begin{array}}
\def\ea{\end{array}}

\def\var{${\text{Var}}(X)$}
\def\varm{{\text{Var}}(X)}

\def\lsim{\,\lower2truept\hbox{${< \atop\hbox{\raise4truept\hbox{$\sim$}}}$}\,}
\def\gsim{\,\lower2truept\hbox{${> \atop\hbox{\raise4truept\hbox{$\sim$}}}$}\,}

\begin{document}
\title[Tracer dynamics in one dimensional gases]{Tracer dynamics in one dimensional gases of active or passive particles}

\author{Tirthankar Banerjee$^{1}$, Robert L. Jack$^{1,2}$ and Michael E. Cates$^{1}$}
\address{$^1$ DAMTP, Centre for Mathematical Sciences, University of Cambridge, Wilberforce Road, Cambridge CB3 0WA, United Kingdom}
\address{$^2$ Yusuf Hamied Department of Chemistry, University of Cambridge, Lensfield Road, Cambridge CB2 1EW, United Kingdom}

\begin{abstract}
We consider one-dimensional systems comprising either active run-and-tumble particles (RTPs) or passive Brownian random walkers.  These particles are either noninteracting or have hardcore exclusions. We study the dynamics of a single tracer particle embedded in such a system -- this tracer may be either active or passive, with hardcore exclusion from environmental particles.  
In an active hardcore environment, both active and passive tracers show long-time subdiffusion: displacements scale as $t^{1/4}$ with a density-dependent prefactor that is independent of tracer type, and differs from the corresponding result for passive-in-passive subdiffusion. 
In an environment of noninteracting active particles, the passive-in-passive results are recovered at low densities for both active and passive tracers, but transient caging effects slow the tracer motion at higher densities, delaying the onset of any $t^{1/4}$ regime.
For an active tracer in a passive environment, we find more complex outcomes, which depend on details of the dynamical discretization scheme. We interpret these results by studying the density distribution of environmental particles around the tracer. In particular, sticking of environment particles to the tracer cause it to move more slowly in noninteracting than in interacting active environments, while the anomalous behaviour of the active-in-passive cases stems from a `snowplough' effect whereby a large pile of diffusive environmental particles accumulates in front of a RTP tracer during a ballistic run.
\end{abstract}

\maketitle


\section{Introduction}

The problem of a tagged particle in single-file diffusion has been of interest to physicists for more than five decades~\cite{harris,richards,pincus, sadhu-prl,hegde-pandit-dhar-prl,cividini-kundu}, since its inception in the study of transport through ion channels in cell membranes~\cite{hodgkin54}. Applications include colloidal motion~\cite{bechinger}; motion in super-ionic conductors~\cite{richards}; protein translocation along DNA~\cite{Berg-Natphys};  diffusion in zeolites~\cite{kaerger-ruthven}; and interacting Markov chains~\cite{arratia}. The properties of a tagged particle reveal important features about the environment it is embedded in. {Single file motion takes place when particles move through such narrow channels that no overtaking is allowed, and also in cases where nano-scale objects move along one-dimensional ($1d$) filaments~\cite{cell-book}; the initial order of particles is thereby maintained in single file motion.} This leads to the emergence of long-time correlations in the system, which means that the mean squared displacement of the tagged particle grows for large time as $\sqrt{t}$, in contrast to the linear scaling for freely diffusing particles~\cite{sadhu-prl,bechinger}.  However, this subdiffusive displacement still follows a Gaussian distribution~\cite{bechinger}. Heuristically, one can explain this scaling by saying that at time $t$, the number $n$ of particles in diffusive contact scales as $\sqrt{t}$, while the single-file rule enslaves all the particles to their centre of mass, whose diffusivity is reduced by a factor of $n$ from that of a free particle; thus $d\langle r^2\rangle/ dt \sim t^{-1/2}$. Tagged particle studies have also appeared in problems related to biased one dimensional systems such as asymmetric exclusion process~\cite{shamik} or random average processes~\cite{rajesh-satya}.

Properties of non-interacting {\em active} particles, such as run-and-tumble particles (RTPs)~\cite{Berg-ecoli,mike-jullien-prl-2008,weiss-randomwalk}, typically become indistinguishable from those of non-interacting passive Brownian particles at long times, for suitably chosen parameters~\cite{tirtha-pre}.
A free RTP runs ballistically at speed $v$ in a certain direction, tumbles and chooses a new random direction for the next run; this motion approximates that of
 {\em E. coli} bacteria~\cite{Berg-ecoli}. 
 This leads to temporal correlations in the particle velocity, 
 with a range 
 of the order of the flip time $\gamma^{-1}$~\cite{tirtha-pre}. This finite persistence time leads to several interesting features.  For example, at the single-particle level one finds non-Boltzmann steady states in confining potentials~\cite{dhar-pre,mike-epl}, first order dynamical transitions~\cite{gradenigo-satya}, and as recently shown in \cite{rtp-cond}- in $d$ dimensions, a {\em condensation} transition (between a fluid phase and a condensed phase) where the order of the transition itself can be tuned upon varying $d$ and a given velocity distribution. The behaviour of RTPs in disordered potentials~\cite{kardar-disorder-rtp} and with stochastic resetting~\cite{evans-satya-rtp-reset} have also been studied. Collective motion has been studied at the level of two interacting RTPs~\cite{slowman-prl, yketa, satya-greg1,satya-greg2} and, in more phenomenological terms, for the full many-body problem where hard-core exclusion leads to motility-induced phase separation (MIPS)~\cite{mike-MIPS}.

The problem of tagged active particles 
in crowded environments has also been entensively studied; see~\cite{bechinger-RMP} for a review. Some important works in this context include RTPs interacting via harmonic forces in $1d$~\cite{tagged-rtp,anupam20}, and exclusion in $2d$~\cite{tagged-rtp}; a tagged RTP embedded in a system of passive particles executing symmetric exclusion process~\cite{voituiriez-njp}; and discrete lattice models for bacterial motion~\cite{thompson-mike,soto-golestanian}. Distribution functions of tagged particle positions have also been calculated for systems with smooth potentials~\cite{Szamel}.

In the context of active particles, a hardcore tracer is somewhat like a mobile boundary wall, in the sense that the environmental particles cannot cross it. While the pressure exerted on confining walls has been studied~\cite{fily,mike-natphys}, the question of how hardcore passive and active tracers behave in a sea of RTPs still remains open.  That is the question that we address here.  

To fix nomenclature, we refer to a single {\em{tracer particle}} in a sea of {\em environmental particles}.
We identify eight separate problems (see Table~\ref{table}),
%
depending on whether the environmental particles are hardcore or noninteracting, whether they are active (RTP) or passive (Brownian), and whether the hardcore tracer is itself active or passive. Several of these cases have been studied before,  but to the best of our knowledge, they are yet to be systematically compared under a common umbrella. This is the main goal of the current paper. To make the comparisons manageable we always choose the long-time free diffusivities of active and passive particles to be equal, and start from an equispaced initial condition of a large (but finite) packet of particles, in an unbounded domain, with the tagged particle at the centre.  All particles are {\em pointlike}, so the most important control parameter is initial density of the environmental particles, nondimensionalized by the RTP run length $v/\gamma$.  If all particles are passive then the run length has no meaning, and the density merely rescales the time variable; see Sec. \ref{recap} below.

We use numerical simulations to analyse and compare the eight problems shown in Table~\ref{table}, which we identify by a three-letter short-hand code whose first letter refers to the environment particles as active (RTPs, {\em A}) or passive (Brownian walkers, {\em P}); the second letter does the same for the tracer; and the third labels whether the environment particles are mutually non-interacting ({\em N}) 
or hardcore ({\em H}). We will use lowercase $x$ as a wildcard character when discussing more than one case at a time (e.g. {\em xxN} means all four cases with a noninteracting environment). 

Table~\ref{table} cites prior literature arising in various different contexts, some close to the present one and some not. For instance {\em PPH} has been a cornerstone for tagged particle studies like ours~\cite{sadhu-prl,hegde-pandit-dhar-prl,cividini-kundu}, whereas {\em AAH} has been studied in 
two dimensions~\cite{tagged-rtp}, with periodic boundaries~\cite{soto-golestanian}, or with pre-set maximum occupancy per lattice site~\cite{sepulveda}. {Tracer motion in the context of RTPs under single-file constraint with a finite interaction-range was recently studied in \cite{dasgupta}.}
The {\em PAH} case has been investigated on a lattice in one dimension~\cite{voituiriez-njp}. Cases {\em APH}, {\em APN}, {\em PAN}, like {\em PAH}, all involve mixtures of active and 
passive particles; these are well studied in terms of phase separation~\cite{pritha-mixture, stenhammar-mike} but not single-file diffusion. Similarly {\em APH} and {\em APN} belong to the large family of problems on passive probes in active media, see, {\em e.g.}~\cite{tirtha-plasma, NirGov1, NirGov2}. For non-interacting RTPs and Brownian particles there is of course a huge literature; however there are relatively few studies on hardcore tracers in these systems (cases {\em AAN} and {\em PPN}). Note also that for tracer diffusion, {\em PPN} is identical to {\em PPH} for reasons given in Sec.~\ref{recap} below~\cite{hegde-pandit-dhar-prl, cividini-kundu,Sadhu-Derrida-JStat,redner-book}.

\begin{table}
 \begin{center}
  \begin{tabular}{ |p{2.5cm}|p{2.5cm}|p{2.5cm}|p{2.7cm}| }
  \hline
  Code & {\em Environment}  & {\em Tracer}  & {\em Interaction} \\
  \hline
  {\em PPH}~\cite{sadhu-prl,hegde-pandit-dhar-prl,cividini-kundu} & passive & passive & hardcore\\
\hline
{\em PAH}~\cite{voituiriez-njp,pritha-mixture,stenhammar-mike} & passive & active & hardcore \\
\hline
   {\em AAH}~\cite{tagged-rtp,soto-golestanian,sepulveda} & active & active & hardcore \\
\hline
{\em APH}~\cite{pritha-mixture,stenhammar-mike,tirtha-plasma,NirGov1,NirGov2} & active & passive & hardcore \\
\hline
{\em PPN}~\cite{hegde-pandit-dhar-prl,cividini-kundu,Sadhu-Derrida-JStat}  & passive & passive & non-interacting \\
\hline
{\em AAN} & active & active & non-interacting \\
\hline
{\em APN}~\cite{pritha-mixture,stenhammar-mike,tirtha-plasma,NirGov1,NirGov2,christ-prl,omer-yariv-julien} & active & passive & non-interacting \\
\hline
{\em PAN}~\cite{pritha-mixture,stenhammar-mike} & passive & active & non-interacting \\
\hline
  \end{tabular}
\caption{ Table summarizing alphabetic system codes used in the text. There are eight possible cases, all of which we discuss.}
\label{table}
 \end{center}

\end{table}

Among the main results of our work are the following: 
(i) In hardcore active environments for both active ({\em AAH}) and passive ({\em APH}) tracers, the typical displacement scales as ${t}^{1/4}$ at late times, with a prefactor that scales differently with density from the known {\em PPH} case~\cite{sadhu-prl}; (ii) In non-interacting but dense active environments ({\em AxN}), the typical displacement of a tracer shows deviations from the $t^{1/4}$ scaling law for both passive and active tracers for all times we can access (Fig.~\ref{nonint-var-full}); (iii) An active tracer in a passive environment ({\em PAx}) can have a much larger long-time typical displacement than in the corresponding active medium ({\em AAx}); (iv) In an active environment, adding hardcore interactions to the environment particles ({\em AxH}) can {\em increase} the motion of the tracer compared to the noninteracting counterpart ({\em AxN}). We connect many of these results with nontrivial evolutions of the density profile of particles in the medium around the tracer position, including the sticking of particles to the tracer (in active noninteracting environments, {\em AxN}) and, separately, the `snowploughing' of particles by an active tracer (in passive environments, {\em PAx}).

The rest of the article is organised as follows: In Sec.~\ref{models} we introduce the different interacting models, the numerical schemes used to study them, and the observables. Sec.~\ref{recap} recapitulates known results for the two all-passive cases  ({\em PPx}). Results for {\em AxH}, along with a comparison with {\em PPH} are presented in Sec.~\ref{int-sec}. Then we study the corresponding non-interacting baths with both passive and active tracers ({\em AxN}), and compare with {\em PPN},  in Sec.~\ref{non-sec}. We present our results for an active tracer in passive environments ({\em PAx}) separately in Sec.~\ref{PAH+PAN}. After a discussion in Sec.~\ref{discu} we summarize in Sec.~\ref{summa}. Appendices addressing some technical aspects follow thereafter.
\section{Numerical Strategy and Observables}\label{models}

\begin{figure}
\centering
\includegraphics[width=8cm]{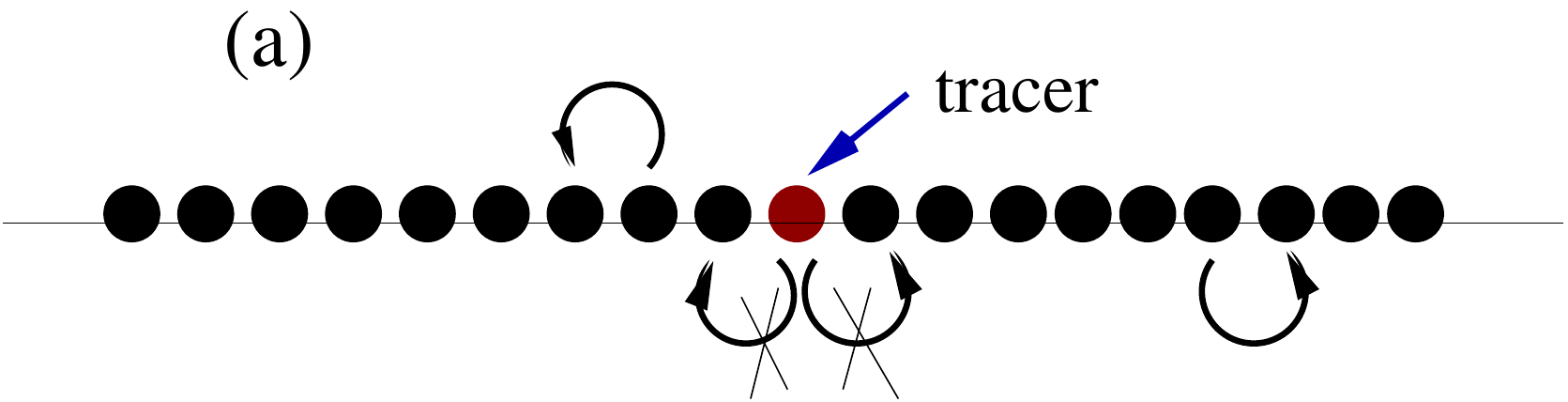}
\includegraphics[width=8cm]{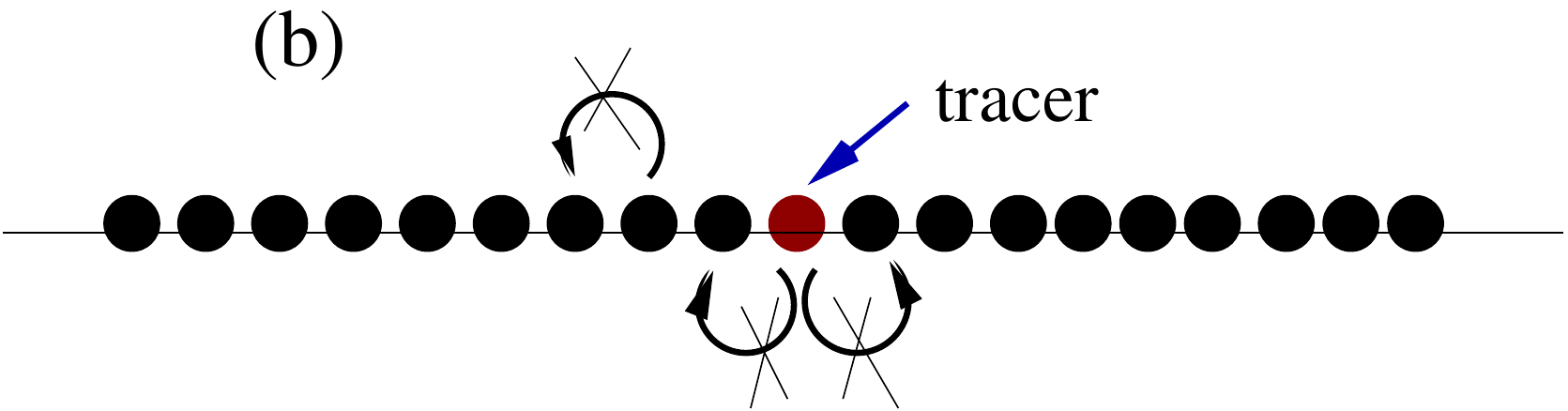}
\caption{Schematic representation of the initial configurations for : (a) non-interacting and (b) hardcore environments. The environment (black particles) as well as the tracer (central red particle) can be either passive or active. The tracer always interacts with its environment through hardcore exclusion. For environments made up of hardcore particles, the initial ordering of particles is maintained forever. In non-interacting environments, the particles can pass through each other but cannot cross the tracer.}\label{model-fig}
\end{figure}

For each case in Table~\ref{table} we investigate the statistics of the position of the tracer. In all our studies, the tracer is the central one among $N+1$ point particles residing initially in a packet spanning $\pm L$ on the infinite line (with the tracer at the origin, $x = 0$). 
The initial condition is fixed (quenched) with all particles 
equispaced at time $t=0$~\cite{sadhu-prl,tirtha-pre,derrida-mft}. The terminology of annealed and quenched initial conditions is borrowed from studies on disordered systems, as a mathematical correspondence exists in averaging over fluctuating ({\em annealed}) or fixed ({\em quenched}) initial configurations/disorder distributions~\cite{derrida-mft}; indeed long-time results for $1d$ Brownian systems are known to be sensitive to the choice of initial conditions~\cite{sadhu-prl,tirtha-pre,derrida-mft,barkai-leibovich}. A schematic of the set-up is shown in Fig.~\ref{model-fig}. The initial density $\rho={N}/(2L)$ is a parameter of the model. For large $N$ the overall density around the tracer does not evolve much on the time scale of our simulations, but finite-size effects due to the packet spreading are detectable in some cases. We discuss these as they arise. 

Our systems evolve by a stochastic (Monte Carlo) dynamics with a discrete time step $\Delta t$. {This might be interpreted as an Euler-type discretization of an underlying continuous-time dynamics, that accounts for the ballistic runs of the RTPs and the hardcore interactions. Alternatively, one may think of active particles whose natural motion involves discrete (finite) steps, so that $\Delta t$ becomes a parameter of the model itself (as opposed to a numerical time step).  These points are discussed further in the later part of this Section.}
Each passive Brownian particle has diffusivity $D$, and we propose MC updates for their positions by drawing Gaussian-distributed random numbers with zero mean and variance $2D\Delta t$. 
Each RTP moves with speed $v$ and its orientation is $\tilde{\sigma}=\pm1$; hence its MC update has proposed displacement {of} $\tilde{\sigma} v\Delta t$. The orientation changes at a Poisson rate $\gamma$~\cite{tirtha-pre}. In the initial state, $\tilde{\sigma}$ is assigned randomly $\pm1$ for each RTP. 
As noted above, we match the long-time diffusivity of free active and passive particles, which means $D={v^2}/(2\gamma)$~\cite{tirtha-pre}.
Particle coordinates are updated sequentially in a fixed cyclic sequence ($1...N+1$). This choice {is more convenient compared to random sequential updates and} in most but not all cases, we find similar results with random sequential updating, see \ref{ran-seq}.

If a proposed MC update does not respect the hardcore interactions of the particles, a collision rule is required.  A collision happens if an MC update causes a particle to move beyond one of the neighbours with which it interacts.
For systems with hard-core environments ({\em xxH}), collisions can take place between any particle $i$ and its neighbours $i\pm 1$ (except that particles $1,N+1$ each have only one neighbour, of course).
For non-interacting environments ({\em xxN}) a tracer update may cause a collision with one of its neighbouring environmental particles; an update of environmental particle can lead to a collision with the tracer (but not with another environmental particle).

In all these cases, the collision rule operates as follows:
%
(i) If the collision happens while updating a passive particle, this particle treats its hardcore neighbours as hard walls and reflects off them. Thus the proposed displacement vector for time increment $\Delta t$ is folded backward upon reaching a hardcore neighbour until the residual part is short enough to require no further folding. (For a particle with only one hard neighbour, there is at most one fold.)  The resultant displacement in $\Delta t$ does not cross any hardcore neighbour. This scheme is chosen because it respects detailed balance for a system of passive Brownian hardcore particles : {If a Monte Carlo move takes a particle from a position say $x_s$ and to say $x_e$ at one time step, then the {\em reverse} move at any other time step would take the particle from $x_e$ to $x_s$.} Another possible scheme is to interchange particle labels as they meet~\cite{hegde-pandit-dhar-prl}; also see~\cite{Rozenfeld-pla}; (ii) If the collision happens while updating an active particle, there is no detailed balance requirement: the displacement of an RTP is  taken (in general) as the minimum of the proposed update and the distance to the nearest hard neighbour in the direction of travel. Hence, when two hardcore RTPs meet head-to-head, the particle being updated stops adjacent to the obstacle particle. Both particles then remain stationary until one or both of them flip(s) direction. If just one particle flips, then both particles subsequently move in this new direction at the same speed. From now on we refer to this as two RTPs {\em sticking} to each other. Instead importantly, by a somewhat different mechanism, an RTP can also stick to a passive particle for as long as the attempted displacements of the latter remain smaller than $v\Delta t$. Since these displacements scale as $(D\Delta t)^{1/2}$ the lifetime of such a stuck pair can depend on the discretization time-step. Such dependences influence the tracer dynamics in certain cases, as we elaborate in Sec.~\ref{tims}. {We note here that our model is tailored to systems where the particles move in discrete (finite) steps in 1D, as happens on actin filaments~\cite{cell-book} or microtubules~\cite{Natbio-mtubl}. As a result, Galilean invariance is broken and there is no concept of momentum or force balance: one left-moving particle can arrest the motion of any number of right-movers piled up to its left. Also we address cases where the natural dynamics involves a discrete timestep, associated for instance with the `walking step' of a myosin motor along an actin filament~\cite{cell-book}.  Although in many regimes the limit of small timestep is well-behaved and one can think of our model as representing a continuous time process, in some cases the timestep remains relevant even when small. Physically, this corresponds to situations where particles' collective motion retains a dependence on the size of their discrete jumps.  In such cases it is important to consider $\Delta t$ as a parameter of the model, and not as an integration time-step for an underlying continuous process.}


To characterise the tracer behaviour, we focus on three observables: (i) the variance of the tracer displacement $X$ as a function of time,
\begin{equation}
\varm \, = \langle X(t)^2 \rangle -  \langle X(t) \rangle^2 ,
\end{equation} which is also its mean squared displacement in symmetric systems where $\langle X \rangle =0$; (ii) the average density profile $\rho(x,t)$ of particles of the medium; and (iii) the probability distribution $P(X)$ of $X$ for late times $t  \gg \gamma^{-1}$.

\begin{figure}
\centering
\includegraphics[width=7cm]{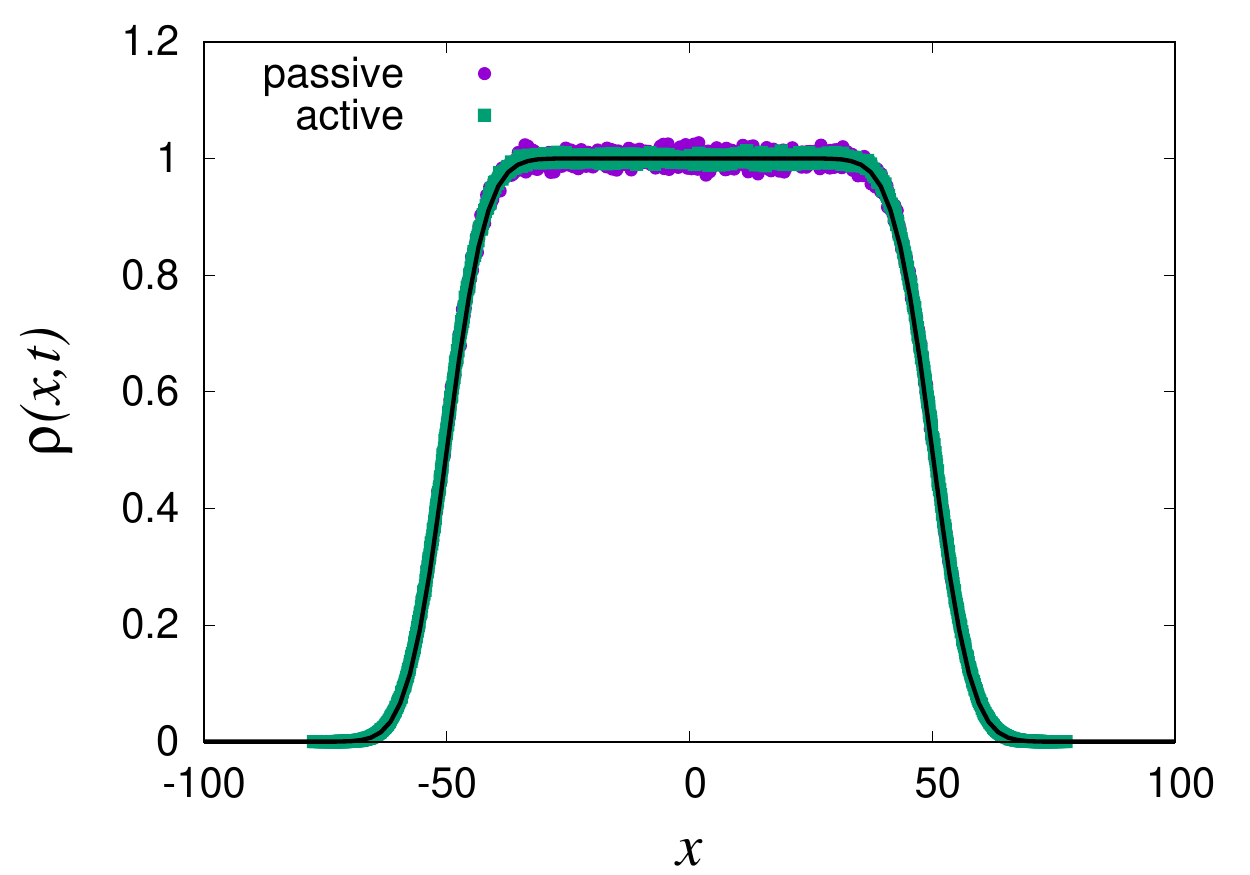}
\caption{Evolution of a top-hat density profile (Eq.~\ref{top-hat}) for free passive ($D=0.5$, purple circles) and active ($v=1=\gamma$, green squares, green dotted lines) particles obtained at $t=40 \gg {\gamma}^{-1}$. Other parameters are $N=100, L=50$. Active and passive results coincide. Solid black line represents Eq.~\ref{top-hat-sol}. The data for the run-and-tumble particles has been averaged over $5 \times 10^5$ realizations, and that for passive Brownian particles over $1\times 10^6$ realizations.}\label{top-hat-free-plot}
\end{figure}

It is useful to recall how a top-hat initial density $\rho(x,0)$ spreads in time for free Brownian particles. This is found by solving the diffusion equation for density $\rho(x,t)$ with initial condition
\begin{equation}\label{top-hat}
\rho(x,0) = \rho_0 ; \,\, \, \, -L \leq x \leq L
\end{equation}
and $\rho(x,0)=0$ otherwise. The solution is 
\begin{equation}\label{top-hat-sol}
\rho(x,t) = \frac{\rho_0}{2}\left[\textrm{erf}\left(\frac{x+L}{\sqrt{4 D t}} \right) - \textrm{erf}\left(\frac{x-L}{\sqrt{4 D t}} \right) \right],
\end{equation}
where $\textrm{erf}(z) \equiv \frac{2}{\sqrt{\pi}} \int_0^{z} e^{-y^2} dy$. In Fig.~\ref{top-hat-free-plot}, we compare the behaviour of free RTPs with  free Brownian particles,  at a late time $t \gg {\gamma}^{-1}$. (The diffusivities are matched via $D={v^2}/2\gamma$, as previously stated.) For both dynamics we have a top-hat initial density on $-50<x<50$ with $\rho(x,0)=1$, describing $N+1=101$ equispaced particles. The results for the active and passive systems are statistically the same, as expected. 

Having specified fully the interactions and discretization strategy, we next study tracer statistics for the various cases enumerated in Table~\ref{table}. We present new results for the active cases in Secs.~\ref{results},\ref{PAH+PAN} below, but first briefly review and numerically confirm some well-known results for the all-passive cases ({\em PPx}), which will serve as a useful benchmark later on. 

\section{Review of Passive Tracers in Passive Environments}\label{recap}

\subsection{PPH: passive tracer in a hardcore passive environment} 

The problem of a tagged particle within a system of hardcore Brownian particles has been extensively studied~\cite{harris, richards, pincus, sadhu-prl, hegde-pandit-dhar-prl, hodgkin54}. The central result is that the variance \var \, of the tracer displacement $X$ scales subdiffusively as $t^{1/2}$. The large-deviation theory for $P(X)$ was developed within the framework of Macroscopic Fluctuation Theory (MFT) in \cite{sadhu-prl}, which showed the robustness of the power law, but also the sensitivity of the prefactor  to the choice of initial conditions. Following \cite{sadhu-prl} and \cite{redner-book} we have, for a `quenched' ({\em i.e.}, equispaced) initial condition of the particles,
\begin{equation}\label{Xvar-sadhugen}
\varm = \frac{1}{\rho}\sqrt{\frac{2 D t}{\pi}}  \equiv \xi_{\text{PPH}}(D,\rho)\;\sqrt{t}\, ,
\end{equation}
thereby defining $\xi_{\text{PPH}}$ as the `coefficient of subdiffusion' for case {\em PPH}.

It is notable that (\ref{Xvar-sadhugen}) is obtained in~\cite{sadhu-prl} using a general theory for tracer motion in a diffusive fluid environment.  The assumptions are those of MFT: On sufficiently long time scales, the collective diffusion of the environment is fully described by two functions which are a diffusivity $D(\rho)$ and a mobility $\sigma(\rho)$.  The physical idea is that the tracer motion is slaved to the collective motion of the environment: combined with the hard core constraint, this leads to long-time subdiffusive motion with a coefficient $\xi$ that depends on $D(\rho)$ and $\sigma(\rho)$.  For Brownian point particles then $D(\rho)=D$ (the free particle diffusion constant) and $\sigma(\rho)=2\rho D$,  in which case Eq.~(18) of \cite{sadhu-prl} reduces to Eq.~(\ref{Xvar-sadhugen}) upto a factor of $\sqrt{2}$ that arises due to the quenched initial condition.

Eq.~\ref{Xvar-sadhugen} holds only at sufficiently late times, because for short times, before meeting another particle, the tracer diffuses freely with $\varm  = 2Dt$. By equating this with $\xi_{\text{PPH}}\,\sqrt{t} $ we  identify the crossover time from early to late regimes as
\begin{equation}\label{PPH-crossover-time}
t_{\text {cr}} \approx \frac{1}{2 \pi \rho^2 D}\,.
\end{equation}
Since these are point particles, diffusion has no timescale of its own, and any change in density $\rho$ can be absorbed by rescaling the time unit (at fixed $D$). This is only true when there is no active particle in the system as holds here for {\em PPx}, but not the other six cases in Table~\ref{table}. 

\begin{figure}
\centering
\includegraphics[width=7cm]{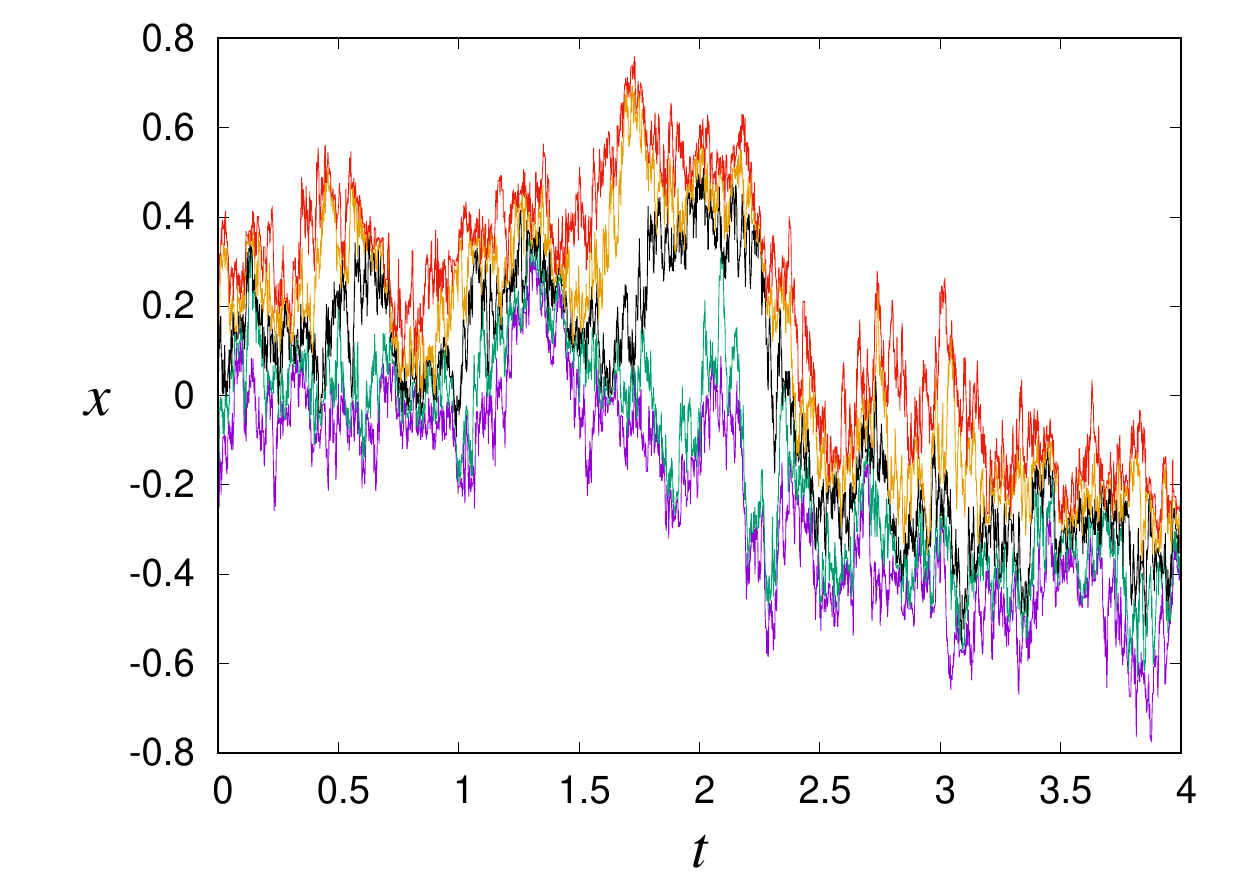}\includegraphics[width=7cm]{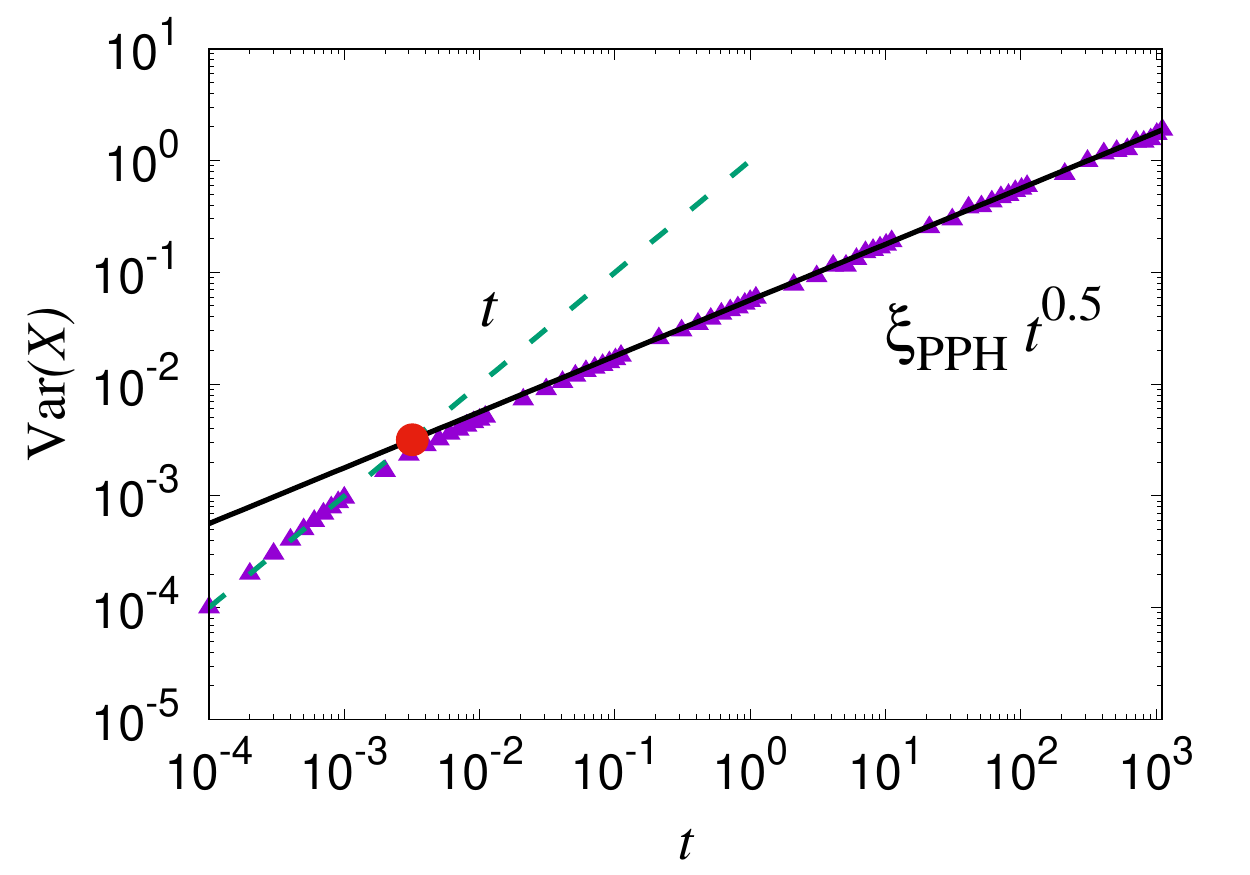}
\caption{(Left) Trajectories for five consecutive diffusive particles within a much larger ($N=1000$) {\em PPH} system. Hardcore exclusion ensures these trajectories never cross each other. (Right) Variance of the tracer position \var \, as a function of time $t$. Solid triangles are the simulation results. The black solid line is Eq.~\ref{Xvar-sadhugen}. The red disc marks the {typical} crossover time $t_{\text {cr}} $ from free diffusion to subdiffusion. Parameters are $D=0.5, N=2000, \rho=10$, $\Delta t=0.0001$. For $t\leq 1$, data averaged over $10^4$ realizations; for $t>1$, data averaged over $900$ realizations.}\label{trajectory-hard-diff}
\end{figure}

To allow later comparisons with these other cases, in Fig.~\ref{trajectory-hard-diff}  we plot trajectories for five consecutive hardcore Brownian particles in the middle of a much larger {\em PPH} system of $N+1=1001$ particles (with the remaining $N-4$ trajectories not shown). We also plot  \var \, for the (central) tracer particle against time, showing the crossover from free diffusion to subdiffusion.

\subsection{PPN--PPH correspondence} 
Case {\em PPN} comprises a system of passive point particles in which the tracer particle cannot cross its neighbours but they can cross each other. However, in {\em PPN} one can choose to relabel the particles' worldines as if they had not crossed (but instead had had a hardcore collision) without changing the Brownian trajectories or their statistics in any other way~\cite{hegde-pandit-dhar-prl, cividini-kundu, Sadhu-Derrida-JStat,redner-book}. Therefore, the environment presented to the tracer in case {\em PPN} is identical to that for case {\em PPH}, so that the tracer statistics are themselves identical at all times. In particular, $\xi_{\text{PPN}} = \xi_{\text{PPH}}$. Put differently, for the two fully passive systems, the only hardcore interaction that matters for tracer dynamics is the one between tracer and environment, which is present in both cases. {We also note that this equivalence of world-lines argument for these two fully passive systems is only exact for the continuous time process (i.e., in the limit of $\Delta t \rightarrow 0$).}

\subsection{Finite-size effects} \label{FSE}
At extremely large times, $Dt \gg L^2$, any finite packet of particles will eventually spread out and reduce in density. 
Since the the non-crossing constraint requires the tracer to occupy the median position among $N+1$ particles, the tracer position ultimately
scales in time like the centre of mass of $n=N$ neighbours. The tracer thus shows free diffusion with diffusivity $\sim D/N$. Hence, there is a second crossover time (dependent on $N$) beyond which linear time scaling for \var \, is restored. In this paper we avoid accessing such timescales as far as possible: we focus on behaviour that is representative of the limit $N,L\to\infty$ (at fixed $\rho$), so the long-time diffusive regime appears as a finite-size effect (because our simulations have finite $N$).  We mention these finite-size effects when necessary to understand our simulation data, with further discussion in \ref{very-late-time-nonint}.

\section{Results for Active and Passive Tracers in Active Environments}\label{results}
In the absence of external forcing or couplings to reservoirs, a collection of passive particles respects local detailed balance; if confined to a finite domain, the system would reach Boltzmann equilibrium. In contrast, if at least one RTP is present, detailed balance is broken. This leads to new phenomena which are absent in the {\em PPx} cases. Our main goal in the rest of this paper is to elucidate the six cases in Table \ref{table} that are not purely passive.  We will compare these with the two all-passive cases just reviewed, wherever this is helpful. 
We first consider hardcore active environments in Sec.~\ref{int-sec}, and subsequently present results for different tracers in non-interacting active media in Sec.~\ref{non-sec}. Then in Sec.~\ref{PAH+PAN} we show how the two cases with an active tracer in a passive environment {\em qualitatively} differ from all the others.

Introducing one or more pointlike RTPs into (or instead of) a set of passive Brownian diffusers introduces two new dimensionless parameters. The first is the diffusivity ratio of passive to active species, $p={2 \gamma D}/{v^2}$, which we always set to unity as stated previously. (This is purely to limit the scope of our study; the dependence on $p$ is certainly interesting in principle.)
 The second is the number of environment particles per RTP run-length, $\tilde\rho \equiv \rho v/\gamma$. This physical density scale prevents us from absorbing $\rho$ by a rescaling of time, as was possible for the all-passive cases  in Sec.~\ref{recap}.

Since we are not interested in either $\gamma\to \infty$ (a limit where RTP particles become Brownian) or $\gamma\to 0$ (where the RTP run length is infinite), then within the subspace $p=1$ we can, without loss of generality, set $v = 1 = \gamma$ for all active particles present. This amounts to a specific choice of units of length and time, which we make from now on. In this case $\tilde \rho = \rho\,; D = 0.5$ and $t_{\text {cr}}^{-1} = \pi \rho^2$ via Eq.~\ref{PPH-crossover-time}. With these choices, the only physical control parameter stemming from activity is simply the density $\rho$. Alongside this we must retain as parameters $N$ or $L$ (governing finite size effects), and $\Delta t$ (governing any time-discretization effects that might arise). 

\subsection{Active hardcore environments}\label{int-sec}
We now address cases {\em AxH}, describing active and passive tracers in active hardcore environments, making comparisons between these and {\em PPH}. 

\subsubsection{AAH: active tracer in hardcore active medium.}
\begin{figure}
\centering
\includegraphics[width=7cm]{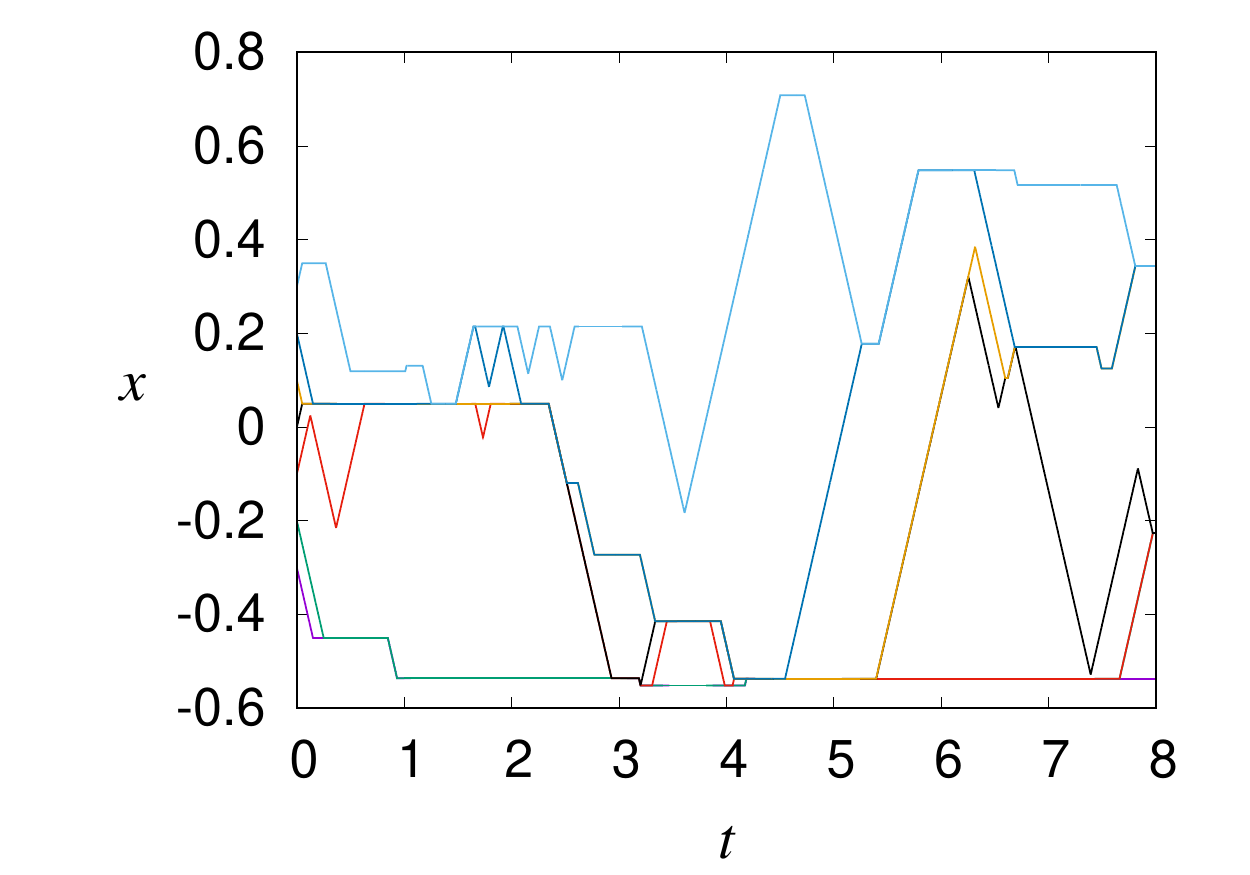}\includegraphics[width=7cm]{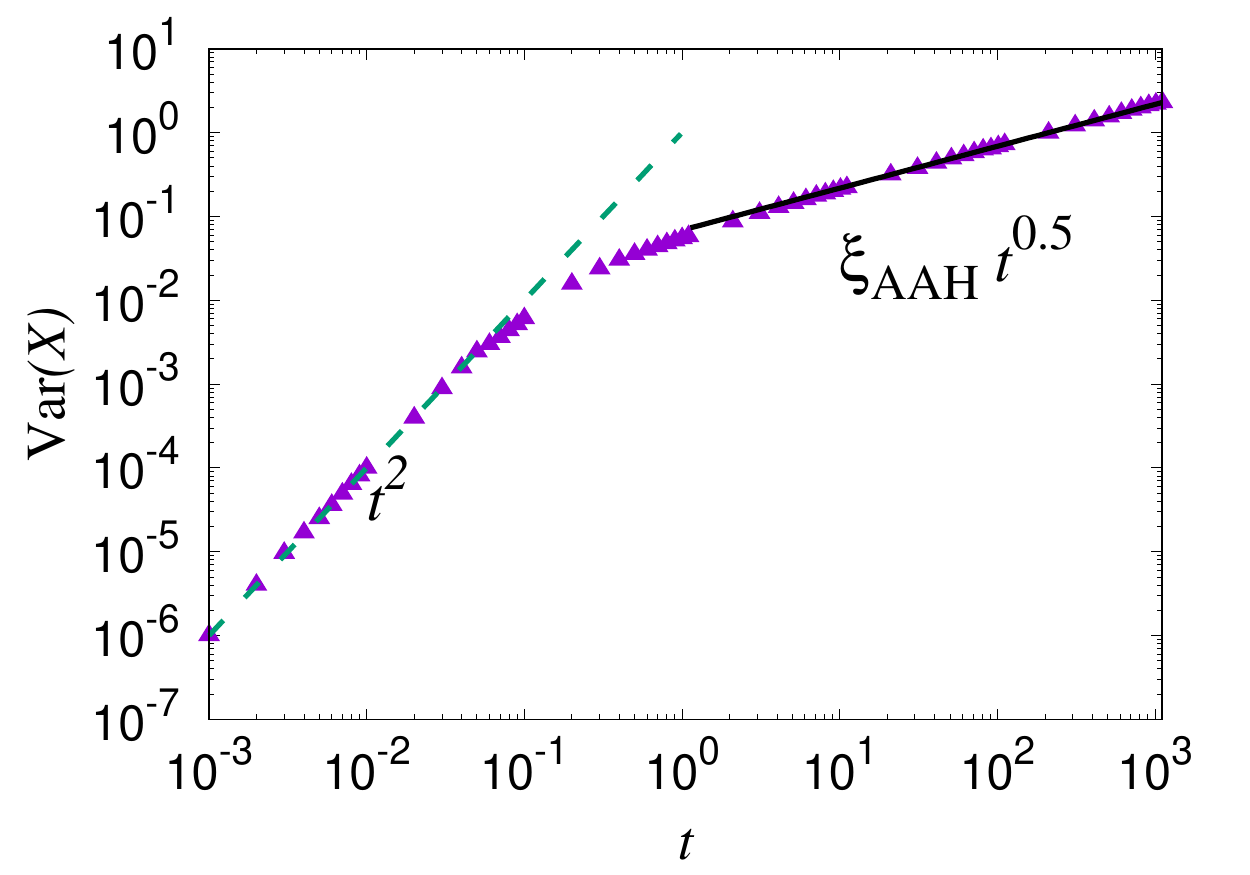}
\caption{(Left) Typical {\em AAH} trajectories for seven consecutive hardcore RTPs (from a total of $N+1=1001$ particles at $\rho = 10$; others not shown). (Right) \var \, as a function of time $t$ (purple triangles). Dashed line: $t^{2}$; solid line is the best $\sqrt{t}$ fit at late times, identifying the subdiffusion coefficient $\xi_{\textrm{AAH}}$. Points are simulation results. The $\sqrt{t}$ behaviour sets in at timescales $t > {\gamma}^{-1}$. Parameters are $\rho=10, N=2000$ and $\Delta t=0.001$. Data averaged over $1.5 \times 10^4$ realizations. Length and time units are chosen so that $v=1, \gamma=1$ (see text); this silently applies also in all subsequent figures.}\label{trajectory-hard-aia}
\end{figure}

The {\em AAH} case describes a collection of $N+1$ point RTPs with mutual hardcore exclusion, where the central particle is the tracer.
Typical trajectories of seven consecutive hardcore RTPs (from the centre of a system with $N = 1000$) are shown in Fig.~\ref{trajectory-hard-aia} (Left). Like the {\em PPH} case in Fig.~\ref{trajectory-hard-diff} these trajectories respect the no-crossing hardcore rule. However, in contrast to {\em PPH}, hardcore RTPs tend to stick to each other on contact. As discussed in Sec.~\ref{models} above, this is due to their finite persistence time: when two RTPs meet head on, they remain stationary until one or both of them flip(s) direction. 

The tracer variance \var ~as a function of time is shown in Fig.~\ref{trajectory-hard-aia} (Right) for a system of $N=2000$ with density $\rho=10$ . At early times, the tracer neither sees its neighbours, nor does it have a significant probability to flip its direction of motion. Thus it moves ballistically and $\varm \sim t^2$. At times longer than the persistence time, $t > {\gamma}^{-1}$, the variance crosses over to the subdiffusive $\sqrt{t}$ scaling, as for {\em PPH}, but with a different prefactor, see Sec.~\ref{prefactors} below.

\subsubsection{APH: passive tracer in hardcore active medium.}\label{superd}
\begin{figure}
\centering
\includegraphics[width=7cm]{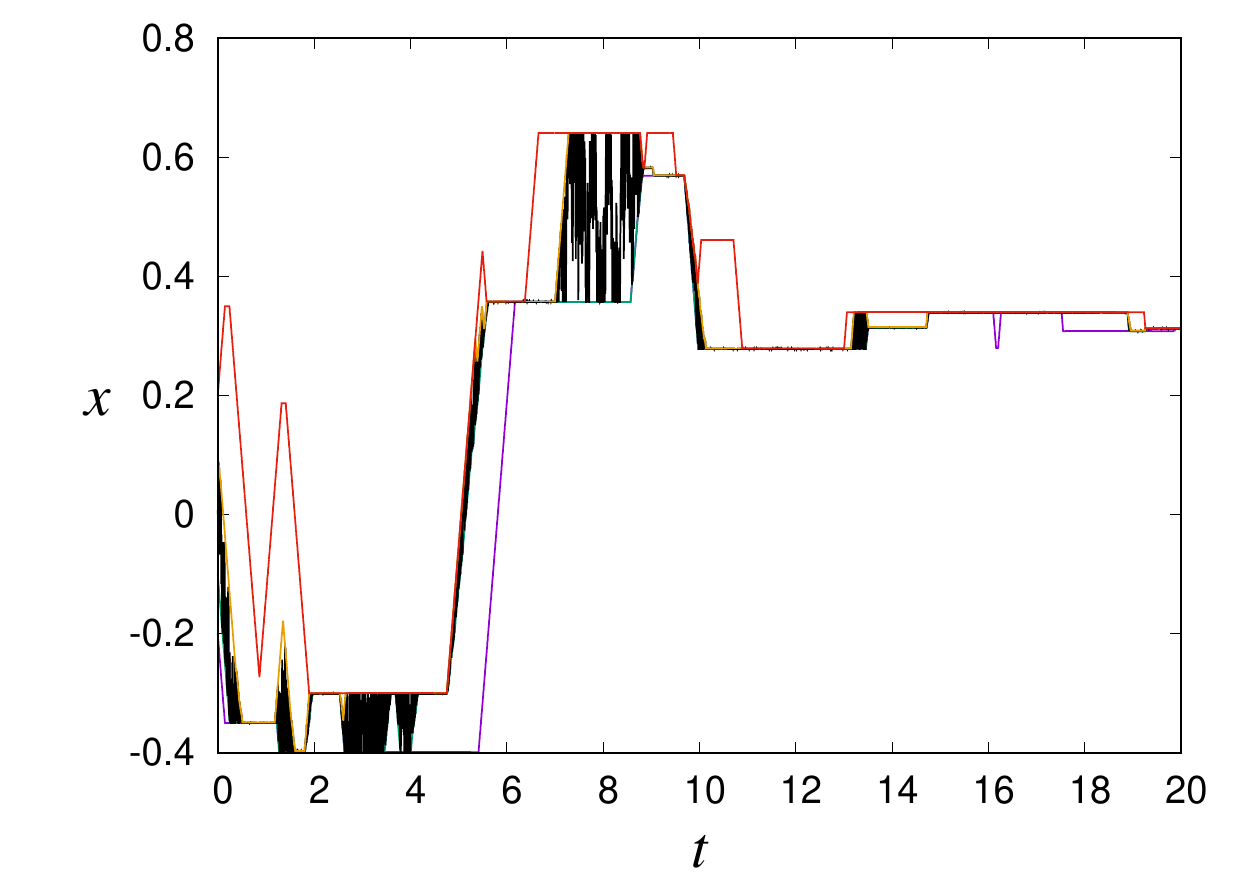}\includegraphics[width=7cm]{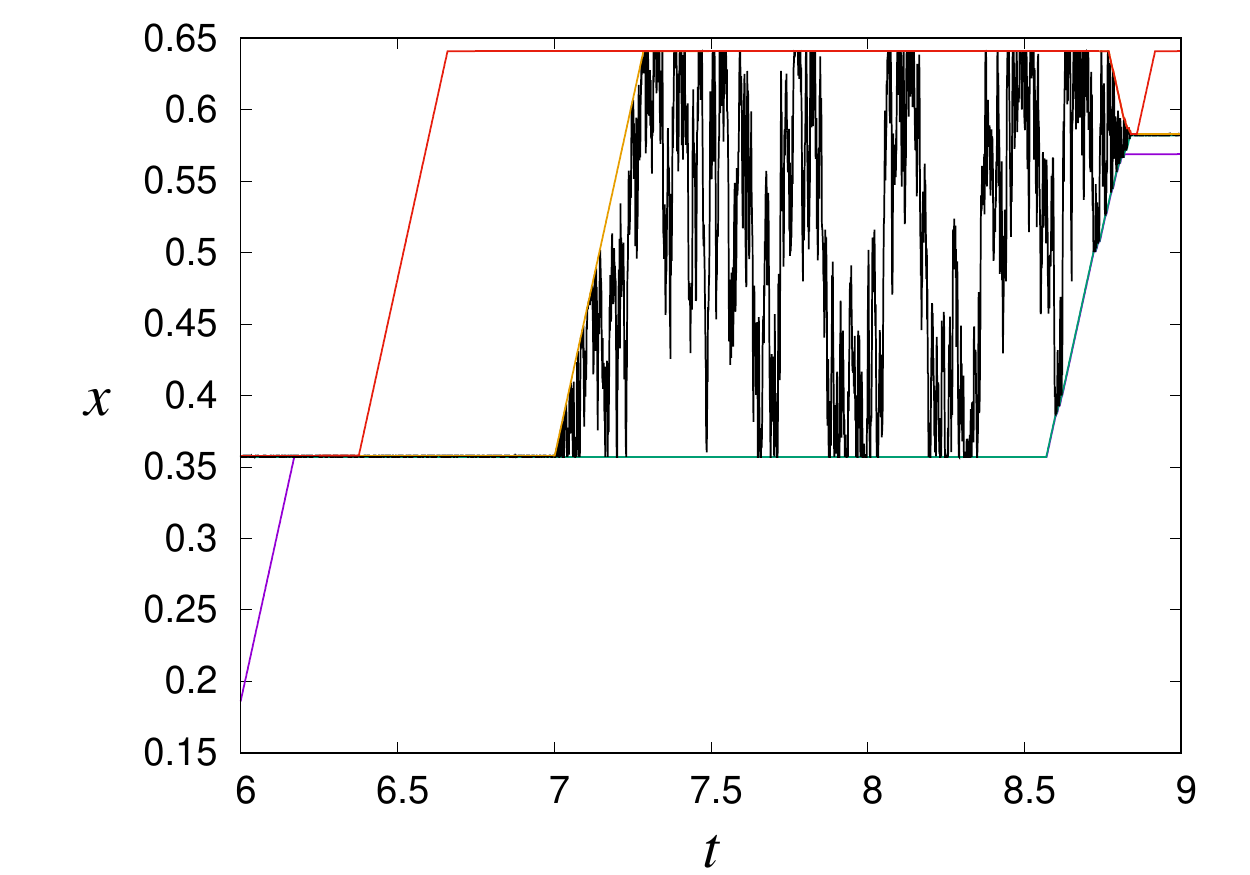}
\caption{(Left) Typical trajectories for the central five particles in an {\em APH} system comprising $N=1000$ RTPs and a centrally located passive tracer, at a density $\rho=10$. (Right) Zoom-in on an episode of free tracer diffusion confined by two RTP particles that are each stationary because of a head-to-head encounter with another RTP. Simulation time-step $\Delta t=0.001$.}\label{trajectory-hard-rtps}
\end{figure}

Typical trajectories for five central particles in a system of $N = 1000$ hardcore RTPs, hosting a centrally placed diffusive tracer (black solid line), are shown in Fig.~\ref{trajectory-hard-rtps} (Left). 
Once again the worldlines do not cross, in accord with the hardcore condition. However, the RTP particles tend to {\em travel together} for significant fractions of the time, including periods when they are static because one or more rightmoving particle lie(s) directly to the left of one or more left movers. Significantly, the passive tracer, while not directly subject to the same `stickiness', can easily become trapped between either head-to-head or comoving RTPs, in which case its Brownian fluctuations are suppressed. On the other hand, there are also epochs where the tracer has a pair of static (head-to-head) RTPs on its left, well separated from another pair on its right, in which case the tracer diffuses within this confining box,~Fig.~\ref{trajectory-hard-rtps} (Right).

\begin{figure}
\centering
\includegraphics[width=7cm]{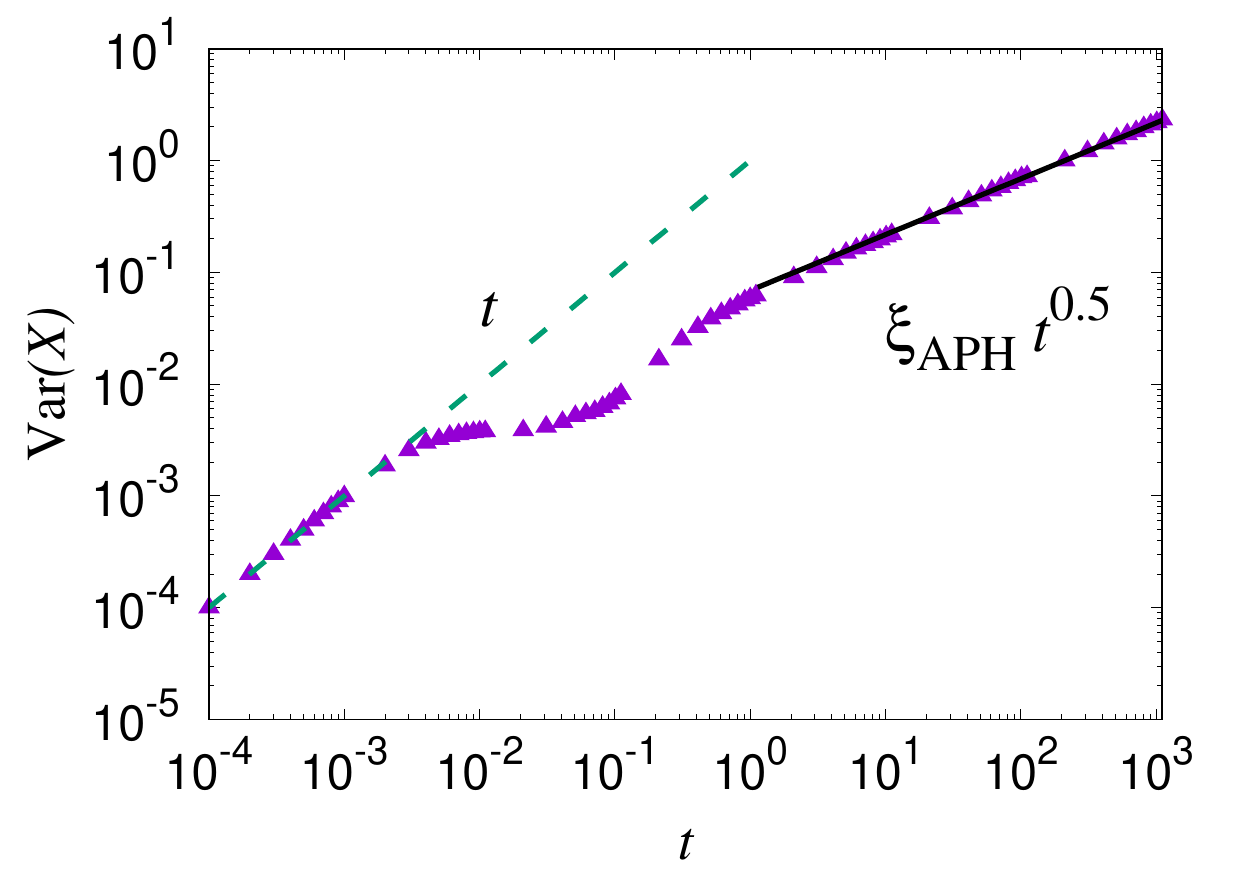}
\caption{Tracer variance \var \,  as a function of time for {\em APH} (a passive tracer surrounded by hardcore RTPs) at density $\rho=10$. Points represent simulation results. At late times, \var \, aproaches $\xi_{\textrm{APH}}\sqrt{t}$. The subdiffusion coefficient $\xi_{\textrm{APH}}$ is a best fit (solid line) to the late-time data. Parameters: $N=2000, v=1=\gamma, D=0.5$. $\Delta t=0.0001$ for $t \leq 10$ and $\Delta t=0.001$ for $t>10$. (Overlap of data for both $\Delta t$ values checked for $t$ in the range $[1,10]$. Smaller $\Delta t$ value taken to extract early-time free diffusive behaviour of the passive tracer.) Data averaged over $10^4$ realizations.}\label{rtp-tracer} 
\end{figure}

Fig.~\ref{rtp-tracer} plots \var \, against time for the passive tracer in case {\em APH}, at density $\rho = 10$. At very early times the tracer is a free Brownian particle: $\varm =2Dt$. Upon encountering its RTP neighbours, which either move ballistically or are stationary until $t \gg \gamma^{-1}$, the diffusive tracer is slowed down drastically. This causes the plateau-like region in Fig.~\ref{rtp-tracer} and illustrates {\em transient caging} of a passive particle by its two active neighbours. The length of the plateau increases with the density of the bath particles, as the passive tracer then encounters its neighbours earlier but still has to wait until $t\sim\gamma^{-1}$ for a neighbour to flip before it can diffuse further. 
The end of the plateau at $t\sim\gamma^{-1}$ thus marks the end of an early time regime during which free diffusion and then caging have both occurred. 
(Note that this plateau indicates a {\em transient} caging effect, in that the size of the cage is primarily determined by the equispaced initial condition.  If the mean-squared displacement is measured instead between times $t_0$ and $t_0+t$ then the results depend significantly on the lag time $t_0$, see \ref{t0}.)

After escaping this transient cage, the tracer movement becomes strongly correlated with the environment particles; at late times, $t\gg \gamma^{-1}$, the latter increasingly resemble hard core diffusers like the tracer itself. Accordingly \var \, scales as $\sqrt{t}$ as found for {\em PPH} (Sec.~\ref{recap}). Similar to {\em AAH}, the coefficient of subdiffusion differs from the passive case, this is discussed next.

\subsubsection{Subdiffusion in AAH, APH and PPH: a comparison.} \label{prefactors}

\noindent {Figs.~\ref{trajectory-hard-diff},\ref{trajectory-hard-aia} and \ref{rtp-tracer} show that 
$\varm  \sim \sqrt{t}$ at large times 
for each of {\em PPH}, {\em AAH} and {\em APH}. We understand this convergence of power laws by arguing that at late times ($t \gg {\gamma}^{-1}$) each RTP performs an effective diffusion between its neighbours; the microscopic details of individual particle motions get lost, 
 and the RTP environment behaves as a diffusive fluid in the sense of MFT, as discussed in Sec.~\ref{recap}.
In this case the arguments of~\cite{sadhu-prl} provide a natural explanation of this subdiffusive behavior.
However, the collective diffusion of the active environment is not the same as for the passive case, so the coefficient of subdiffusion $\xi$ differs.
In Fig.~\ref{hardcore-variance-compare} we plot $\xi_{\textrm{AAH}}$ and $\xi_{\textrm{APH}}$, found by fitting the late time data to $\sqrt{t}$, as a function of $\rho$. These quantities precisely coincide with each other, but not with $\xi_{\textrm{PPH}}$.
{The fact that $\xi_{\textrm{AAH}}=\xi_{\textrm{APH}}$ is consistent with the idea that the tracer motion is controlled by the collective dynamics of the environment -- this is essential for the MFT analysis of~\cite{sadhu-prl} which relates the tracer motion to the environmental diffusivity and mobility.  (We will show below that the {\em PAH} system differs strongly from {\em AAH} in this respect.)

Moreover, the deviation between {\em PPH} and {\em AxH}
is not a constant multiplier, but instead a factor that itself depends on density, which presumably originates in the mobility and diffusivity of the active environment.}  Indeed from Eq.~\ref{Xvar-sadhugen},
$\xi_{\textrm{PPH}}\propto {\rho}^{-1}$, and while this scaling is recovered for $\rho \ll 1$ (to be precise $\frac{\rho v}{\gamma} \ll 1$) where RTPs reach a diffusive limit before encountering any neighbours at all, the plot of $\xi_{\textrm{AAH,APH}}(\rho)$ shows appreciable downward curvature throughout the studied range {($0.1 < \rho \leq 40$)}. This quantitative difference can be intuitively understood by considering the dimensionless quantity $\frac{\rho v}{\gamma}$ which can in turn be interpreted as a ratio of two competing timescales, $t_{\textrm{flip}}=\frac{1}{\gamma}$ and $t_{\textrm{int}}=\frac{1}{\rho v}$. The diffusive ({\em PPH}) limit is recovered when $t_{\textrm{flip}} \ll t_{\textrm{int}}$. For $t_{\textrm{flip}} \sim t_{\textrm{int}}$ ballistic hard-core collisions arise (and persistence comes into play) between RTPs before their diffusive regime gets established. If $t_{\textrm{flip}} \gg t_{\textrm{int}}$, the system will effectively be jammed for long observational time-windows(we do not access such limits in this paper).

Given these differences, it is notable that, as exemplified in Fig.~\ref{hardcore-variance-compare}, in the late time regime the distribution $P(X)$ for the {\em AAH} and {\em APH} cases is Gaussian (to within numerical error), just as it is known to be (exactly) for {\em PPH}~\cite{sadhu-prl}. The emerging picture is therefore of equivalence between these three cases at long times, up to a density-dependent renormalization of the coefficient of sub-diffusion in the hardcore active environment. This renormalized sub-diffusion coeffcient that results from an interplay of the density $\rho$ and the active run length $\frac{v}{\gamma}$ (or equivalently between $t_{\textrm{flip}}$ and $t_{\textrm{int}}$), is a {\em collective} effect and cannot be understood from a single-particle picture (as the inter-changing worldlines argument for the PPH case also does not work for two hardcore RTPs). This reinforces the idea that in fully hardcore active systems the dynamics of a tracer is enslaved to the environment so that its own properties are almost immaterial. It is important to recall here the Percus relation which works quite generally for hardcore {\em passive} Brownian particles : the mean-squared displacement of a particle in the hardcore system is equal to the ratio of the average displacement of the corresponding free particle to the density of the hardcore environment~\cite{percus,barkai-percus}. In an active environment composed of hardcore RTPs, we infer from Fig.~\ref{hardcore-variance-compare} that the Percus relation no longer remains valid.

\begin{figure}
\centering
\includegraphics[width=7cm]{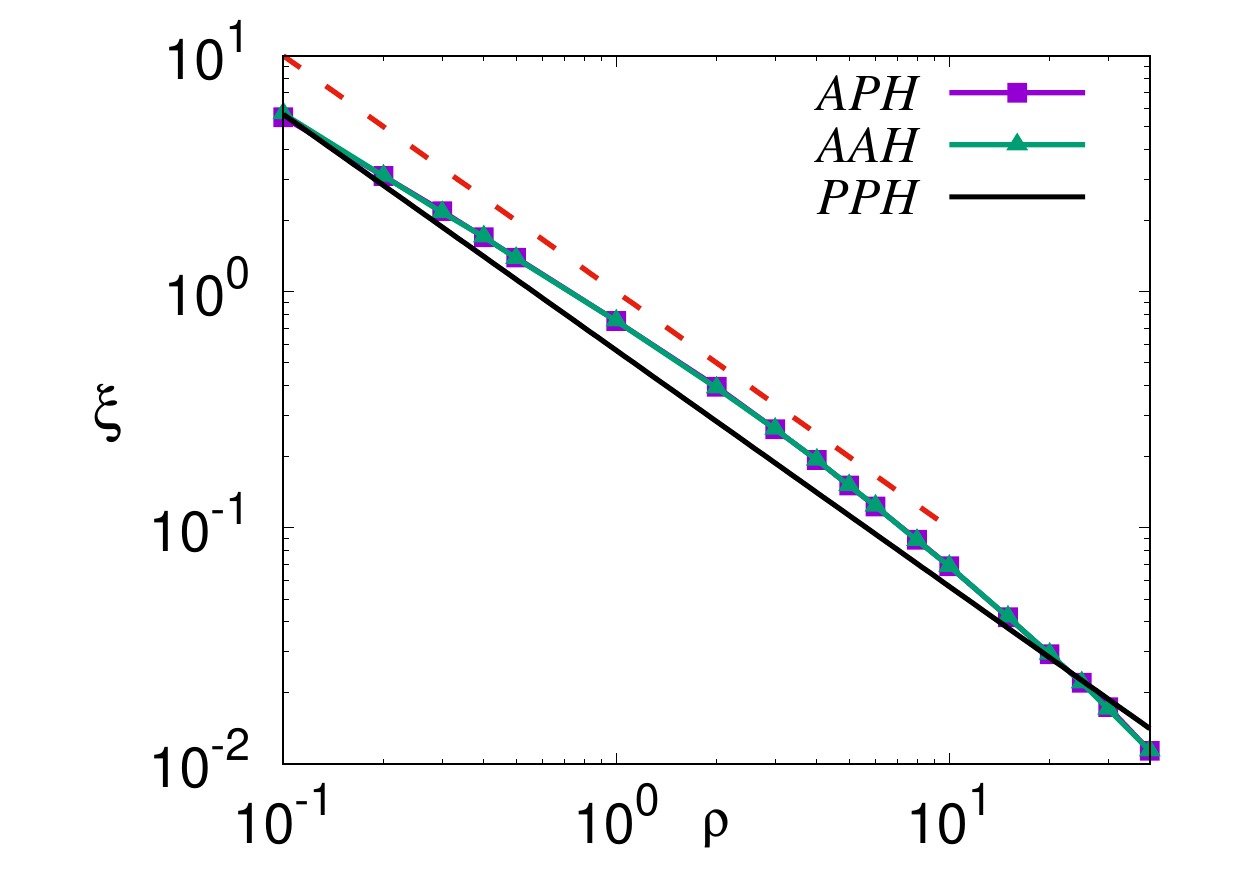}\includegraphics[width=7cm]{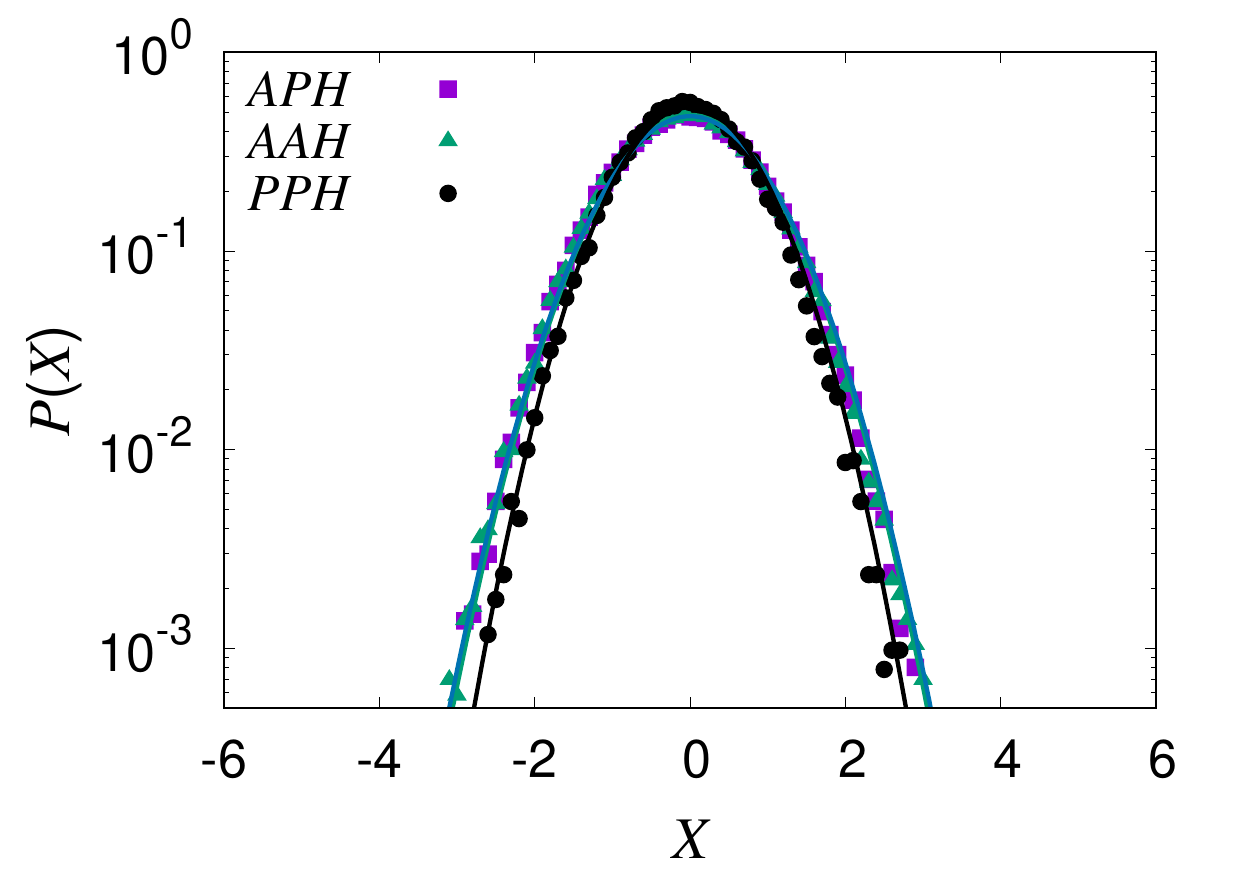}
\caption{(Left) Comparison of simulation results for the coefficient of subdiffusion $\xi$ in {\em APH} and {\em AAH} cases.  $D=0.5, N=2000,v=1=\gamma, \Delta t=0.001$. All data averaged over at least $1.5 \times 10^3$ realizations. Solid line: $\xi_{\textrm{PPH}}$ from Eq.~\ref{Xvar-sadhugen} with $D=0.5$. Dashed line $\propto\frac{1}{\rho}$ is to guide the eye. (Right) Histogram (on a semi-log scale) of tracer position $X$ at late time $t=100$ with $\Delta t=0.001$ and $N=2000$ environment particles at initial density $\rho=10$. {\em AAH}, {\em APH} and {\em PPH} data averaged over $8.6 \times 10^4$, $8.8 \times 10^4$ and $9 \times 10^4$ realizations, respectively. The solid lines (with the same colour codes as points) are Gaussian fits to each case with the respective variances calculated numerically.}\label{hardcore-variance-compare}
\end{figure}

\begin{figure}
\centering
\includegraphics[width=7cm]{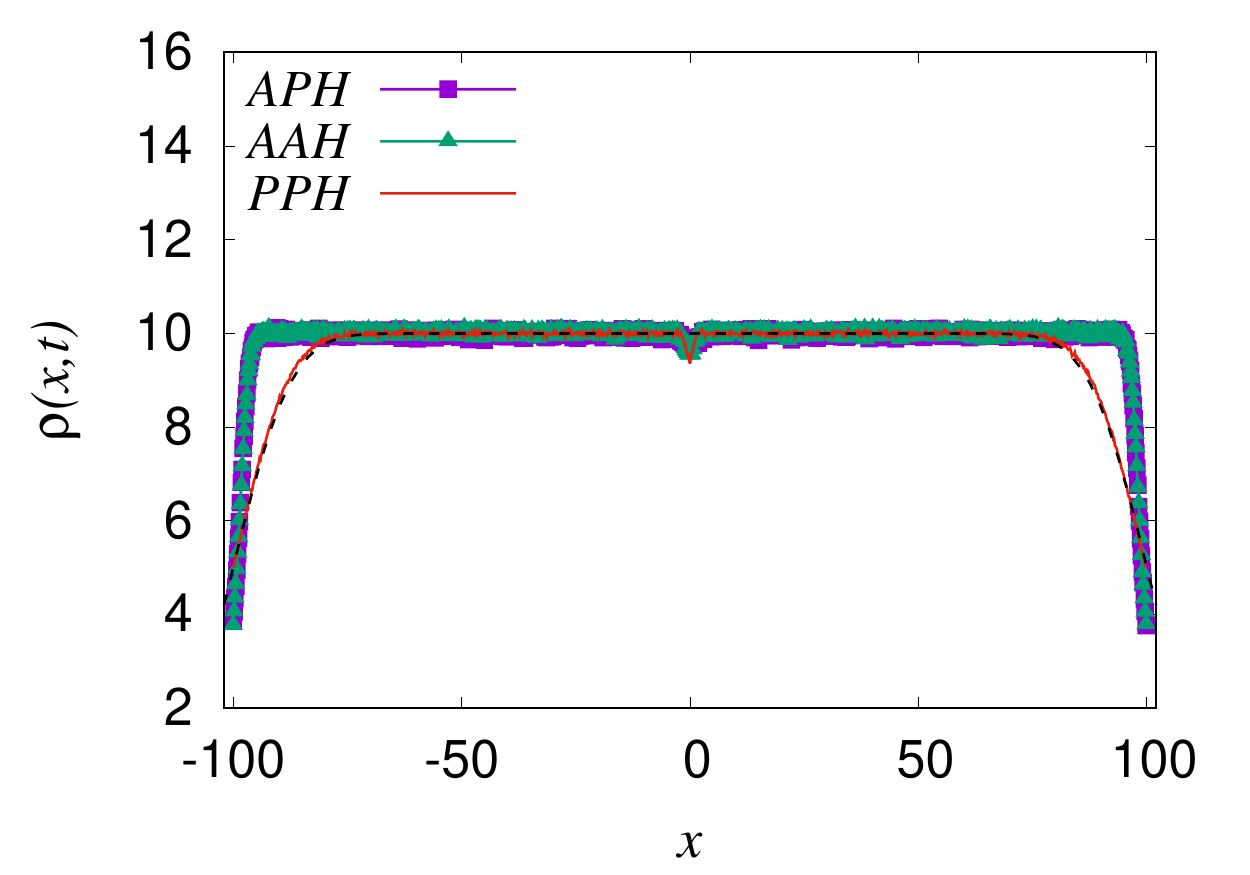}
\caption{Density profile of environment particles ($N=2000$) at a late time $t=100$ for cases {\em PPH}, {\em AAH}, and {\em APH}. $v=1=\gamma, D=0.5, \Delta t=0.001$. Initially we have $L=100$, therefore $\rho(x,t=0)=10$. While the {\em AAH} and {\em APH} density profiles are nearly indistinguishable, they are markedly different from the {\em PPH} profile. The {\em PPH} density profile converges on the result in Eq.~\ref{top-hat-sol} (dotted black line, mostly overlaying {\em PPH} data; also see Fig.~\ref{top-hat-free-plot}). {\em AAH}, {\em APH} and {\em PPH} data averaged over $1.7\times 10^5$, $1.6 \times 10^5$ and $5\times 10^4$ realizations, respectively. The data was sampled only in the range shown in the plot. }\label{hardcore-density-compare}
\end{figure}

With this in mind, we now focus on the environment particles, to mechanistically elucidate the tracer statistics shown in Fig.~\ref{hardcore-variance-compare}.  Density profiles of the environment particles are shown in Fig.~\ref{hardcore-density-compare} at a late time $t \gg {\gamma}^{-1}$, in the subdiffusive regime where \var \, scales as $\sqrt{t}$. Consistent with our arguments, the results for the {\em AAH} and {\em APH} cases coincide, while the {\em PPH} density profile remains distinct.
The latter matches Eq.~\ref{top-hat-sol} modulo a residual small dip at the origin (as the position of the tracer has been ignored). This outcome arises because, as discussed in Sec.~\ref{recap}, the hardcore interaction does not alter the worldline statistics in a 1d gas of purely Brownian point particles. 

However, this changes drastically for an environment comprising RTPs, whose orientational persistence can lead to jamming and causes nontrivial density correlations throughout the system. These correlations directly affect the late-time {\em collective motion} of the environment particles. 
For the chosen initial density ($\rho=10$) in Fig.~\ref{hardcore-density-compare}, we obtain a
 much slower spreading of the initial top-hat density profile when compared to  {\em PPH} case. 
(This effect can be understood by considering the rightmost particle that has orientation $\tilde{\sigma}=-1$: this particle acts as a (temporary) hard wall for the remainder of the particles, which tends to confine them, until such time as it flips its orientation.  By contrast, passive particles are in constant diffusion motion, so they are much less effective in blocking other particles.  This effect hinders spreading; it may be interpreted as a macroscopic analogy of the transient caging effect described above.)

{We also note that while our {\em AAH} situation bears close resemblance to the single-file RTP system studied in \cite{dasgupta}, our point-like particles are strictly hardcore, in contrast to those of \cite{dasgupta} where they have a finite interaction-range. Thus in our case, we do not have spatially extended clusters. Note also that the particles in \cite{dasgupta} interact by physical forces within a Langevin description, which leads to significantly different dynamics for particle clusters, compared to this work (recall Sec.~\ref{models}). This, we believe, leads to a different scaling for our coefficient of sub-diffusion with density, as compared to the one in \cite{dasgupta}.}


\subsection{Active noninteracting environments}\label{non-sec}
Having dealt with active hardcore environments we next address cases {\em AxN} in which the environment comprises active but noninteracting particles. (As always, the tracer has hardcore interactions with its environment and the active and passive diffusivities are matched.) We will make comparisons with the all-passive counterpart {\em PPN} -- which as explained in Sec.~\ref{recap} is equivalent with {\em PPH} for our purposes.

The removal of interactions among active environment particles raises important questions. First,  does the $\sqrt{t}$ law for the variance of a passive tracer continue to hold?  Second, what if the tracer is itself active? And third, if the $\sqrt{t}$ law does hold, then how does the coefficient of subdiffusion get modified in each case?

\subsubsection{Typical trajectories.}

To start to answer these questions, we show typical trajectories of environmental and tracer particles in {\em AAN} (Fig.~\ref{AAN-traj}) and {\em APN} (Fig.~\ref{APN-traj}), preceded for comparison by the relevant passive benchmark {\em PPN} (Fig.~\ref{PPN-traj}). In each case we show results for two different densities, $\rho=1$ (left panel) and $\rho=10$ (right panel). As usual we give trajectories for a few ($n = 5$ or $7$) particles from the middle of a larger ($N = 1000$) system with the tracer particle in the centre. Note however that these cease to be the middle $n$ particles over time because the ordering of the particles is no longer fixed by the hardcore interaction: environment worldlines can cross each other (but of course {\em not} that of the tracer).

\begin{figure}
\centering
\includegraphics[width=7cm]{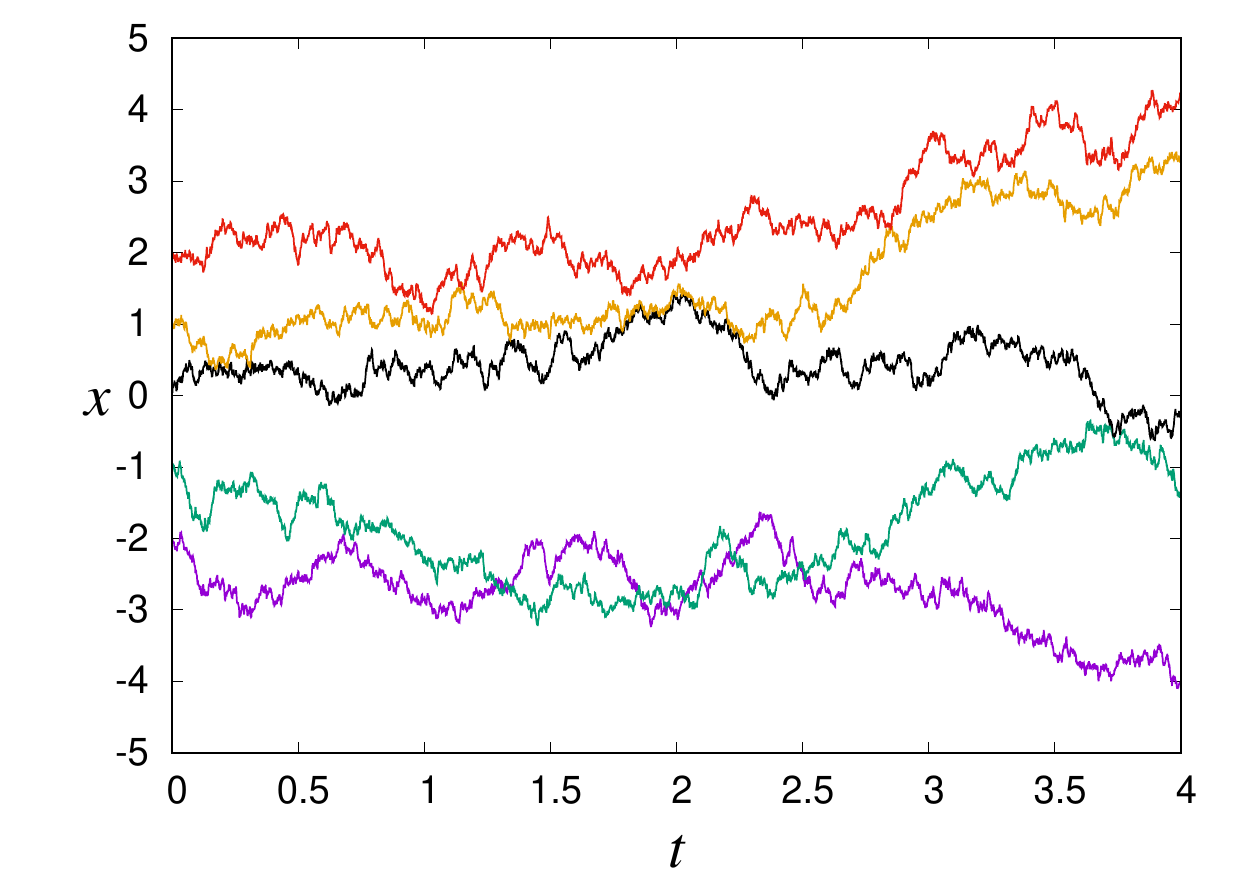}\includegraphics[width=7cm]{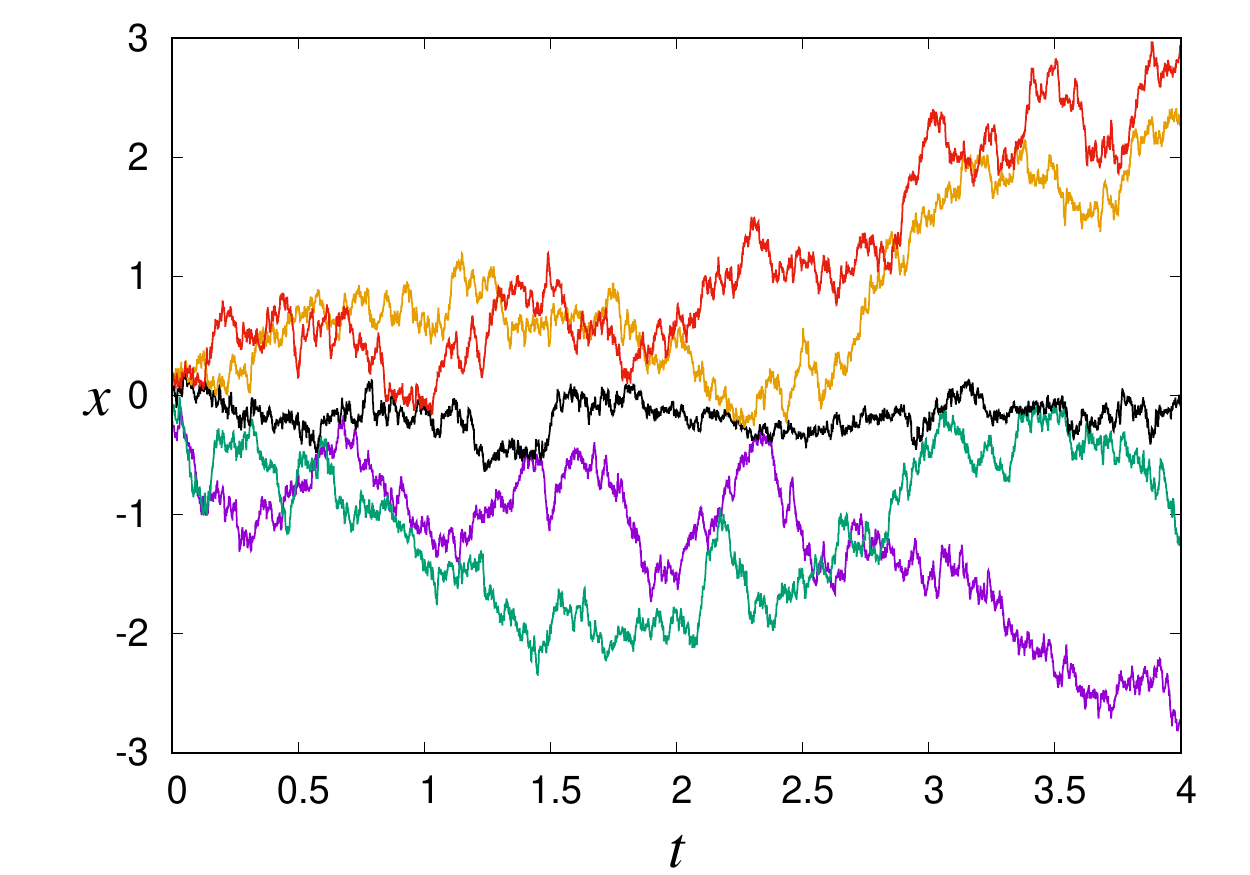}
\caption{Trajectories for five central particles in the {\em PPN} case with $N=1000 , \Delta t=0.001$, at density $\rho=1$ (Left) and $\rho=10$ (Right). The central black line is the tracer trajectory.  Since the environment particles have zero persistence, the passive tracer is able to make large excursions, even at high density, unlike in dense active environments (compare Fig.~\ref{APN-traj} (Right)).}\label{PPN-traj}
\end{figure}

\begin{figure}
\centering
\includegraphics[width=7cm]{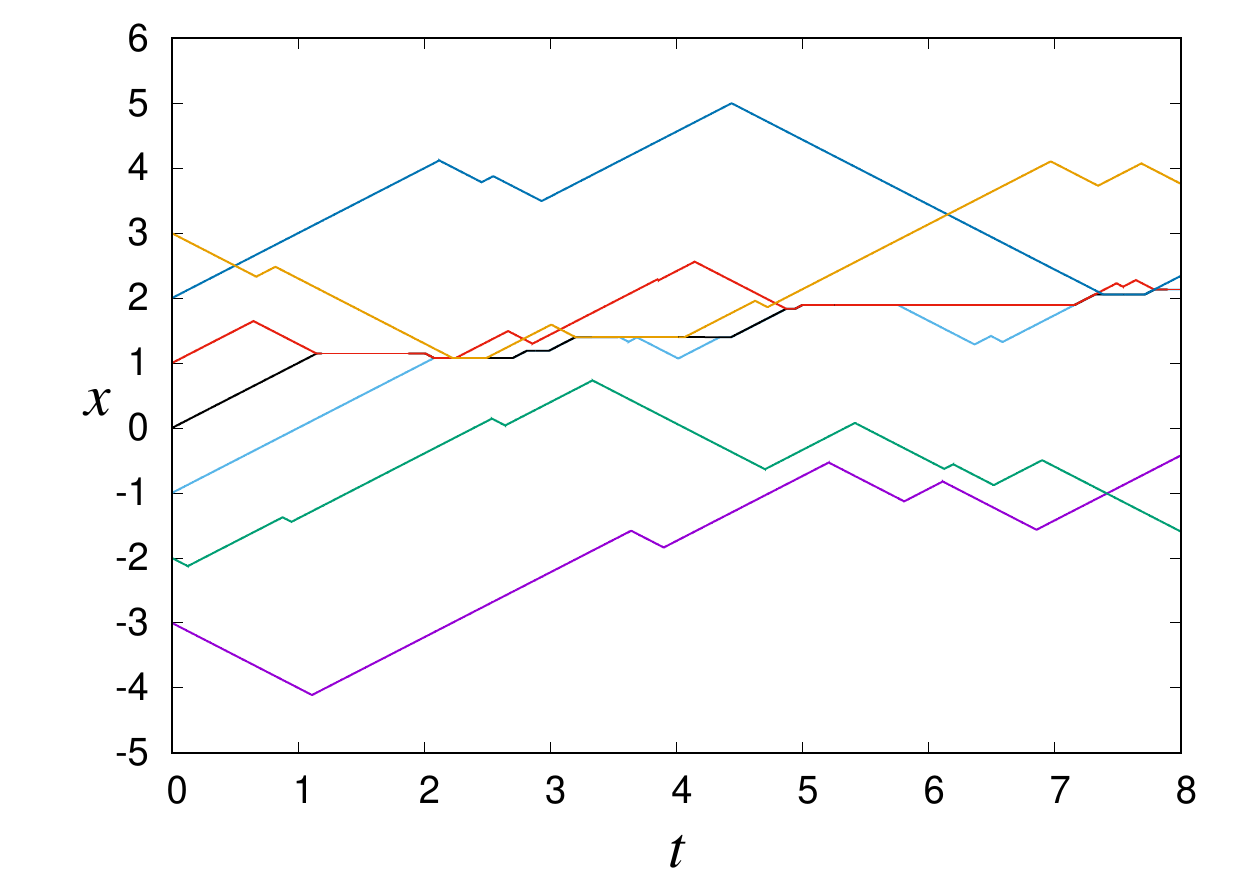}\includegraphics[width=7cm]{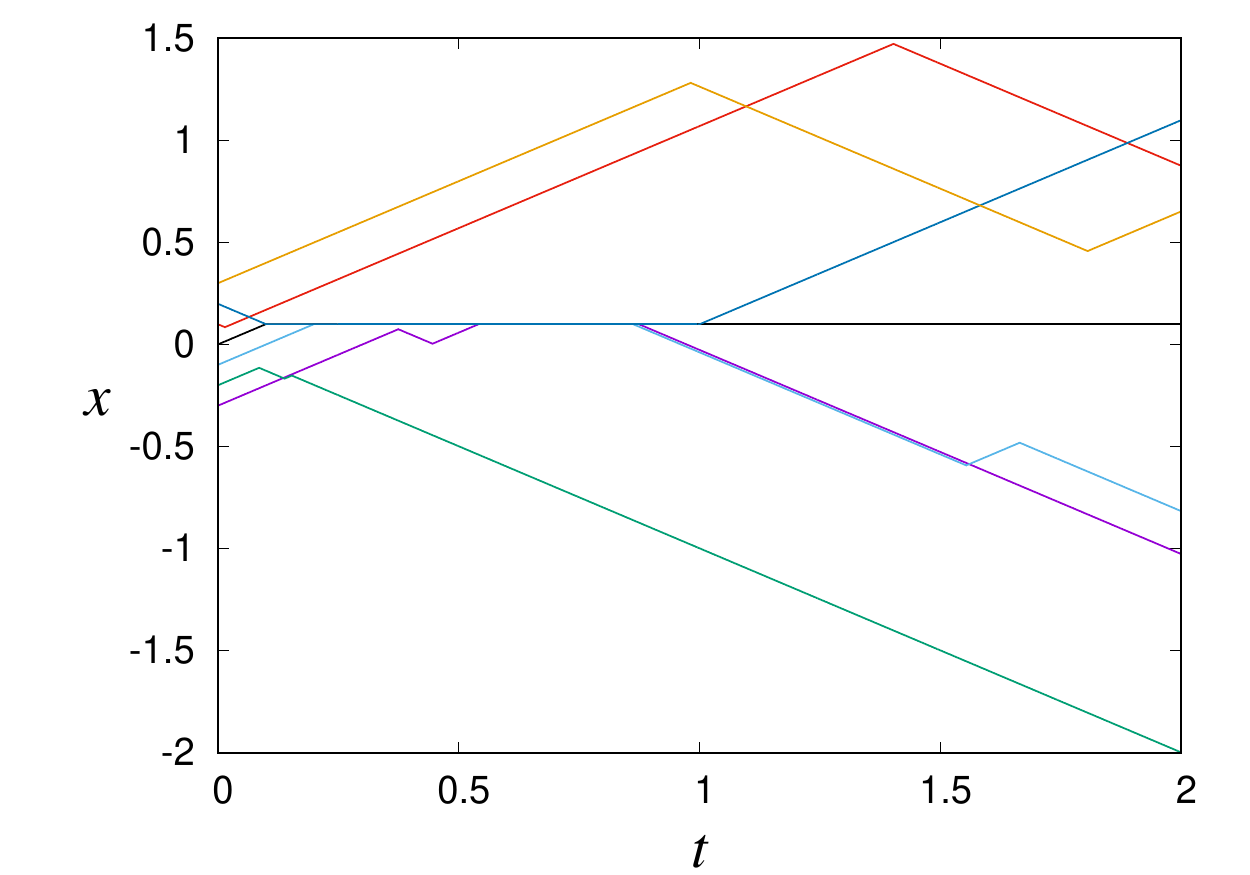}
\caption{Trajectories for seven central particles in the {\em AAN} case with $N=1000, \Delta t=0.001$ at density $\rho=1$ (Left) and $\rho=10$ (Right). The central black line is the tracer trajectory. The active tracer can make long excursions at low environment density, but in a highly dense active environment (Right), becomes increasingly immobile due to frequent encounters with another persistent particle.}\label{AAN-traj}
\end{figure}

\begin{figure}
\centering
\includegraphics[width=7cm]{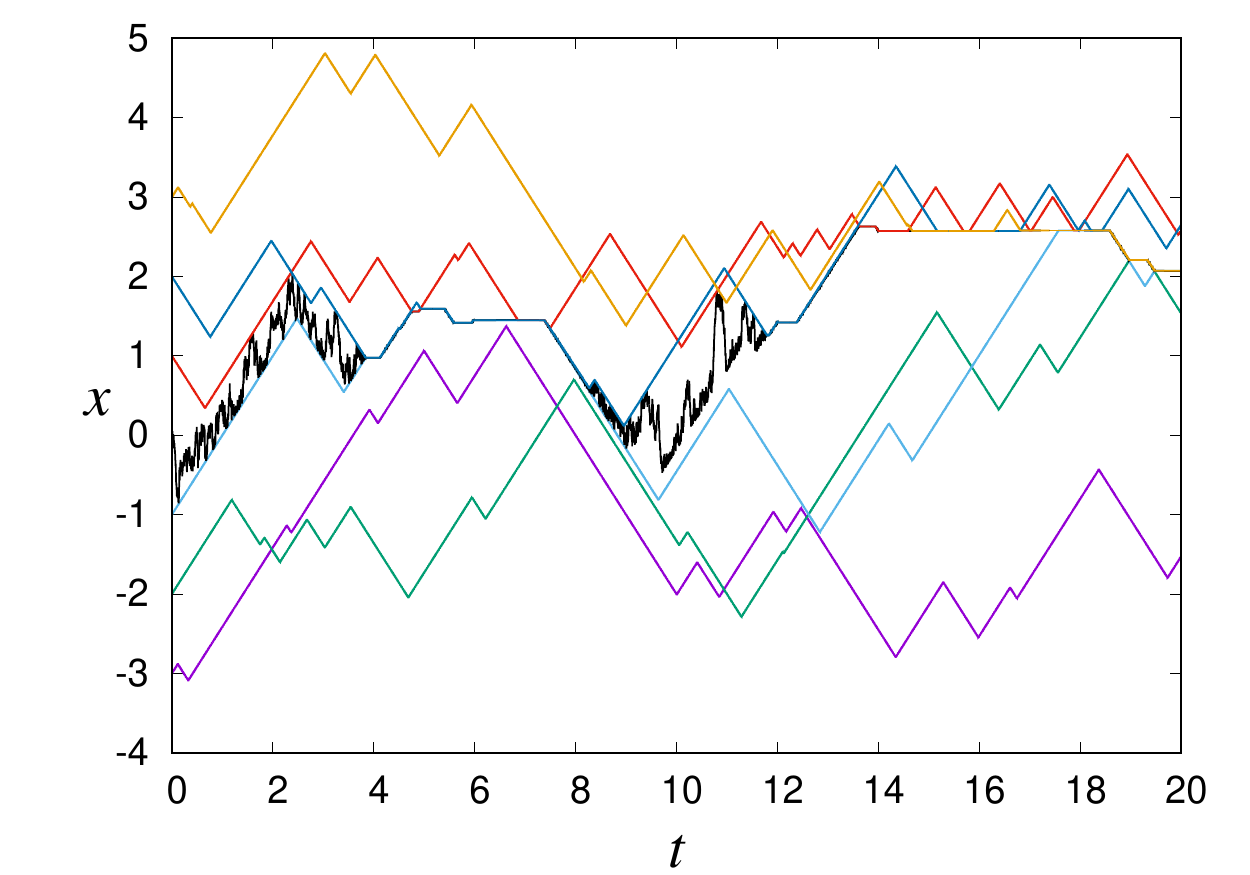}\includegraphics[width=7cm]{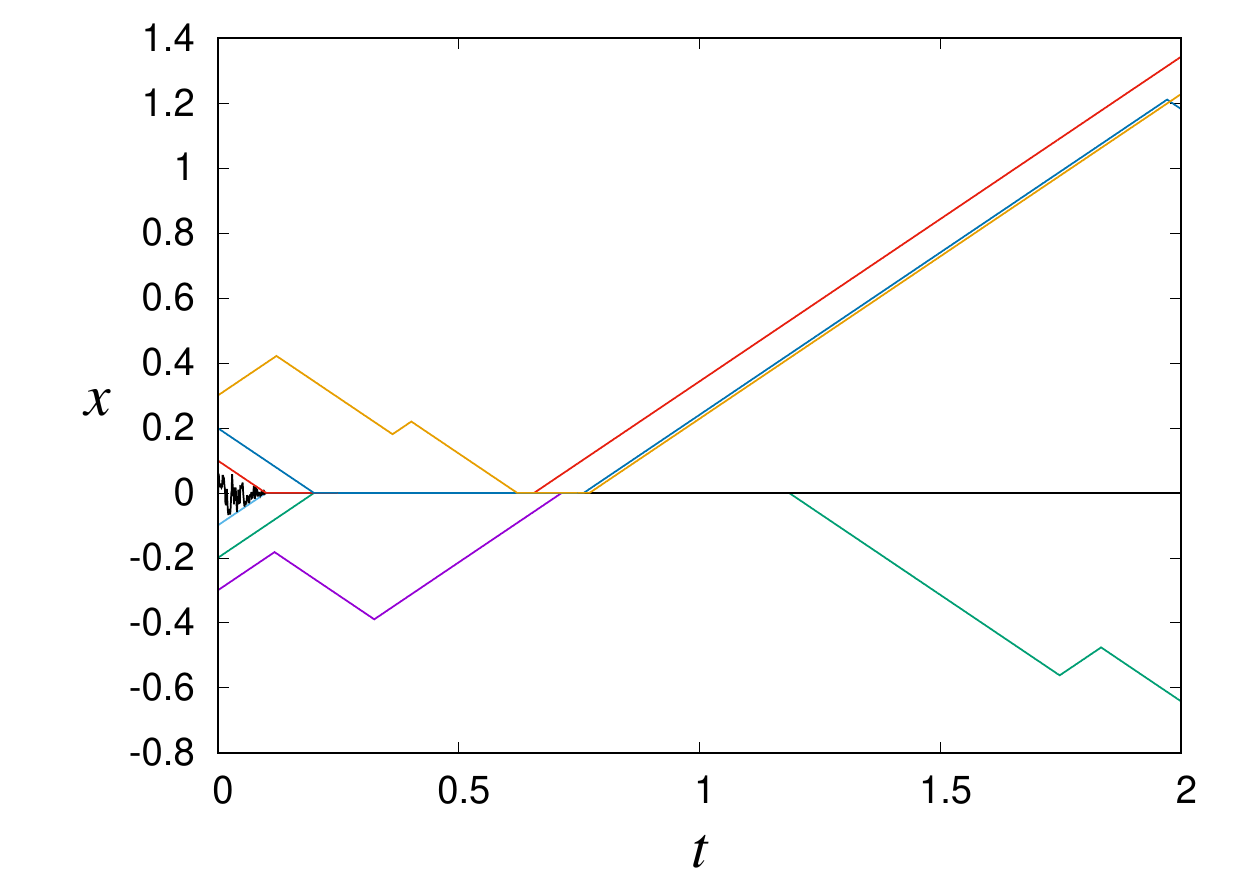}
\caption{Trajectories for seven central particles in the {\em APN} case with $N=1000,\Delta t=0.001$ at density $\rho=1$ (Left) and $\rho=10$ (Right).  Similar to the {\em AAN} situation in Fig.~\ref{AAN-traj}, clustering of RTPs around the passive tracer makes it effectively immobile for long durations in dense active environments.}\label{APN-traj}
\end{figure}

Interestingly, in the active environment, density has a more pronounced effect on both active and passive tracers, compared with {\em PPN}.
 That is, comparing Figs.~\ref{AAN-traj},\ref{APN-traj} with Fig.~\ref{PPN-traj}, the tracers in the active environments make comparatively longer excursions at low environment density (left panels) but become nearly immobile for high density (right panels). This is because the active particles of the environment can stick to the hardcore tracer; an active tracer stops moving when head-to-head with an active environment particle whereas a passive tracer gets stuck between two head-to-head neighbours. As the density is raised, the probability of such encounters increases and the tracer's movement becomes more and more restricted. This effect is amplified by having free environment particles rather than hardcore ones which would otherwise tend to arrest each other rather than the tracer (as discussed further below). 

\begin{figure}
\centering
\includegraphics[width=7cm]{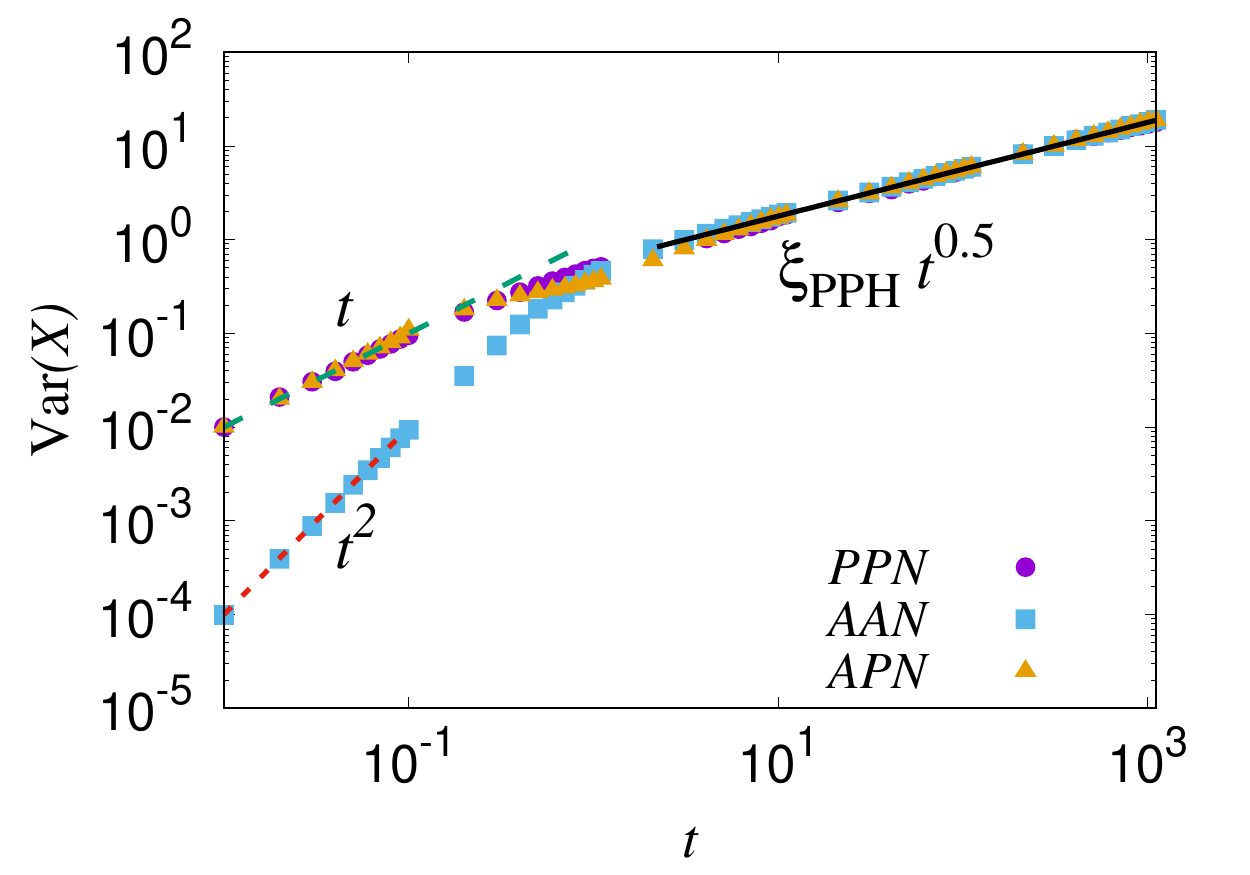}\includegraphics[width=7cm]{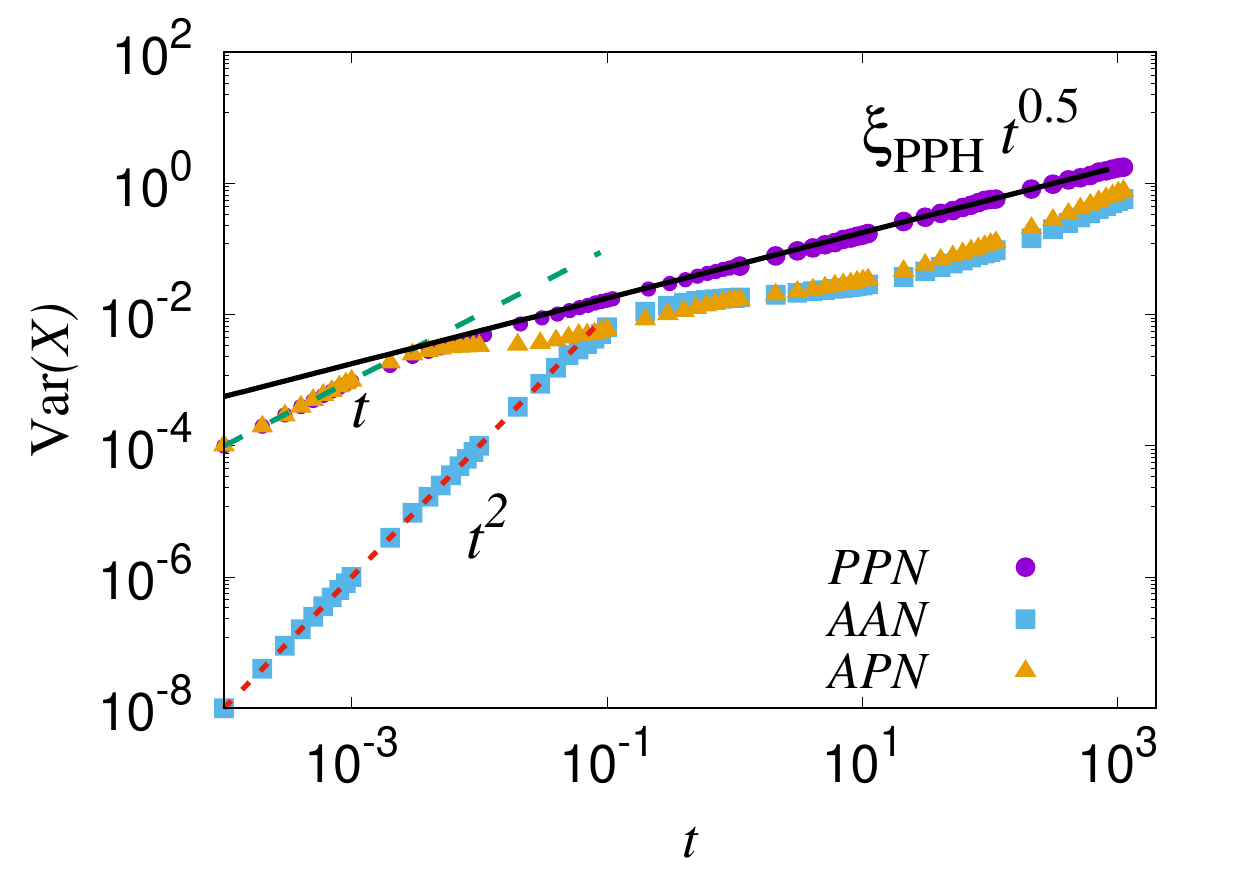}
\caption{Time evolution of \var \, for low (Left panel, $\rho=1$) and high (Right panel, $\rho=10$) densities. $N=2000$ with $\Delta t=0.0001$ for $t \leq 1$ and $\Delta t=0.001$ for $t>1$. (Overlap of data for both $\Delta t$ values checked for $t$ in the range $[1,10]$. Smaller value of $\Delta t$ taken to extract the early-time free diffusive behaviour of passive tracers.) All data averaged over at least $5\times 10^3$ realizations. At low densities, in the long time limit, all cases coincide, with the same subdiffusion coefficient $\xi_{\textrm{PPN}}$ (which is also exactly $\xi_{\textrm{PPH}}$, see Sec.~\ref{recap}).  At high density the active environments deviate from $\sqrt{t}$ behaviour for the observed time windows, owing to sticking of active particles to the tracers. {We have truncated the data before finite-size effects can set in for the chosen value of $N$; see also \ref{very-late-time-nonint}.}}\label{nonint-var-full}
\end{figure}

\subsubsection{Tracer variance.} \label{TVFSE}
Fig.~\ref{nonint-var-full} compares the time evolution of  \var \, in the {\em AAN}, {\em APN} and {\em PPN} systems, for two different densities.  
At very short times, the tracer does not interact with the environment so \var \, scales diffusively ($\sim t$) for a passive tracer and ballistically ($\sim t^2$) for an active one. This changes as the tracer starts to interact with its neighbours; the higher the density, the sooner this is.

When the density is low, the late-time data for  {\em AAN}, and {\em APN} coincide to numerical accuracy with {\em PPN}, which in turn is exactly governed by Eq.~\ref{Xvar-sadhugen} for {\em PPH} as explained in Sec.~\ref{recap}. Thus at low enough density the long-time subdiffusive tracer behaviour is identical to the case of a passive diffusive environment, because any active particle (whether tracer or environmental) is already diffusive on the typical timescales of collision. Any tracer in such sparse  active environments effectively sees a train of passive Brownian particles on either side, and therefore, both {\em AAN} and {\em APN} reduce to the {\em PPN} (and therefore {\em PPH}) case. The same argument applies in principle to hardcore active environments -- but for these, as seen in Sec.~\ref{prefactors} above, $\rho = 1$ is not a low enough density to reach the {\em PPH} asymptote. 

At the higher density, $\rho = 10$, case {\em PPN} continues to exactly track {\em PPH}, as it must, but significant deviations from this arise for both {\em APN} and {\em AAN}. The idea of swapping worldlines that lies behind the {\em PPH}--{\em PPN} equivalence of Sec.~\ref{recap} does not work for the persistent trajectories of RTP particles. The ultimate long-time behaviour remains open: in our simulations, before any convergence onto a subdiffusive, {\em PPH}-like behaviour at late times can be seen, finite-size effects (of the kind described in Sec.~\ref{FSE} above) come into play. See \ref{very-late-time-nonint} for further discussion. 
Given the discussion above, it is plausible, but not certain, that in an infinite system the {\em AAN} and {\em APN} curves would asymptote from below towards the solid line, $\varm = \xi_{\textrm{PPH}}\sqrt{t}$, even though {\em AAH} and {\em APH} do not (see Sec.~\ref{prefactors} above). We return to this question in Sec.~\ref{den-effect} where a wider range of densities is addressed.

At $\rho = 10$, for both {\em AAN} and {\em APN}, there is a sustained region where \var \, falls well below the results for the passive environment, with no clean power law. As reported above for case {\em APH}, this indicates {\em transient caging} of the tracer particle by its active neighbours. Indeed the trajectory data of Figs.~\ref{AAN-traj},\ref{APN-traj} show this caging effect directly. Interestingly, in contrast to the {\em AxH} cases (see Figs.~\ref{hardcore-variance-compare} and \ref{hardcore-density-compare}), we find that at high density {\em AAN} and {\em APN} results do not generally coincide with each other in the long-time regime, $t \gg {\gamma}^{-1}$. This means that, unlike for the hardcore cases, the tracer dynamics is not fully enslaved to that of the surrounding environment in this regime. 
This mismatch between {\em AAN} and {\em APN} cases can also be seen from the respective tracer-position distributions, see Fig.~\ref{noint-act-pass-dist}. Note that in these cases the {\em only} interactions are those directly involving the respective tracers, in contrast to the hardcore active environment ({\em AxH}), whose multiple interactions correlate the tracer dynamics with $n\sim \rho (Dt)^{1/2}$ environmental particles. It is understandable therefore that in the hardcore active environment only, the dynamics of a lone tracer is eventually enslaved and its own character is forgotten.

\begin{figure}
\centering
\includegraphics[width=7cm]{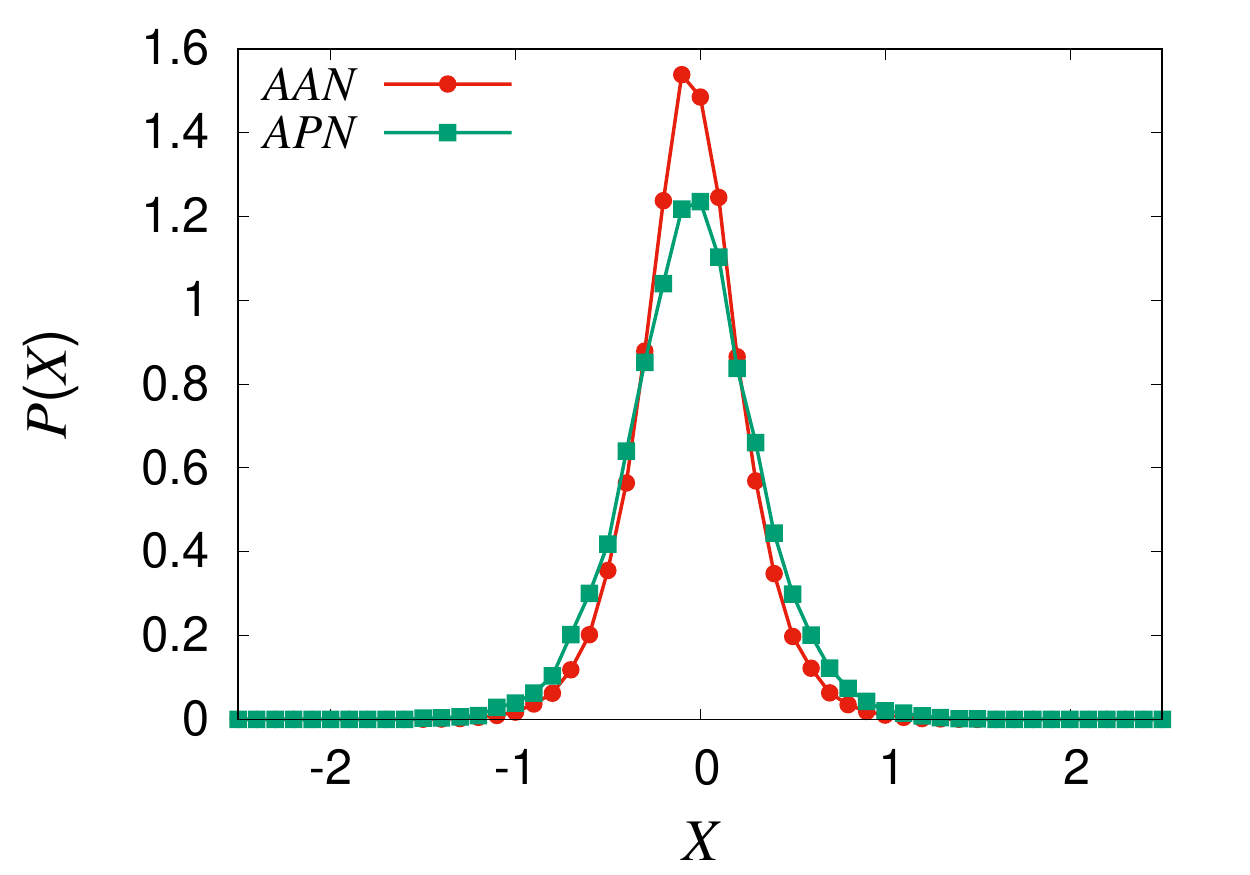}
\caption{Histograms of the passive and active tracer-position $X$ in the non-interacting active environment at time $t=100 \gg \gamma^{-1}= 1$. The active tracer is more localized than the passive tracer in the same active environment. $N=2000, \rho=10$ and $\Delta t=0.001$. Data averaged over $2\times 10^5$ realizations.}\label{noint-act-pass-dist}
\end{figure}

\begin{figure}
\centering
\includegraphics[width=7cm]{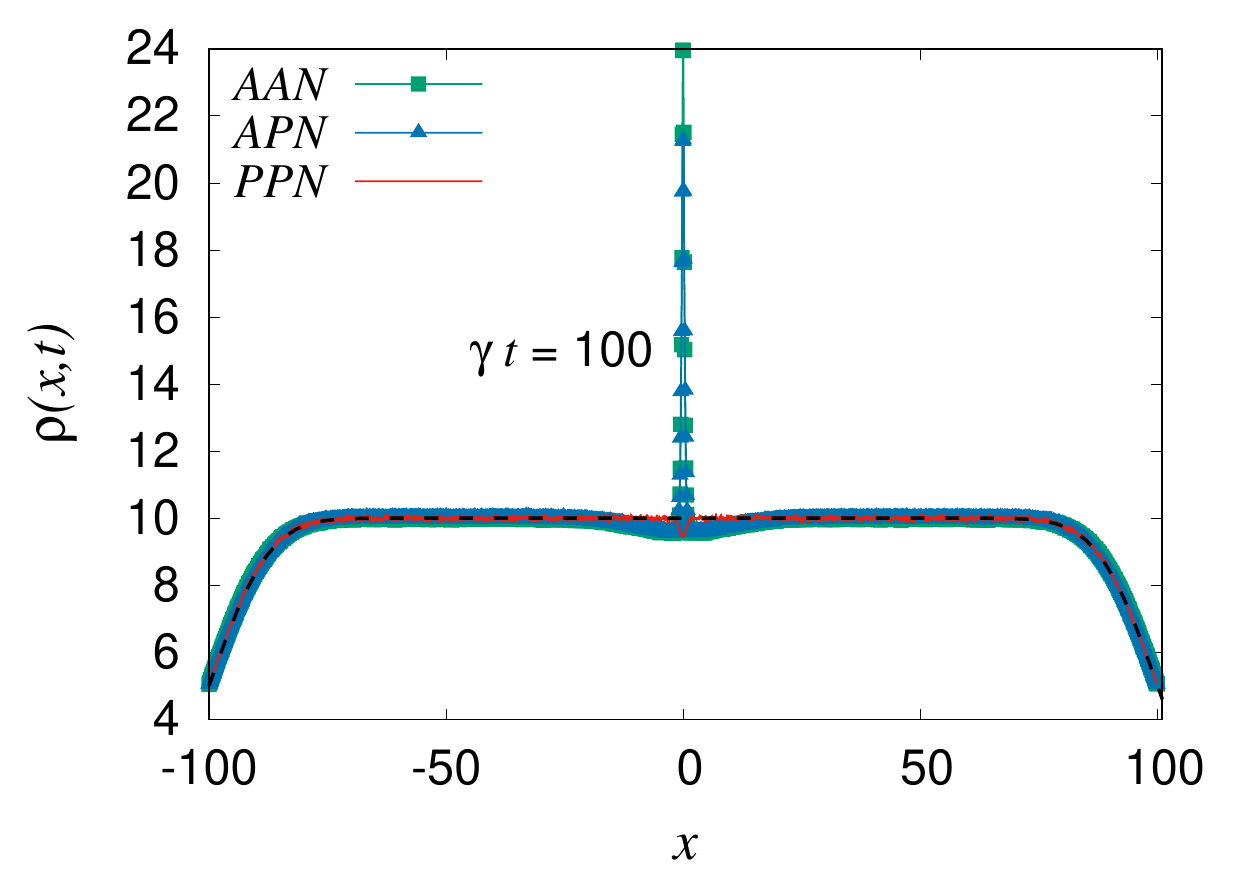}\includegraphics[width=7cm]{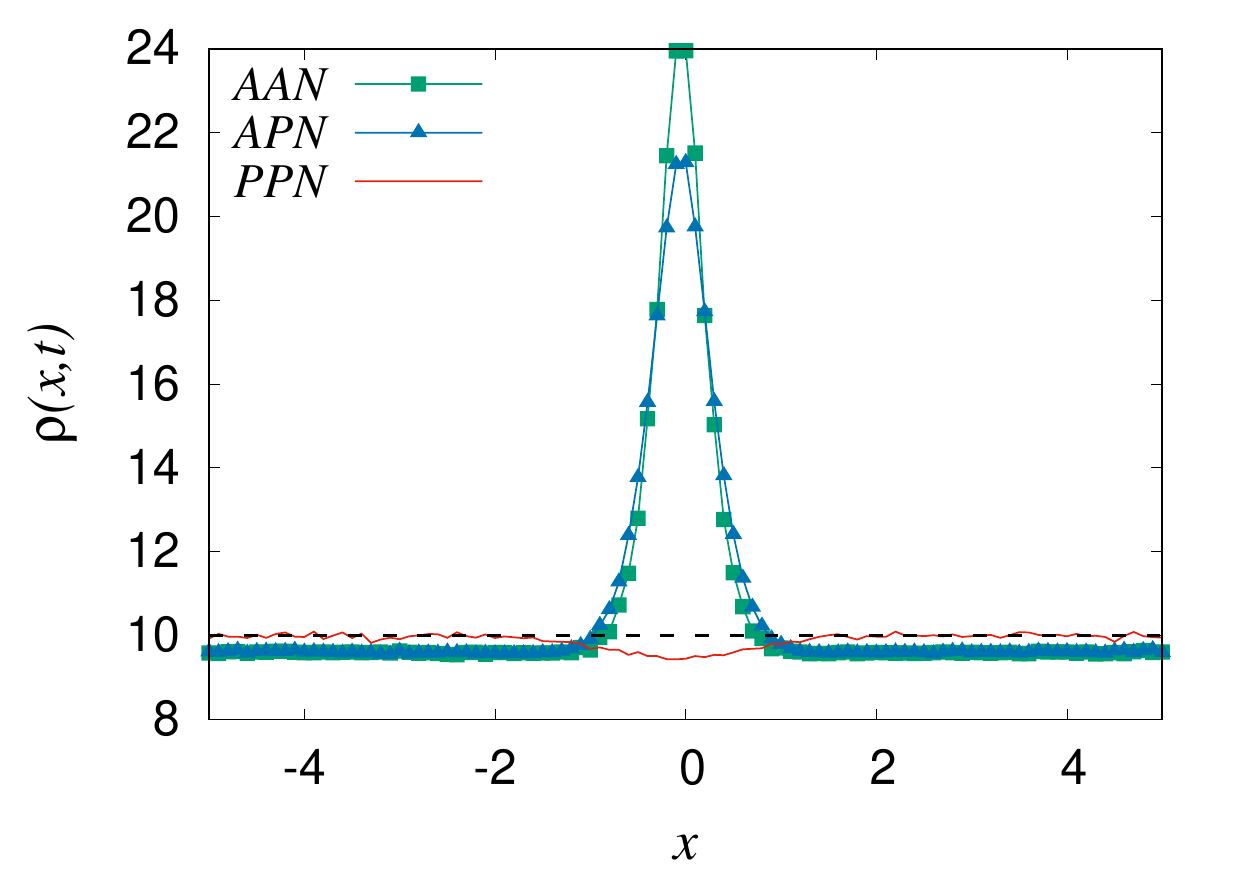}
\caption{(Left) Density distribution of environment particles  for a top-hat initial density of height $\rho=10$ and $N=2000$ (so $L = 100$) at time $t = 100$. $\Delta t=0.001$; {\em PPN}, {\em AAN} and {\em APN} data averaged over $5.8\times 10^{4}$, $3 \times 10^5$ and $3 \times 10^5$ realizations, respectively.  Active particles of the medium cluster around both active and passive tracers. Dashed black line represents Eq.~\ref{top-hat-sol} for  non-interacting passive Brownian particles.  Data sampled only in the range shown in the plot. (Right) Zoomed-in version of the left panel. Though qualitatively similar, we observe more clustering for case {\em AAN} than for {\em APN}.}\label{nonint-density-full}
\end{figure}

To understand better our results for the tracer variance \var \, in dense active media, we show in Fig.~\ref{nonint-density-full} the density profiles of the environment particles around the tracer, sampled at a late time $t=100$, for {\em AAN}, {\em APN} and, for comparison {\em PPN}. Far away from the tracer the density profiles are identical to statistical accuracy, and match the solution given in Eq.~\ref{top-hat-sol}. This is as expected: particles sufficiently far away from a hardcore tracer do not feel its presence, and (whether active or passive) diffuse freely on these timescales. However, close to the tracer, we see that the active environment particles develop a pronounced density peak caused by sticking to the tracer via the mechanism described in Sec.~\ref{models}. The peak height differs somewhat between the cases of an active ({\em AAN}) and a passive ({\em APN}) tracer because the details of the sticking mechanism are tracer-dependent as described previously. There is of course no peak at all in {\em PPN} whose density, by the arguments of Sec.~\ref{recap}, is oblivious to hardcore interactions either with the tracer or in the environment. We note here that our case {\em APN} and the problem studied in reference~\cite{omer-yariv-julien} both address a similar situation : a passive probe immersed in a sea of non-interacting RTPs. However, unlike \cite{omer-yariv-julien} we consider a hardcore, pointlike tracer which leads to sub-diffusion in our model, but which is absent in \cite{omer-yariv-julien}.

\subsubsection{Effects of density and ultimate long-time behaviour.}\label{den-effect}

\noindent {So far we have looked at just two density values, $\rho = 1,10$. At the higher of these, we did not observe \var \, scaling as $\sqrt{t}$ within our investigated time-window. To clarify the long time behaviour we now study a wider range of densities, focussing on the {\em AAN} case. (The qualitative aspects are the same for {\em APN}.) In Fig.~\ref{AAN-den-var-Xvart} we plot \var \, against time for $\rho = 1,2,5,10$. We observe that the $\sqrt{t}$ behaviour sets in later as the density of active particles is increased, preceded by the plateau-like region whose length increases with density. But only at the highest density does the final $\sqrt{t}$ regime fall entirely beyond the observation window (which is limited by finite size effects as previously discussed). Based on this data, we expect there is no true breakdown of the relation $\varm \propto \sqrt{t}$ in the ultimate long-time limit, so long as the limit $N\to\infty$ is taken first to avoid finite-size issues. Indeed such a breakdown would require some sort of critical density at which the asymptotic exponent changes value; we have found no evidence for this.

For a physical understanding of this {\em AAN} case, note that higher densities result in
more environmental particles sticking to the tracer, flattening the plateau in \var \, at intermediate times. At these times the active tracer is mainly trapped, and only freed by rare events in which {\em all} of the stuck particles on its forward side start moving away. This effect is even more pronounced at higher densities than considered here, when these individual events can appear as sudden jumps in the log-log plots unless there is sufficient averaging over runs. 
 
\subsection{Comparison of noninteracting and hardcore active environments}\label{comp1} 
It is clear from Figs.~\ref{hardcore-variance-compare} and \ref{nonint-var-full} that $\varm_{\textrm{AAH}}$ and $\varm_{\textrm{AAN}}$  have generically different behaviour at equal density $\rho$, even at quite late times ($t\gg\gamma^{-1}$). Interestingly, depending on the density and time, $\varm_{\textrm{AAH}}$ can be larger than $\varm_{\textrm{AAN}}$. This disproves the idea that adding hardcore repulsions somewhere in the system will always slow the tracer motion; instead, since this also alters the statistics of sticking, more subtle factors can come into play. 
In particular, $\xi_{\textrm{AAN}}$, the coefficient of subdiffusion for the {\em AAN} case, appears from Fig.~\ref{AAN-den-var-Xvart}(Left) to be the same within statistical error as $\xi_{\textrm{PPN}}$ and hence also $\xi_{\textrm{PPH}}$ {for low density ($\rho=1$) as given in Eq.~\ref{Xvar-sadhugen}, but as the density increases, $\xi_{\textrm{AAN}}$ saturates to a value slightly less than $\xi_{\textrm{PPH}}$ 
(see Fig.~\ref{AAN-den-var-Xvart}(Right)). This contrasts with $\xi_{\textrm{AAH}}$ which is larger than $\xi_{\textrm{PPH}}$ in this density range, see Fig.~\ref{hardcore-variance-compare}.}

Likewise for general $\rho$ and $t$, the tracer variances $\varm_{\textrm{APH}}$ and $\varm_{\textrm{APN}}$ are different in general. In fact, $\varm_{\textrm{APN}}$ shows the same qualitative behaviour as $\varm_{\textrm{AAN}}$, and remains less than $\varm_{\textrm{APH}}$ for intermediate densities. (We do not show these results separately.)
To summarize:  the subdiffusion constants for $\varm_{\textrm{AxH}}$ are generically larger than  $\varm_{\textrm{AxN}}$ at intermediate density $0.2 \lesssim \rho \lesssim 15 $.

\begin{figure}
\centering
\includegraphics[width=8cm]{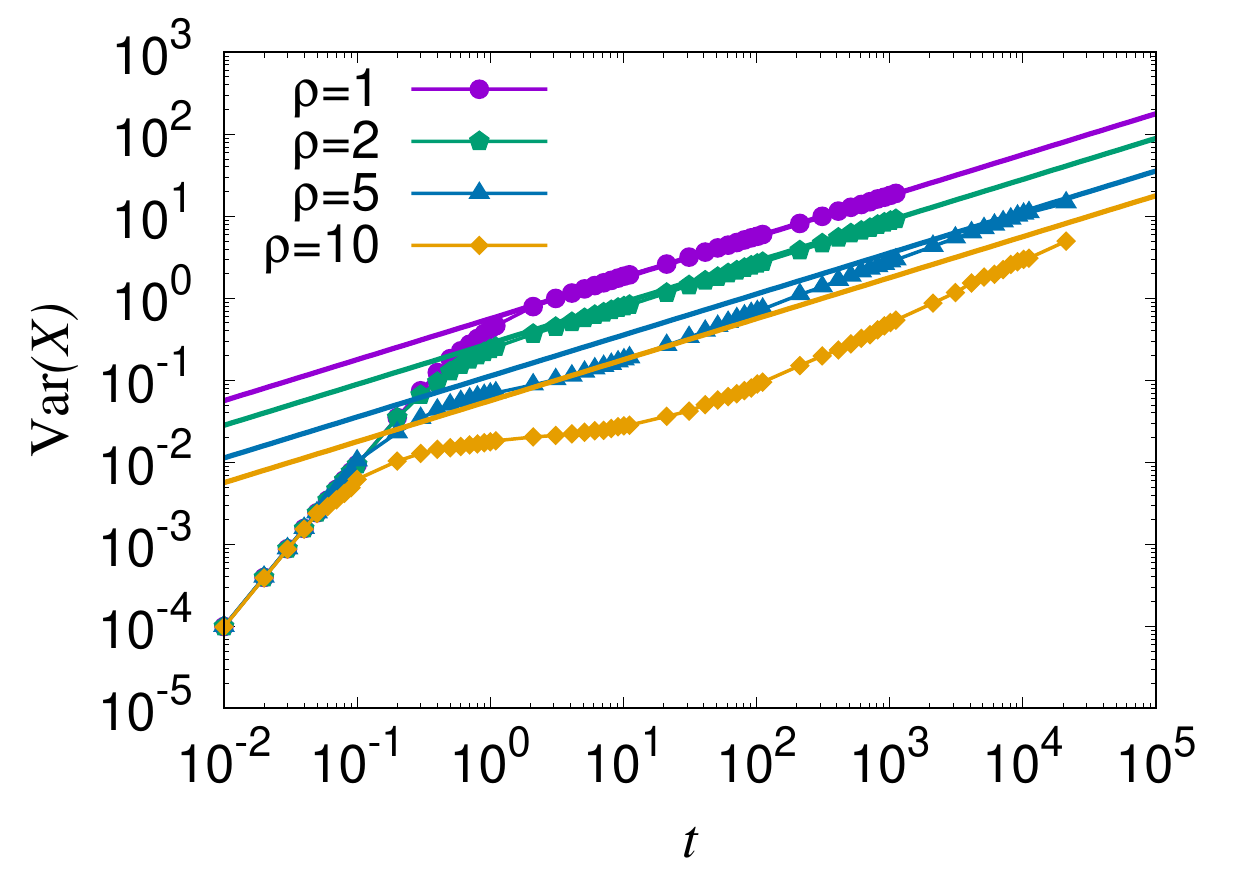}\includegraphics[width=8cm]{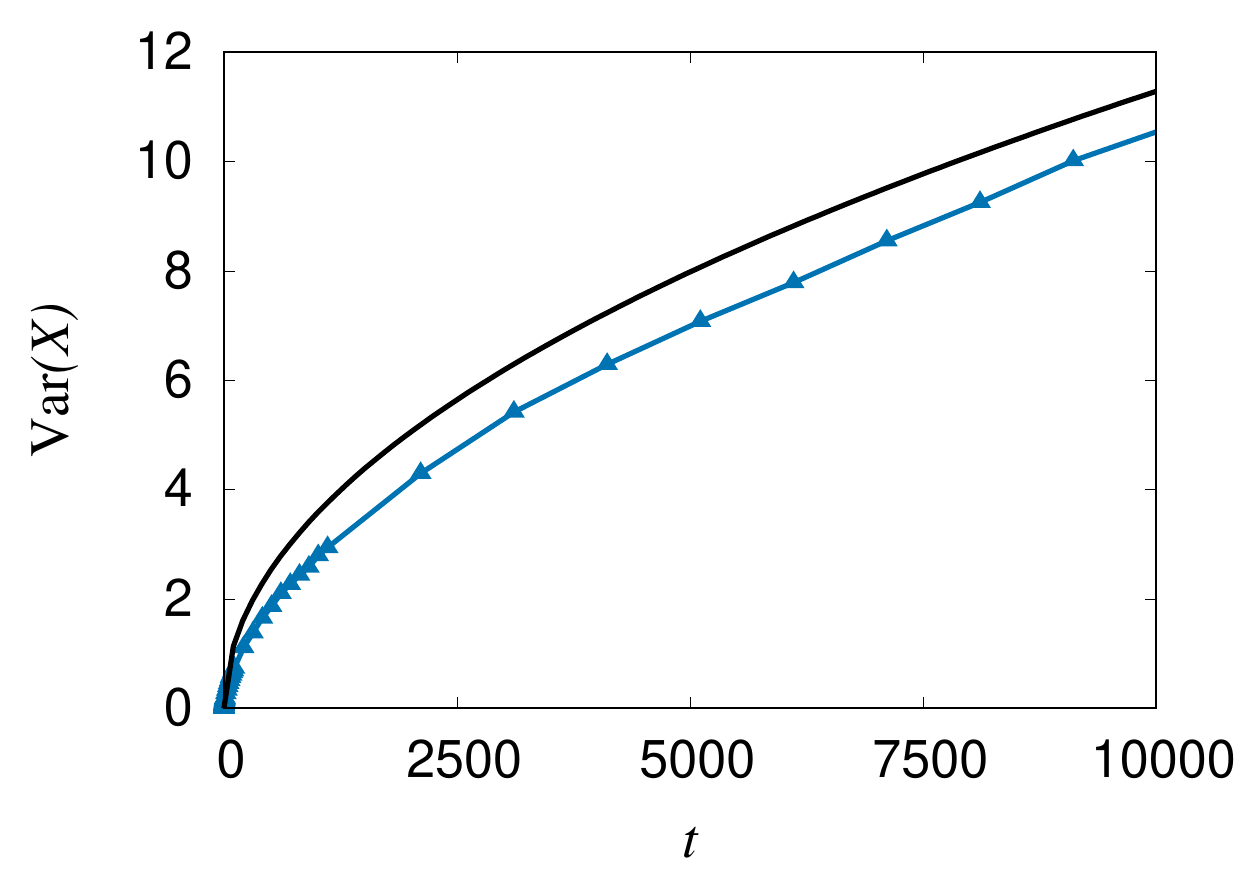}
\caption{(Left) Time evolution of $\varm_{\textrm{AAN}}$ for various densities $\rho$. Points are simulation results. The free solid lines (with same colour codes as points) represent $\xi_{\textrm{PPH}}\sqrt{t}$ for each density (and $D=0.5$, Eq.~\ref{Xvar-sadhugen}). Parameters $N=8000, \Delta t=0.01$ (Note that for cases {\em AAx}, long time results saturate already for $\Delta t=0.01$ for our sequential updates). To observe $\varm\sim \sqrt{t}$ at very high densities ($\rho \geq 10$) much larger time windows and  larger number of particles would be required. We have truncated the data for $\rho=5$ and $10$ at a point where finite size effects set in. Number of realizations :  $10^{4}$ for $\rho=1,2,5$ and $2 \times 10^{3}$ for $\rho=10$. (Right) Time evolution of $\varm_{\textrm{AAN}}$ for $\rho=5$ in the linear scale. Points represent simulation results for $N=8000, \Delta t=0.01$ and $10^4$ realizations. The solid black line corresponds to $\xi_{\textrm{PPH}}\sqrt{t}$, with $\xi_{\textrm{PPH}}$ given by Eq.~\ref{Xvar-sadhugen} (for $D=0.5, \rho=5$).}\label{AAN-den-var-Xvart}
\end{figure}

\subsection{Role of initial conditions}
As noted above, all results presented so far are for an initial condition with equispaced particles, and the mean squared displacement \var \, is measured between times $0$ and $t$.  By contrast \ref{t0} discusses how the results (for active environments) are changed if one instead measures the mean squared displacement between times $t_0$ and $t_0+t$, for some lag time $t_0$ (in particular $t_0=1$).  This affects the transient caging behaviour at short and intermediate time, but the large-$t$ behaviour is the same.


\section{Results for Active Tracers in Passive Environments}\label{PAH+PAN}

\begin{figure}
\centering
\includegraphics[width=7cm]{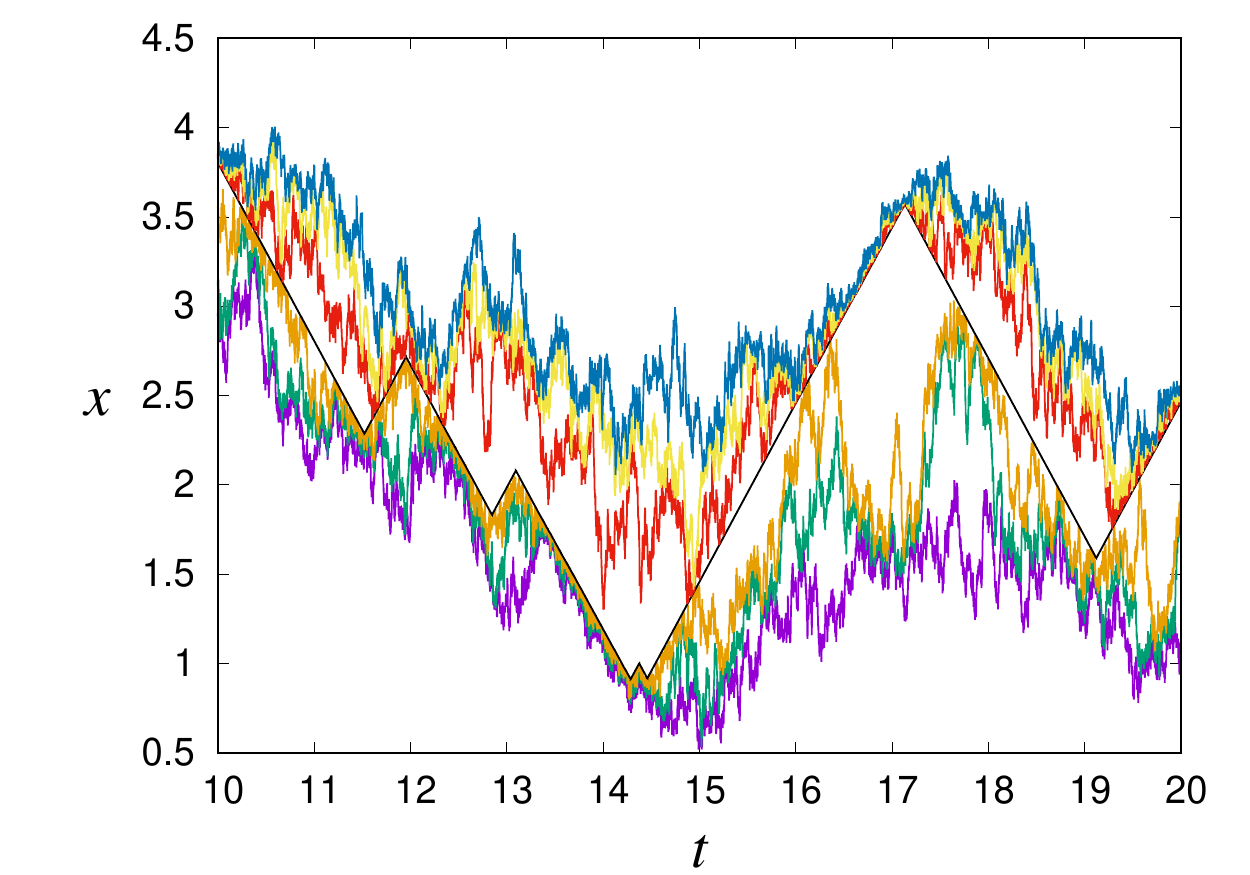}\includegraphics[width=7cm]{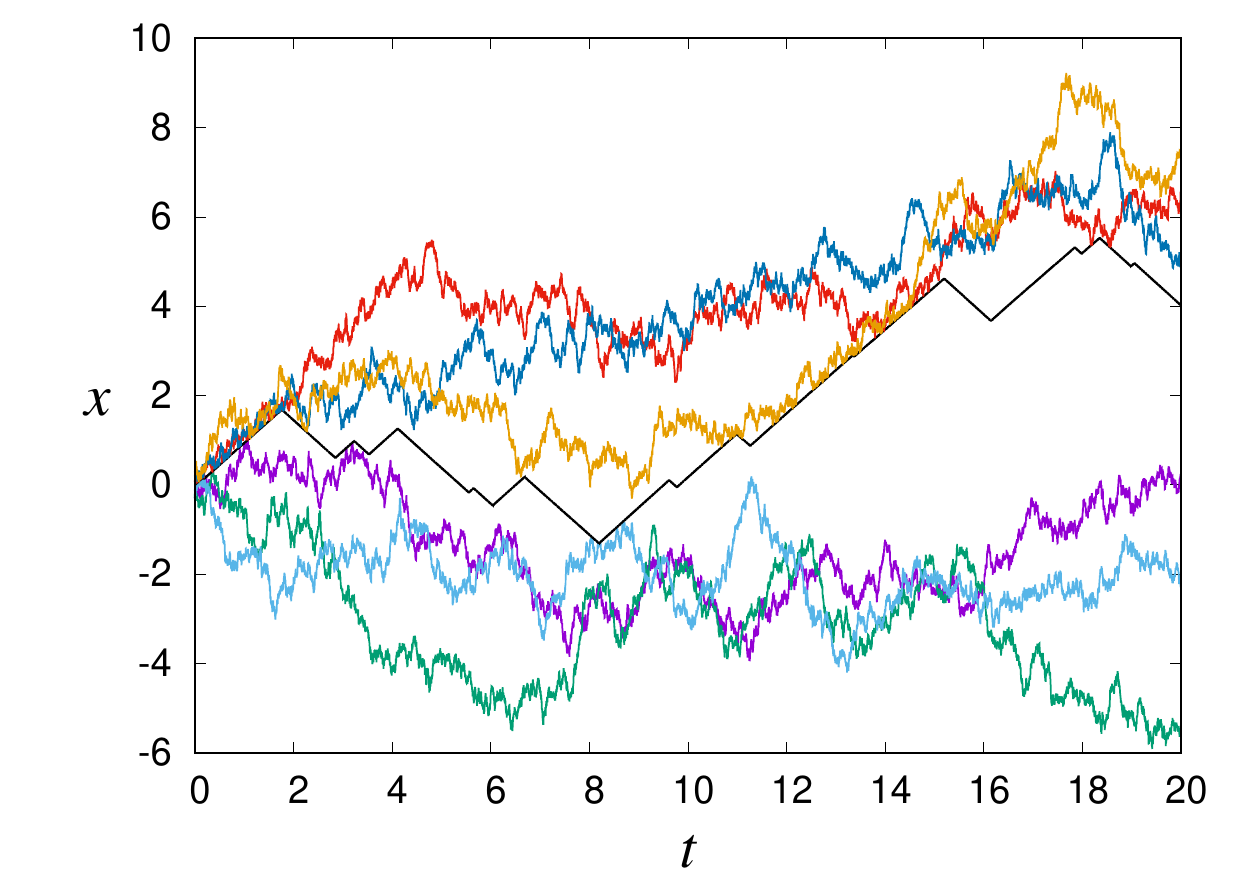}
\caption{Trajectories for seven central particles in the {\em PAH} (Left) and {\em PAN} (Right) cases; $N=1000$ and $\rho=10, \Delta t=0.001$. The central black line represents the RTP tracer. The active tracer often creates a {\em snowheap} of passive particles in front of itself.}\label{PAN-traj}
\end{figure}

Having reviewed the two passive-in-passive cases ({\em PPx}) in Sec.~\ref{recap}, and explored the four cases ({\em Axx}) of active or passive tracers in active environments with and without hardcore forces in Sec.~\ref{results}, there remain two cases so far unexplored in Table~\ref{table}. These are {\em PAH} and {\em PAN}, respectively describing an active RTP tracer surrounded by free or hard-core passive Brownian particles. 
In this Section we show how the qualitative behaviours of {\em PAH} and {\em PAN} are similar to each other, but differ substantially from all the other cases presented so far. 

Typical trajectories for the two cases are shown in Fig.~\ref{PAN-traj}, for seven central particles, including the active tracer, in the middle of a large system ($N = 1000$).  These show qualitatively similar tracer motion although the environment worldlines differ considerably due to the presence (in {\em PAH}) or absence (in {\em PAN}) of the no-crossing constraint among them. (As always, no particle can cross the tracer.)  

To explain why these active-in-passive systems are different from the other cases considered so far, we note from 
%
%
Sec.~\ref{models} that a purely ballistic tracer moving with speed $v$ moves freely, unless it encounters a particle within a distance $v\Delta t$. A low density environment of passive particles will slightly reduce the mean forward speed but this effect is small unless the (local) density is of order $(v\Delta t)^{-1}$.  However, when a ballistic tracer moves forward in such an environment, it tends to accumulate a pile of passive particles in front of itself, by a `snowplough' effect.  This acts to increase the {local} density in front of the tracer, which may eventually exceed $(v\Delta t)^{-1}$, even if the average density of the system is small.  We will show that this leads to results that depend significantly on the {parameter $\Delta t$} of our numerical algorithm, in contrast to active-in-active systems. (While this argument assumes a perfectly ballistic tracer, it is also relevant for RTP tracers, if the run length is long enough that the pile of passive particles has time to form between tumbles. Note that a pile-up of passive particles in one dimension due to {\em driven} (not active) tracers were discussed earlier in~\cite{emil,evans-DTP}.)

It is important for this argument that  if a passive particle encounters the active tracer, it reflects from it as described in Sec.~\ref{models}, so there is almost always some space ahead of the tracer, and it continues to make progress.  For {\em PAH} the passive neighbours are also restricted by their own hardcore passive neighbours. In {\em PAN} this restriction is lifted and the tracer has comparatively more freedom to move forward.

\begin{figure}
\centering
\includegraphics[width=7cm]{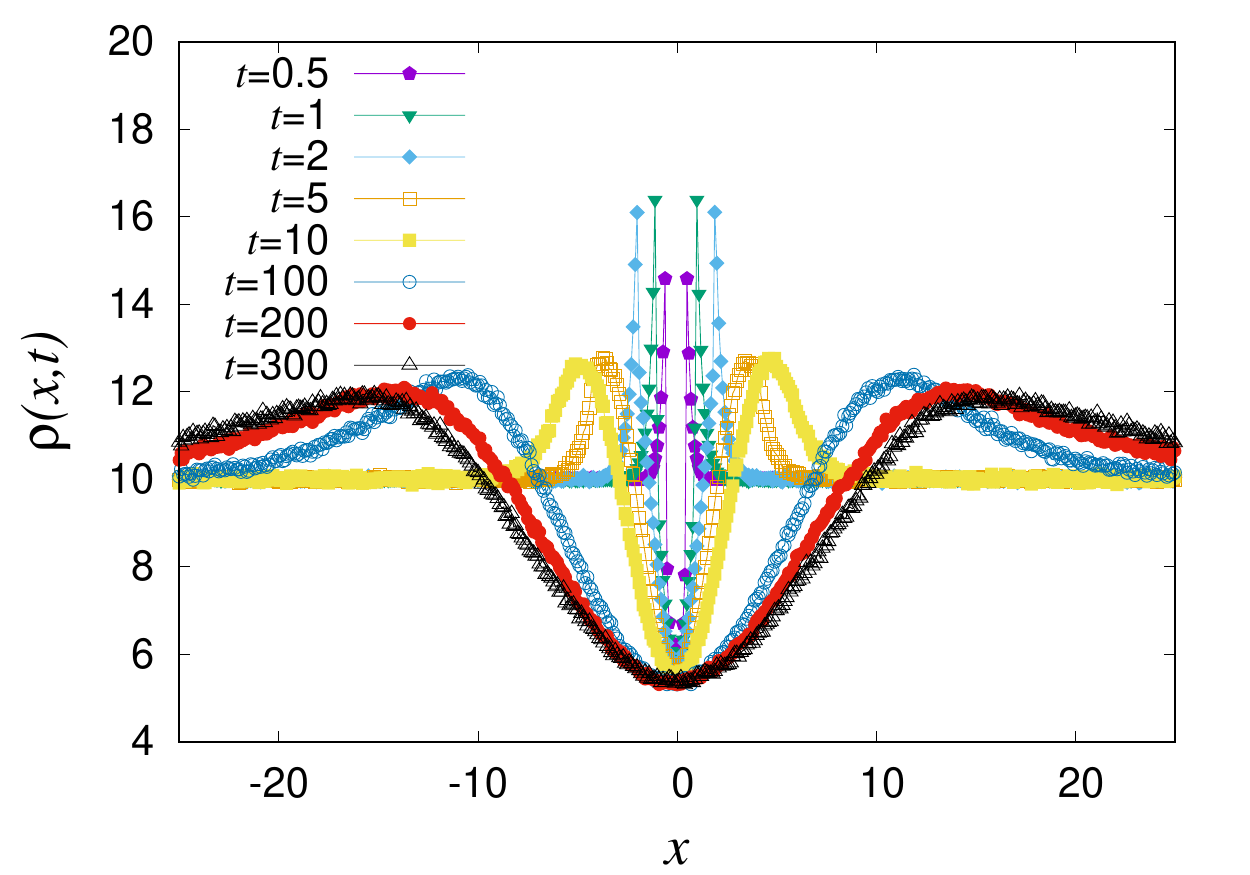}\includegraphics[width=7cm]{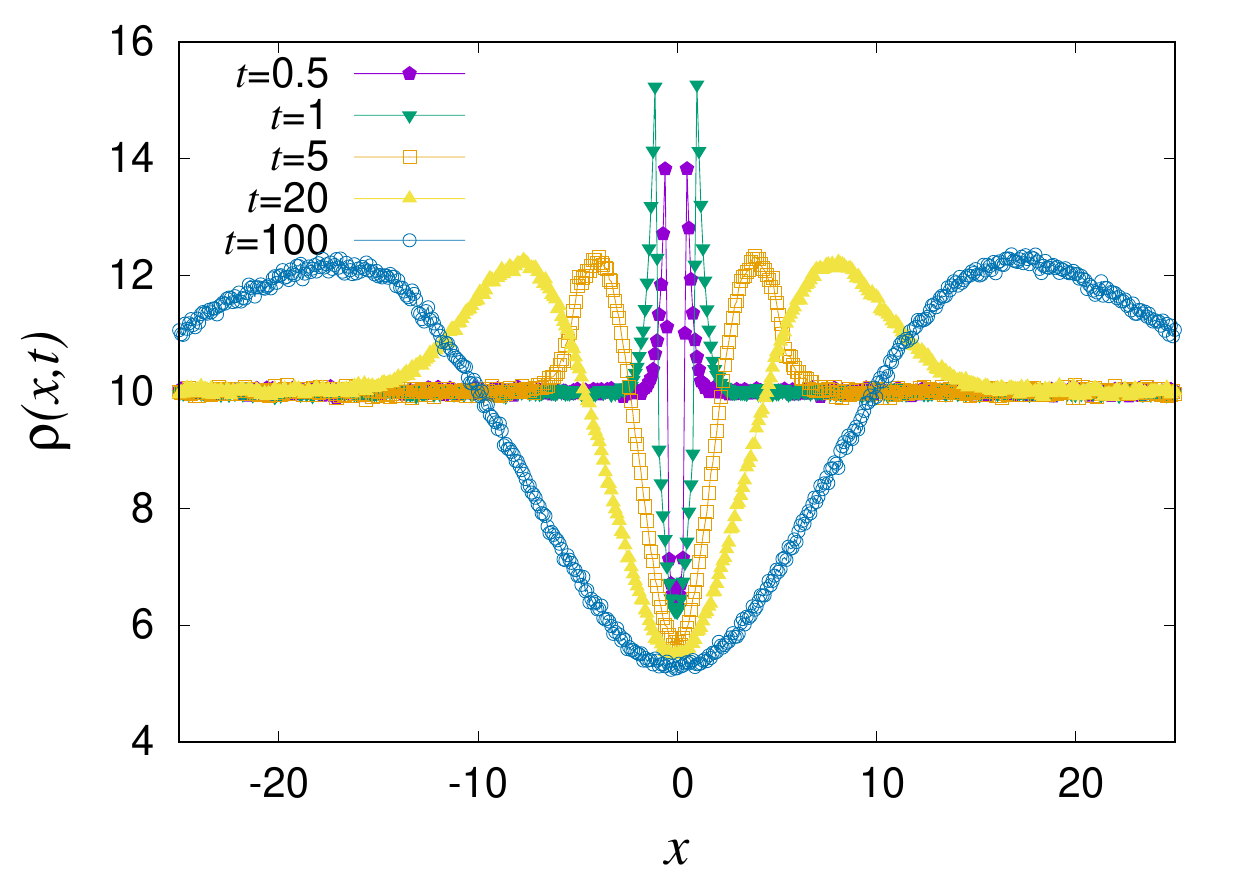}
\caption{Time evolution of average density profiles around the tracer for {\em PAH} (Left) and {\em PAN} (Right). $N=2000, L=100, \Delta t=0.001$. All {\em PAH}({\em PAN}) data averaged over $1.3 \times 10^4$ ($2.5 \times 10^{4}$) realizations or more.}\label{evolve-PAX-density}
\end{figure}

To illustrate this effect, the time evolution of the density of passive particles around an active (RTP) tracer is shown in Fig.~\ref{evolve-PAX-density} for {\em PAH} and {\em PAN}. At very short times the tracer is ballistic, and non-interacting. However as it encounters the environment, the tracer continues to move ahead, creating a pile of passive particles in front of itself, just as a purely ballistic particle would. This continues until $t\sim \gamma^{-1} = 1$ when the tracer flips direction. It then pushes the passive particles that were previously {\em behind} it, until it flips again. Thus the tracer keeps piling up passive particles on both sides and shunts them away from its initial position. 
The flipping of the tracer also allows each passive pile to spread diffusively half the time. Accordingly, after $t\sim \gamma^{-1}$, {the density peaks broaden in time without decreasing in mass (this argument neglects finite-size effects, as usual)}. 

It is clear from Fig.~\ref{evolve-PAX-density} that the tracer particle strongly perturbs the environmental density around it -- so tracer motion is not enslaved to its environment.  This breaks the argument of~\cite{sadhu-prl} that subdiffusion should be generic in single-file systems, which explains the qualitative difference between {\em PAx} and the other cases considered so far.  An MFT description of the {\em PAH} system might still be possible, but it would require that the ballistic motion of the tracer (between tumbles) is explicitly accounted for, to capture the effect of the tracer on the environmental density, recall~\cite{emil,evans-DTP}.

\subsection{Parameter $\Delta t$ sensitivity}\label{tims}
\begin{figure}
\centering
\includegraphics[width=7cm]{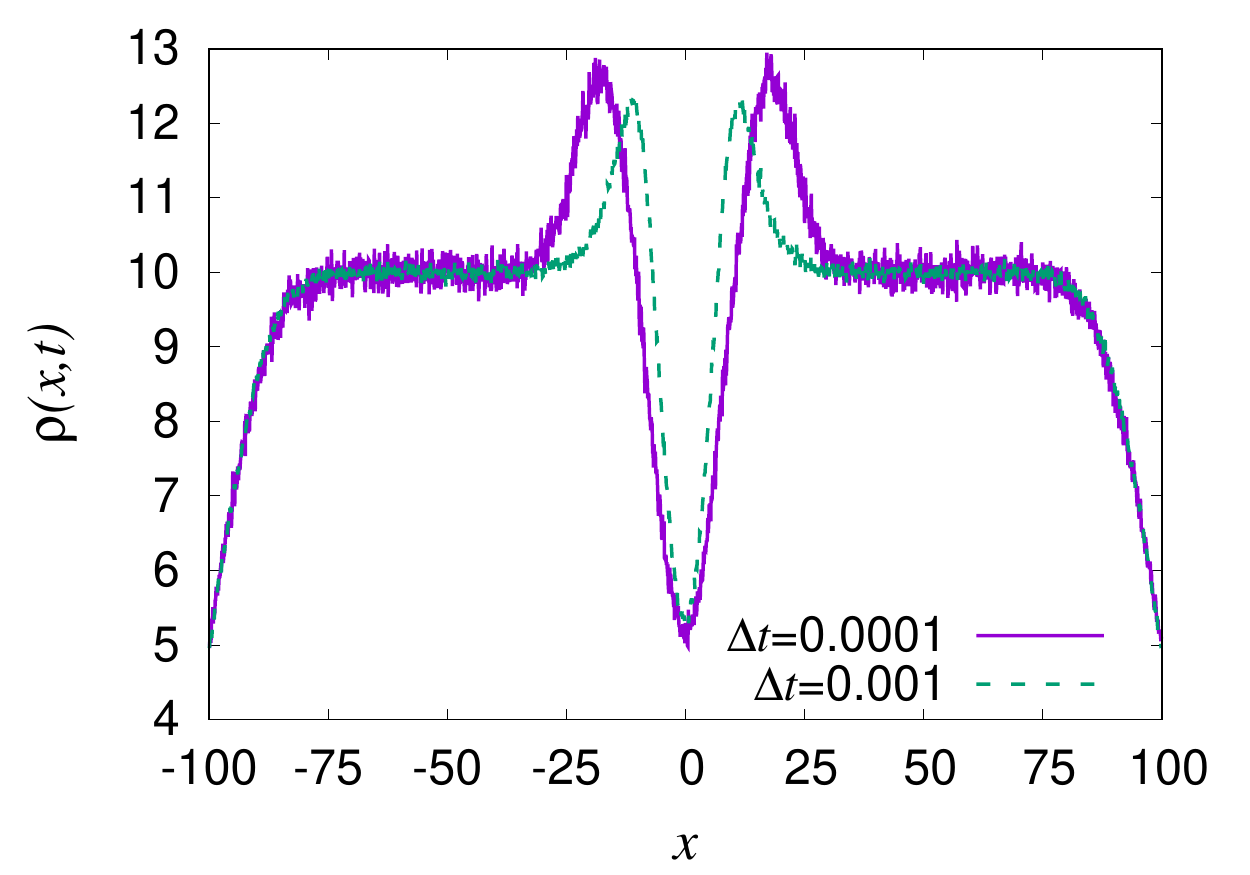}\includegraphics[width=7cm]{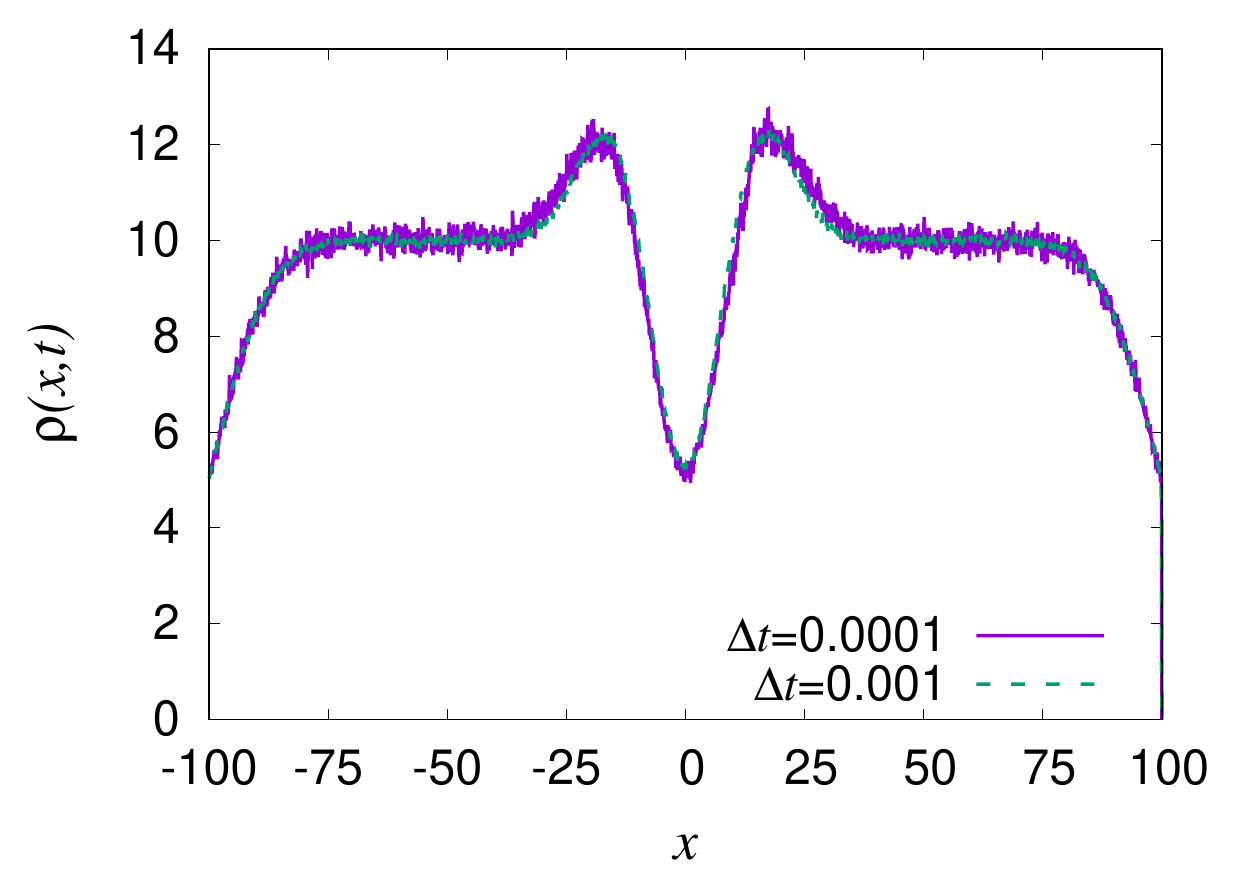}
\caption{Effect of time-step $\Delta t$ on environment particles in cases {\em PAH} (Left) and {\em PAN} (Right). Plot of the density profile of environment particles at a late time $t =100 \gg \gamma^{-1}=1$. We have $N=2000$ passive particles with $L=100$. For {\em PAH}, the $\Delta t=0.001$ data is averaged over $2.3 \times 10^4$ realizations, while that corresponding to $\Delta t=0.0001$ is averaged over $5000$ realizations. For {\em PAN}, $\Delta t=0.001$ data is averaged over more than $2.6 \times 10^4$ realizations, while for $\Delta t=0.0001$, $3.5\times 10^3$ realizations have been used. Other parameters are $v=1, D=0.5$. For smaller $\Delta t$, the active tracer accumulates a more massive snowheap of passive particles on either side. For the chosen values of $t$ and $\Delta t$, this effect is pronounced for the {\em PAH} case and hardly any difference is observed for {\em PAN}. To observe the $\Delta t$ sensitivity for {\em PAN} one needs either larger $t$ or $\Delta t$; see Fig.~\ref{trajectory-PAH} in similar context.  Data sampled only in the range shown in the plots.}\label{tims-PAx}
\end{figure}

\begin{figure}
\centering
\includegraphics[width=7cm]{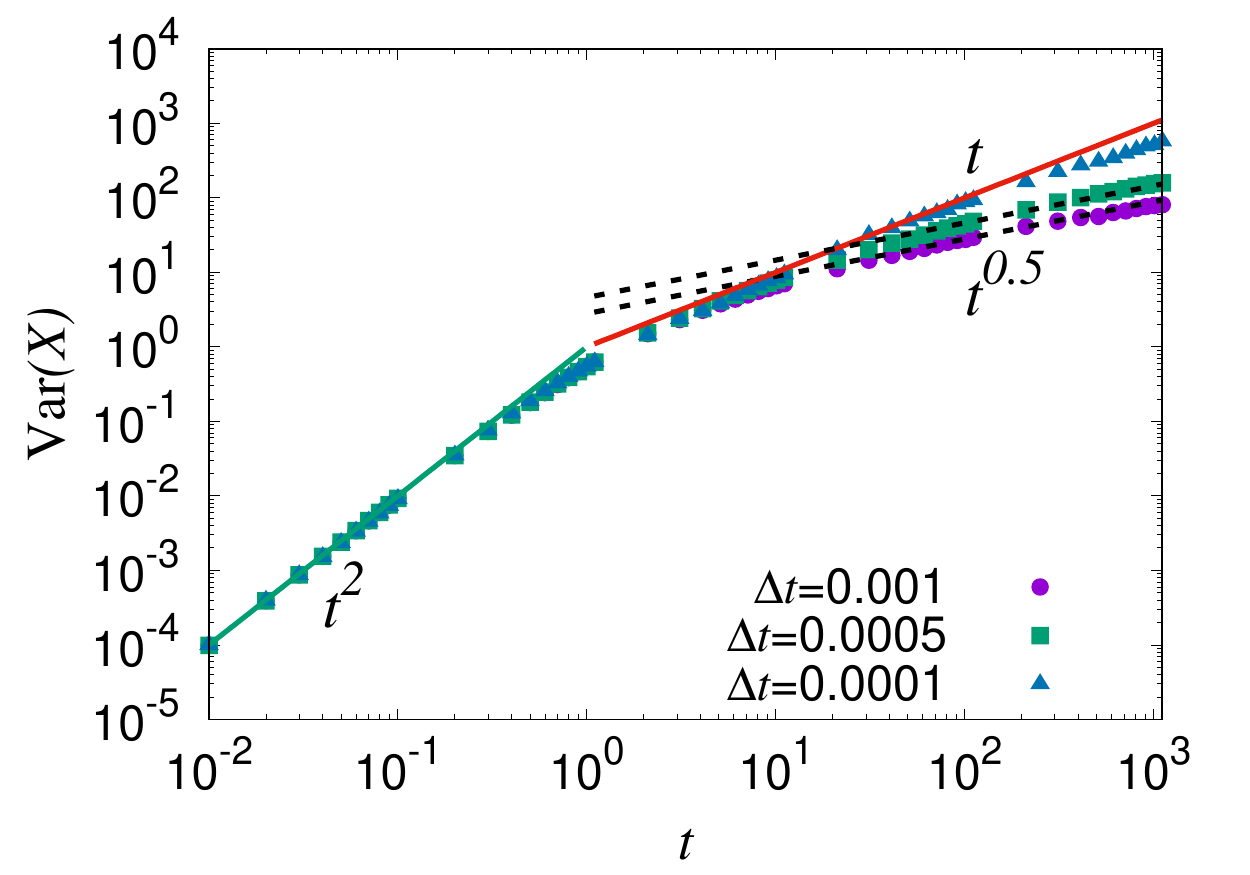}\includegraphics[width=7cm]{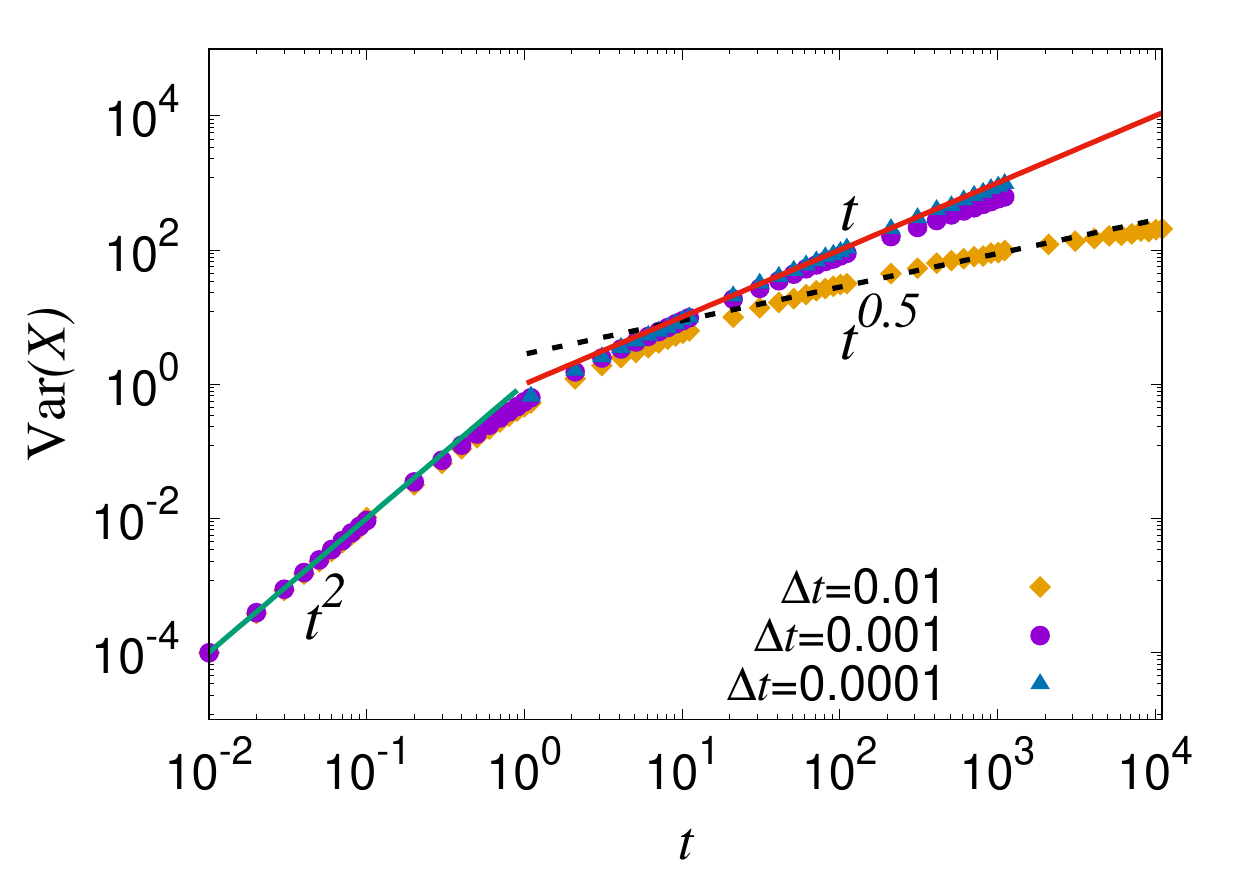}
\caption{Tracer variance \var \, as a function of time in {\em PAH} (Left) and {\em PAN} (Right). $N=2000$, $\rho=10$. Points are simulation results. At short times the tracer is ballistic. For very small $\Delta t$ ($=0.0001$) the tracer shows nearly free diffusive behaviour at moderately long times for both {\em PAH} and {\em PAN} cases. For {\em PAH}, as $\Delta t$ is increased, \var \, starts to show $\sqrt{t}$ scaling behaviour instead ($\Delta t=0.0005,0.001$).{ Within the time-window of observation, $\Delta t=0.001$ is not large enough to reach $\sqrt{t}$ scaling for {\em PAN}, although subdiffusive behaviour starts to appear at late times. $\Delta t=0.01$ indeed shows close to $\sqrt{t}$ scaling for \var \,, but we suspect such a large value of $\Delta t$ can be prone to errors for interacting passive Brownian particle-simulations. Number of realizations : {\em PAH} -- $4.5 \times 10^3$ for $\Delta t=0.001$, $2.3 \times 10^3$ for $\Delta t=0.0005$ and $500$ for $\Delta t=0.0001$; {\em PAN} -- $2.5\times 10^{3}$ for $\Delta t=0.01$, $5000$ for $\Delta t=0.001$ and $500$ for $\Delta t=0.0001$.}}\label{trajectory-PAH}
\end{figure}

\begin{figure}
\centering
\includegraphics[width=7cm]{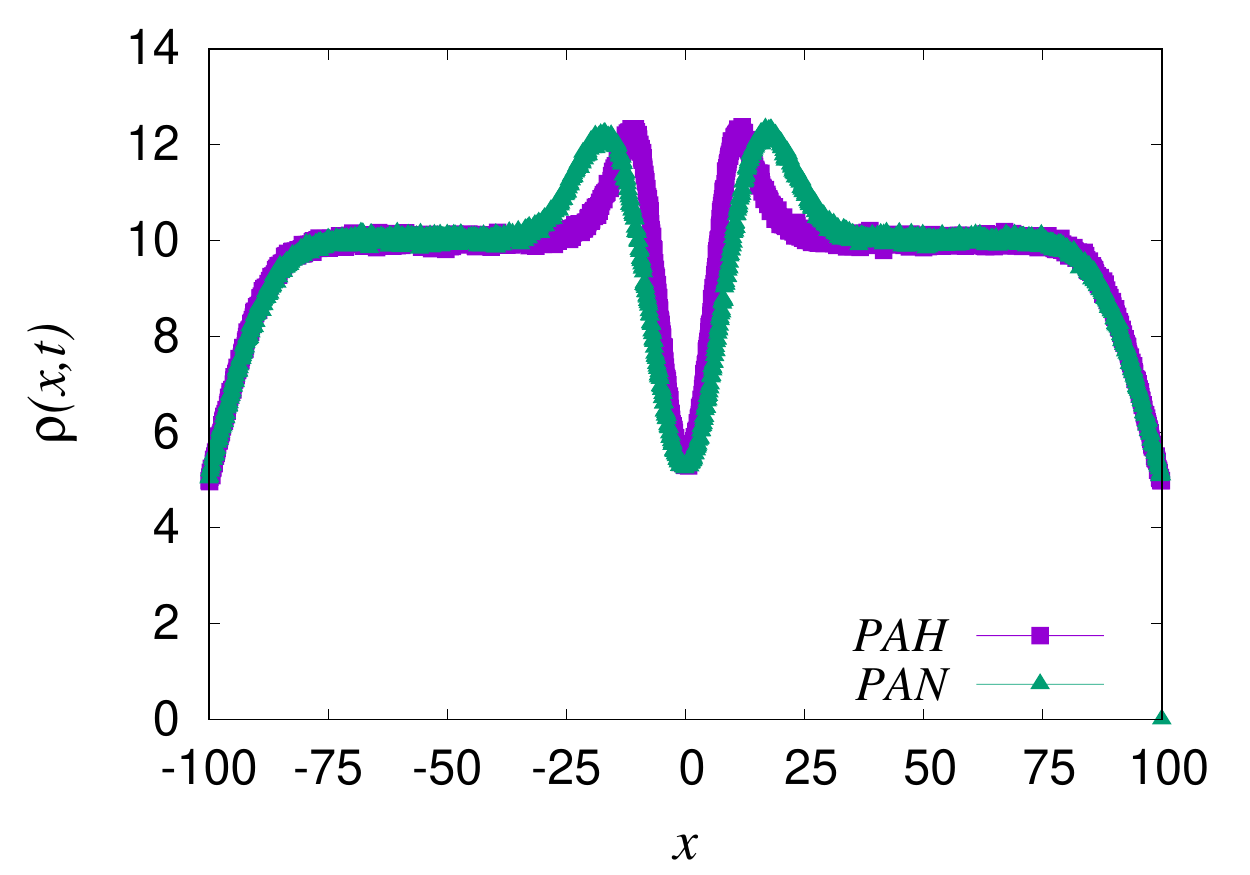}
\caption{ Average density profiles for bath particles in the {\em PAH} ($2.3 \times 10^4$ realizations) and {\em PAN} ($2.6 \times 10^4$ realizations) cases at time $t=100 \gg {\gamma}^{-1} = 1$ with $\Delta t=0.001$, $N=2000, L=100$.  Data sampled only in the range shown in the plot.}\label{PAX-density}
\end{figure}

{For active tracers in passive media, the sensitivity of the snowplough effect to the particle step size $v \Delta t$  means that long-time observables likewise depend on $\Delta t$.} This is illustrated in Fig.~\ref{tims-PAx} for the density $\rho$ of environmental particles.  Also, Fig.~\ref{trajectory-PAH} shows \var \, as a function of time with different values of $\Delta t$, for {\em PAH} and {\em PAN}. For small enough time-step the tracer variance crosses from ballistic at short times ($t\ll\gamma^{-1}$) to diffusive or near-diffusive behaviour at longer ones.\footnote{In contrast to these two active-in-passive cases ({\em PAx}), the dependence of \var \, on the time-step (and also on the update scheme, see below) becomes minimal beyond $\Delta t \sim 10^{-3}$ or $10^{-4}$, for fully active ({\em AAx}) and fully passive ({\em PPx}) cases respectively.}

The {\em PAH} data are consistent with a scenario in which the mounds of passive particles ultimately slow the active tracer down to a subdiffusive $t^{1/2}$ behaviour, which sets in beyond a crossover time $t_{\textrm{sub}} (\rho,\Delta t)$ that decreases with the density $\rho$.  However, smaller timesteps $\Delta t$ result in larger $t_{\textrm{sub}}$, which suppresses the subdiffusive regime. To see this regime clearly for $\Delta t \le 10^{-4}$ might require very long observation times, in turn demanding much larger system sizes $N$ to ensure that finite size effects are pushed out to $t\gg t_{\textrm{sub}}$. The ultimate behaviour at $\Delta t \to 0$, $N,L\to\infty$ at fixed $\rho>0$, and $t\to\infty$ remains open, and may depend on the order in which these limits are taken. Subdiffusive scalings for \var \, are harder to reach in simulations for lower densities than those shown in Fig.~\ref{trajectory-PAH}. We do not show those results here.

We observe very similar trends on varying $\Delta t$ for {\em PAN}; the data is also broadly consistent with the scenario elaborated in the preceding paragraph. Comparing the two cases at the same density and at the same (large enough) timestep, the active tracer to moves further in the non-interacting environment ({\em PAN}), compared to {\em PAH}. 
Fig.~\ref{PAX-density} compares the two density profiles involved and shows more space for tracer motion in the {\em PAN} case, 
indicating (perhaps unsurprisingly) that it is harder for the active tracer to push a collection of hardcore particles than noninteracting ones.

\subsection{Role of update rule}
All results shown above for {\em PAH} and {\em PAN} were obtained with cyclic sequential Monte-Carlo updates as explained in Sec.~\ref{models}. Unlike for the other models, switching to a random sequential scheme (which is more prone to numerical diffusion), leads to significant quantitative changes in the time evolution of  \var \, in both these cases. However, the qualitative trend with respect to $\Delta t$ is independent of the update rule chosen. This issue is discussed further in \ref{ran-seq}. 

\begin{figure}
\centering
\includegraphics[width=7cm]{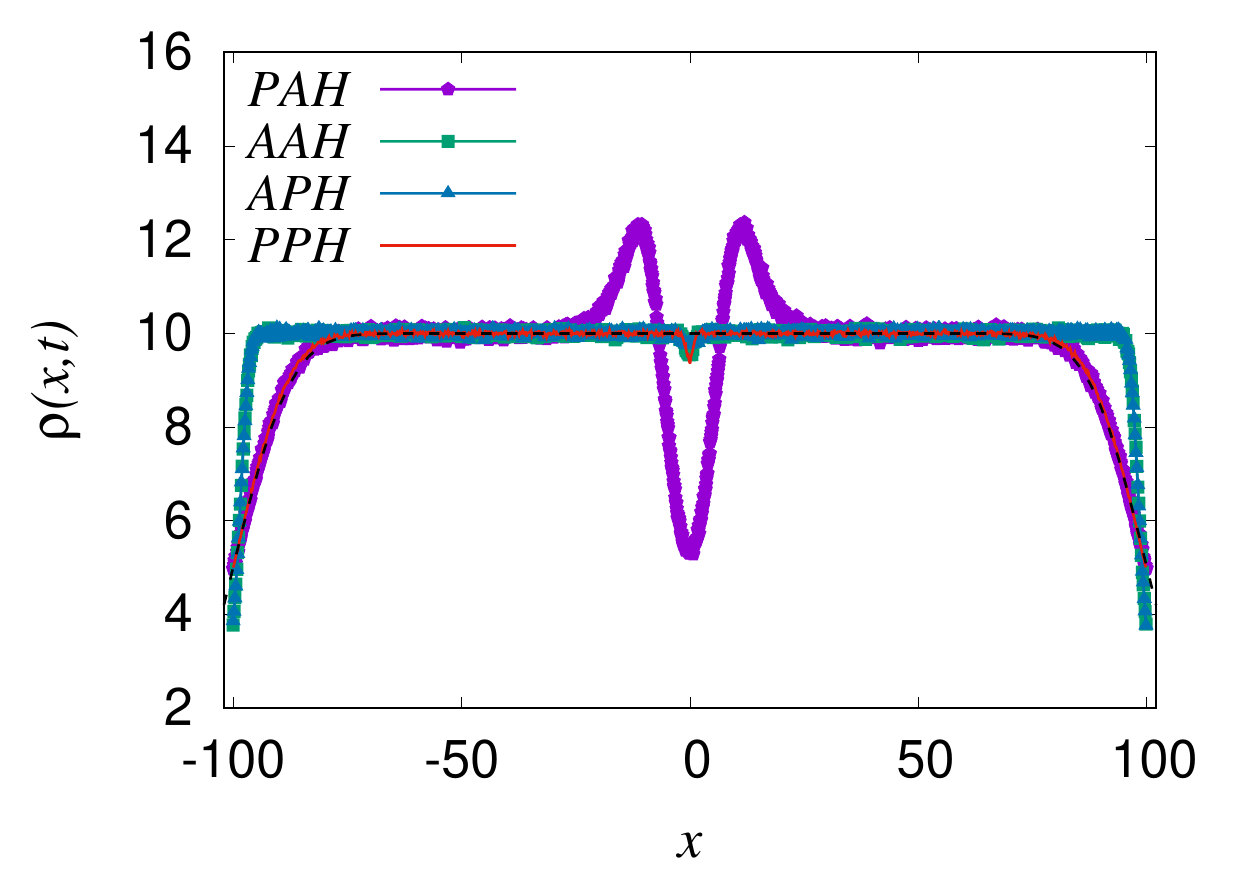}\includegraphics[width=7cm]{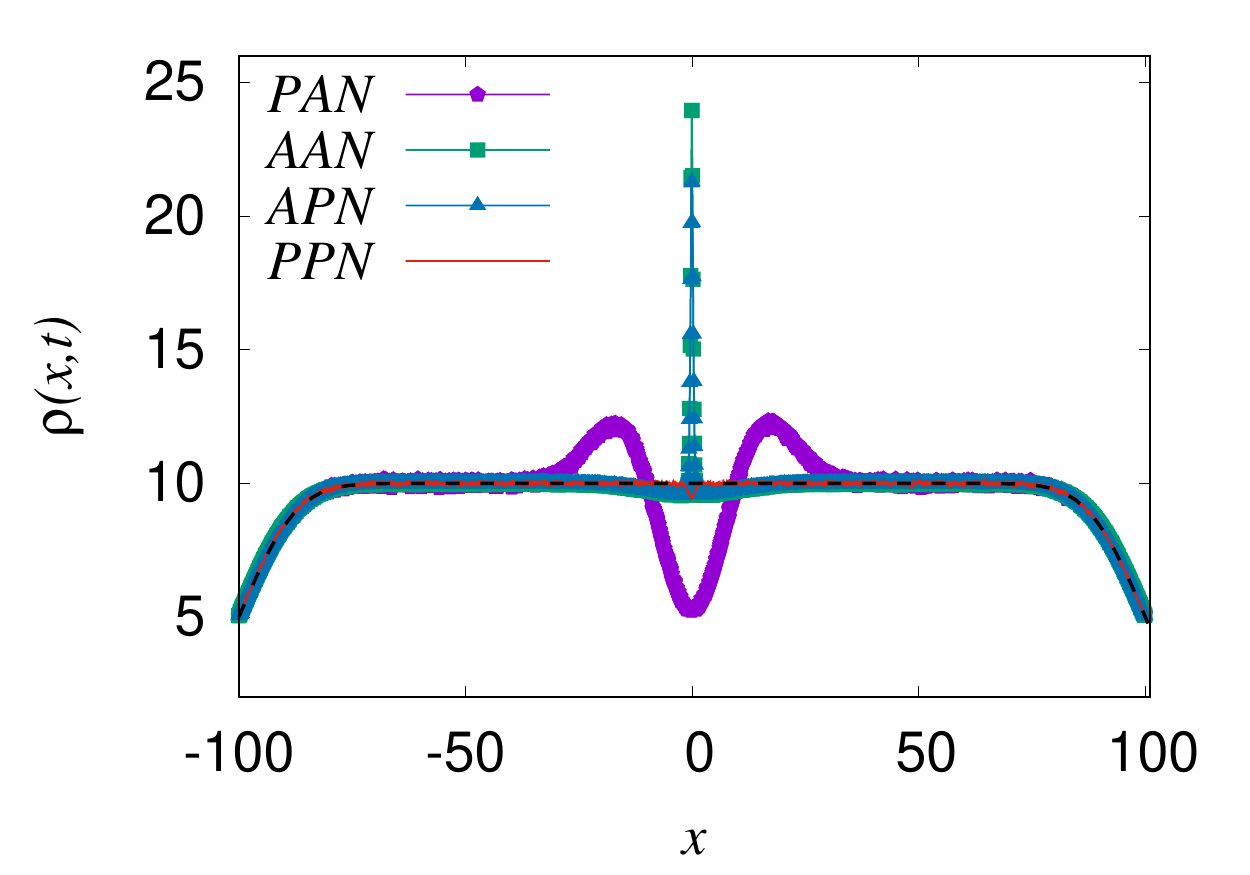}
\caption{Density profile of environment particles at a late time $t=100 \gg\gamma^{-1} = 1$, for hardcore (Left) and noninteracting (Right) environments. For both panels $\Delta t=0.001$, initial density $\rho(x, t=0)=10$ with $N = 2000, L = 100$. At late times, while the {\em AAH} and {\em APH} density profiles are nearly indistinguishable, they are markedly different from the {\em PPH} and {\em PAH} profiles. {\em AAN} and {\em APN} density profiles show clustering of active particles around the initial position of the tracer, while {\em PAN} and {\em PAH} profiles show heaps of passive particles displaced by active tracers. The dashed black line in each panel represents Eq.~\ref{top-hat-sol} which coincides neatly with {\em PPH}/{\em PPN} data (barring the residual small dip at the origin, see Sec.~\ref{prefactors}). {\em AAH}, {\em APH}, {\em PAH} and {\em PPH} data are averaged over $1.7 \times 10^5, 1.6 \times 10^5, 2.7 \times 10^4$ and $ 5 \times 10^4$ realizations, respectively. {\em PAN} data is averaged over more than $2.6 \times 10^4$ realizations. {\em AAN}, {\em APN} and {\em PPN} data are the same as used in Fig.~\ref{nonint-density-full}. Other parameters are $v=1,D=0.5$. Data sampled only in the range shown in the plots.}\label{PAX-AAX-den}
\end{figure}

 \begin{figure}
\centering
\includegraphics[width=7cm]{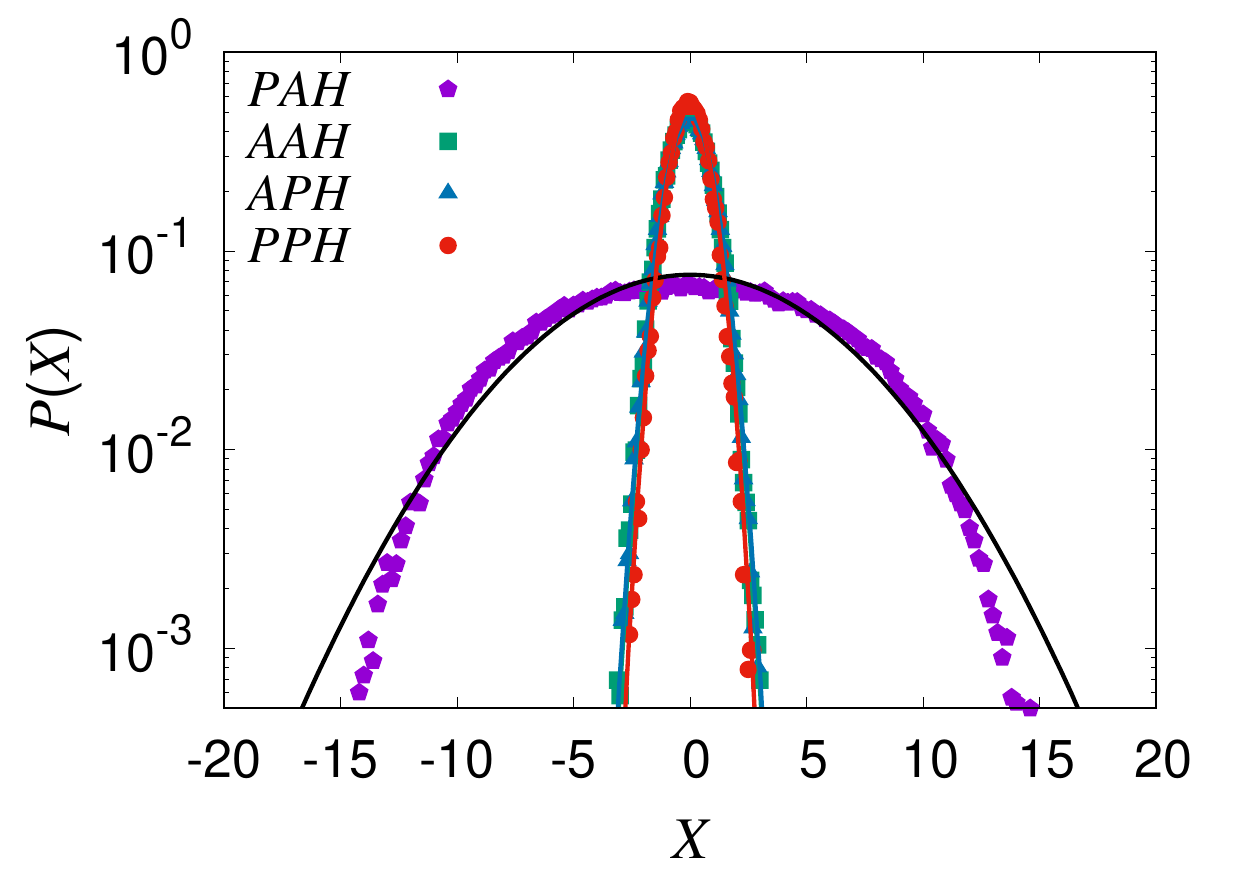}\includegraphics[width=7cm]{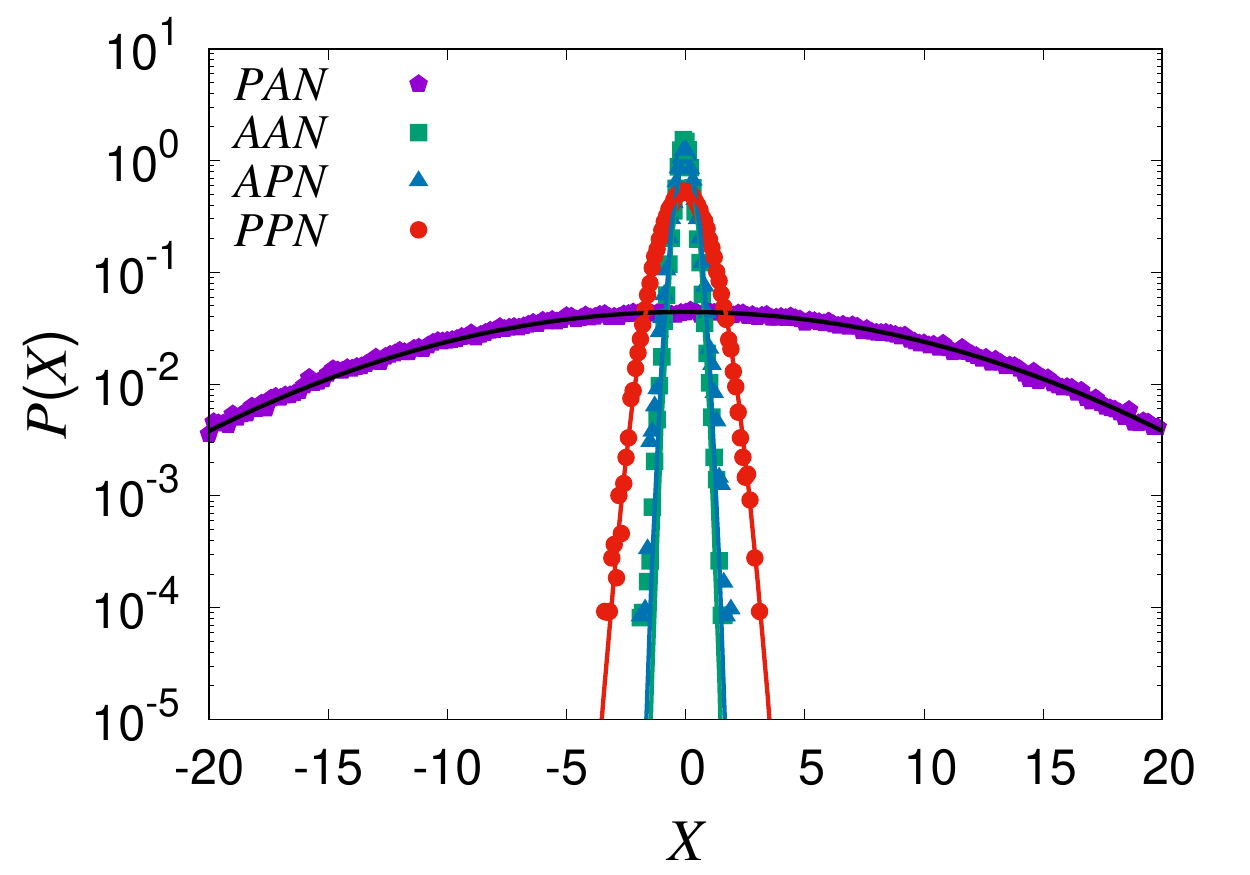}
\caption{Histogram of tracer position $X$ at late time $t=100$  in different hardcore environments ($N = 2000, \rho=10,\Delta t=0.001, v=1=\gamma, D=0.5$) on a semi-log scale. (Left) {\em XXH} and (Right) {\em XXN}. {\em PAX} cases have much larger variances compared to the other six situations of Table~\ref{table}. Solid lines (Black: {\em PAX}, Red: {\em PPX}, Blue: {\em APX} and Green: {\em AAX}) represent Gaussian fits with the respective variances calculated numerically. {\em PPH}, {\em AAH} and {\em APH} data are the same as used in Fig.~\ref{hardcore-variance-compare}(Right). {\em PAH} data is averaged over $1.5 \times 10^5$ realizations. {\em AAN} and {\em APN} data are the same as used in Fig.~\ref{noint-act-pass-dist}. {\em PPN} and {\em PAN} data are averaged over $10^5$ and $1.7 \times 10^5$ realizations, respectively.}\label{PAH-AAH-variance}
\end{figure}

\subsection{Further comparisons}
At long times $t \gg {\gamma}^{-1}$, the density footprints left by an active tracer on passive environments  are very different from those it leaves on active environments. In Fig.~\ref{PAX-AAX-den} (Left) we compare the various hardcore density profiles for $\gamma t = 100 \gg 1, \rho = 10, \Delta t=0.001$. In the passive environments, the active tracer shows a much larger displacement variance, with more freedom to move around  in the `correlation hole' created by pushing the environment particles into mounds further away. This is also seen from Fig.~\ref{PAH-AAH-variance} where the tracer position distribution $P(X)$ is compared for all the four hardcore environments at $\rho=10, \Delta t=0.001$, $t=100$. 

From Figs.~\ref{PAX-AAX-den} and \ref{PAH-AAH-variance}, it is clear that {\em PAH} is qualitatively distinct from the other three {\em xxH} cases due to the correlation hole / snowplough effect at large enough $\Delta t$. Similar analysis holds for {\em PAN}, see Fig.~\ref{PAX-AAX-den} (Right); the active  tracer is more efficient in pushing passive particles away ({\em PAN}) than active ones ({\em AAN}) and enjoys much larger freedom of movement as a result. The same figure also shows the separate phenomenon of sticking of the active environment particles to the tracer in the two $AxN$ cases, as was already discussed in relation to Fig.~\ref{evolve-PAX-density}.

\section{Further discussion of clustering, sticking and caging}\label{discu}

At high densities we have observed two distinct clustering effects caused by activity, neither of which occurs in the purely passive systems ({\em PPx}) that were extensively studied in previous literature. The first effect arises in the case of noninteracting active environments ({\em AxN}), and is caused by an accumulation of stuck particles around an active or passive tracer  (Fig.~\ref{nonint-density-full} (Right)). This type of pile-up is a somewhat universal phenomenon and also seen where free active particles impact on static boundaries~\cite{mike-natphys,walls}.
Interestingly, in our context at least, it is suppressed by hardcore interactions among the environment particles, mainly because these spend too much time sticking to each other to build up a high density next to the obstacle \footnote{Note that Motility-Induced Phase Separation (MIPS) does not arise for our hardcore active environments at the densities and timescales considered here. If present, it would surely cause non-Gaussian statistics for $P(X)$, {\em e.g.} a superposition of two Gaussians corresponding to the two coexisting densities in the phase separation, in contrast to Fig.~\ref{hardcore-variance-compare}.}. (This might well change in dimensions higher than one). A very interesting consequence is that, as seen in Sec.~\ref{comp1}, by making the environment more uniform, the introduction of hardcore interactions to an active environment tends to speed up, rather than slow down, the tracer dynamics.

The second type of clustering is the formation of `snowheaps' of passive environment particles on either side of an active tracer in both {\em PAH} and {\em PAN}. For reasons we have described, 
these effects strongly depend on the discretization time-step and, to a lesser extent, on the update sequence (random {\em vs}.~cyclic). In this sense such effects are much less universal than those related to stuck particles. 
Nonetheless, both types of density inhomogeneity represent interesting signatures of activity that could be the focus of experimental attention. Importantly though, activity alone is not enough to see these inhomogeneities: neither effect is seen for {\em AAH} or {\em APH}.

Via the snowplough effect, the presence of an active RTP in a crowded passive environment also speeds up the transport of passive particles, at least at early times while the two `snowheaps' are being created by the ballistic tracer dynamics. This may  be related to the fact that mixtures of active and passive particles can optimize certain first passage properties in the cytoplasm, as reported in~\cite{hafner-Screports,voituriez-rmp}.

The trapping of an active or passive tracer in an environment of dense non-interacting active particles bears comparison with glassy systems, where a tagged particle with early time ballistic motion undergoes dynamic arrest at intermediate timescales due to caging, before eventually showing diffusive behaviour~\cite{glass1,glass-KA}. A related connection between active and glassy dynamics in dense systems was previously reported in~\cite{glass-ludovic}; also see~\cite{efepl} for links between activity-induced fluctuations and caging phenomenon in living cells. 
These links between hardcore tracer dynamics in single-file active environments and in glasses might be further interrogated by studying standard observables such as the pair-correlation function, but this lies beyond the scope of the present paper. Our results might also be relevant for studies on bacterial cytoplasm where dynamical arrest has been observed by monitoring the mean-squared displacement of cytosolic components~\cite{parry-cell}.

\section{Summary}\label{summa}
\noindent We have studied the behaviour of  tracers in hardcore interacting and non-interacting media in one-dimension for all eight combinations (Table~\ref{table}) of active and passive tracer and environment particles. The interaction between tracer and environment is always via hardcore exclusion. Our active particles are point Run-and-Tumble Particles  (RTPs) of velocity $\pm v$ and tumble rate $\gamma$ (both taken as unity without loss of generality by choice of space and time scales) while our passive particles are point Brownian diffusers of free diffusivity $D = v^2/2\gamma = 0.5$ chosen to match the long-time behaviour of a free RTP. 
The initial configuration is such that all particles are equispaced, often called a `quenched' initial configuration~\cite{sadhu-prl,derrida-mft}, with random assignment of the orientation of each RTP. For each case we investigated the statistics of the tracer displacement (through its variance \var) and the environmental density profile around the tracer. 

In active hardcore environments we observed late-time subdiffusive scaling for {\em AAH} and {\em APH} scenarios with a shared prefactor, $\varm  \to \xi_{\textrm{AxH}}(\rho)\sqrt{t}$. The identical prefactor is explained by enslavement of the tracer to the active interacting environment. This prefactor does not however coincide with that known for a passive tracer in a passive environment,
$\xi_{\textrm{PPH}}(\rho) = \xi_{\textrm{PPN}}(\rho) = (2D/\pi)^{1/2}\rho^{-1}$, under similar initial conditions. Note that {\em only} in the all-passive case are the noninteracting and hardcore environments automatically equivalent for tracer dynamics, via the worldline-swapping argument of Sec.~\ref{recap}. Nonetheless, the density of environment particles around the tracer remains almost uniform in both {\em AAH} and {\em APH}. 

In non-interacting environments at low enough density, the results for {\em PPN}, {\em AAN} and {\em APN} results merge at long times: $\varm \to \xi_{\textrm{PPN}}(\rho)\sqrt{t}$. This is because at low enough density, encounters between particles occur on a timescale beyond the flip time $\gamma^{-1}$ of an RTP which has therefore effectively become Brownian. 
In contrast, for highly dense non-interacting active environments, any $\sqrt{t}$  regime recedes beyond our observable time windows (as set by finite size effects), for both active and passive tracers. This is a consequence of clustering of stuck RTPs around the hardcore tracers, which leads to caging and narrower displacement distributions. Moreover, in dense active environments, long time results for the active and passive tracers do not coincide, in contrast to what we found in hardcore environments at all densities.

The statistics of active tracers in passive environments ({\em PAH} and {\em PAN}) show a lower degree of universality than all other cases, and -- at least in our implementation of the dynamics -- {depend strongly on the particles' step size $v \Delta t$}. At small enough $\Delta t$, the {\em PAH} tracer shows nearly normal-diffusive behaviour for quite long times, and subdiffusive behaviour appears only much later (if at all) compared to the other three hardcore scenarios. Increasing the time-step leads to faster arrival of the subdiffusive regime. 
At equal density and time-step, the active tracer moves further in a noninteracting passive environment than an interacting one, reflecting the relative ease with which the environment particles can be gathered into mounds by a snowplough effect (operating during each ballistic `run' of the RTP), thereby creating space for the tracer to move. 

While we have studied all eight cases numerically, obtaining analytical solutions for the six that involve active particles offers a rich source of challenging open problems, to match those already addressed in purely passive systems (see~\cite{sadhu-prl}). We expect standard MFT techniques, in principle, to work for cases {\em AxH} (as the tracer remains slaved to the environment)~\cite{sadhu-prl}. However, cases {\em AxN} and {\em PAx} require much more careful analytical treatments. A direct extension of our work would be to calculate current fluctuations through the initial tracer position for both interacting and non-interacting active environments. Whether or not one observes dynamical phase transitions in these set-ups remains an interesting question for the future. While our study is restricted to the case of matched diffusivities, relaxing this restriction could yield a range of further interesting regimes particularly when the mismatch becomes strong.  Further, connections between tracer dynamics and interface growth models~\cite{satya-Mustansir} in interacting active systems need to be investigated by applying an external bias on both the tracer and the environment. Also, how correlations between particle positions in single file active systems get affected by a {\em local} bias~\cite{kundu-local}, remains a question for the future.

Although this paper focuses entirely on simple 1d models, the results presented here, include qualitative predictions about eventual subdiffusion, density anomalies, caging at intermediate times, and other activity-induced effects upon single-file diffusion. These might inspire and inform the future experimental exploration of systems with various probe/environment combinations such as in-vitro set-ups mimicking transport processes in the bacterial cytoplasm~\cite{parry-cell}, or a passive probe placed within an active bacterial culture~\cite{pollen1,pollen}.

\section*{Acknowledgements}
We thank Satya Majumdar for initially bringing to our attention the problem of a passive tracer in a sea of repulsive run-and-tumble particles and for suggesting interesting references. Work funded in part by the European Research Council under the Horizon 2020 Programme, ERC grant
agreement number 740269. MEC is funded by the Royal Society.

\appendix 

\section{Comparison of update rules}\label{ran-seq}
Here we compare \var \, results for some of the cases in Table~\ref{table} found through the cyclic sequential update (detailed in Sec.~\ref{models}), denoted S, with those from a random sequential update scheme, denoted RS.

For the RS scheme, each simulation time-step, we draw $N+1$ (bath particles + tracer) integer random numbers between $1$ and $N+1$. Whichever value the random number takes, we update the particle with that index.
The rest of the algorithm is exactly similar to the one described in Sec.~\ref{models}. Here we summarize the comparisons:

$\bullet$ {\em AAH}: Fig.~\ref{AAH-AAN-rseq} (Left) shows excellent agreement of S with RS for $\Delta t=0.001$.

$\bullet$ {\em AAN}: Fig.~\ref{AAH-AAN-rseq} (Right) shows excellent agreement of S with RS for $\Delta t=0.001$.

$\bullet$ {\em PPH}: Fig.~\ref{PPH-rseq} (Left) shows a good agreement of S with RS for $\Delta t=0.0001$.

$\bullet$ {\em PAH}: Fig.~\ref{PPH-rseq} (Right) shows a mismatch at moderately late times between S and RS for both $\Delta t =0.0001$ and $\Delta t=0.001$. However, the trend that \var \, shows with respect to $\Delta t$ is the same for both schemes.

\begin{figure}
\centering
\includegraphics[width=7cm]{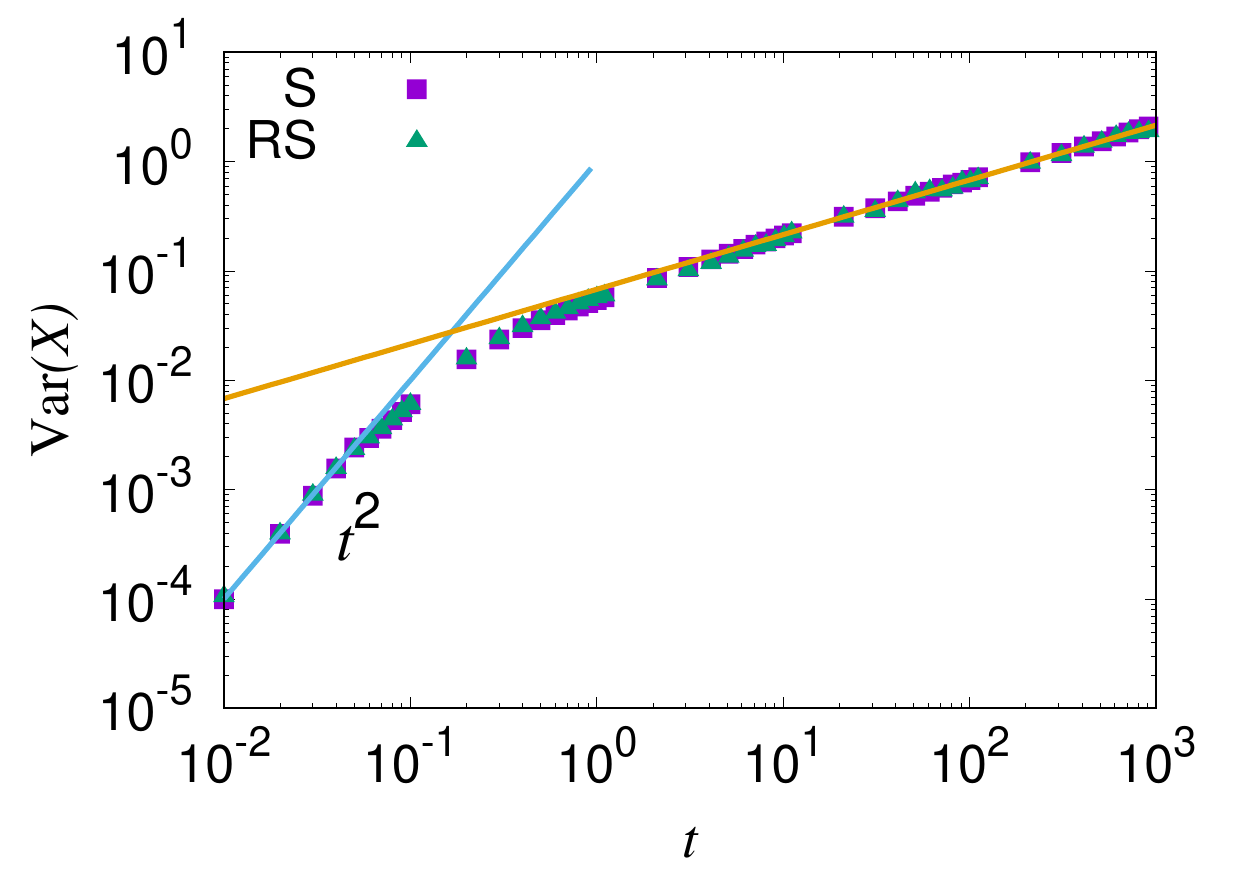}\includegraphics[width=7cm]{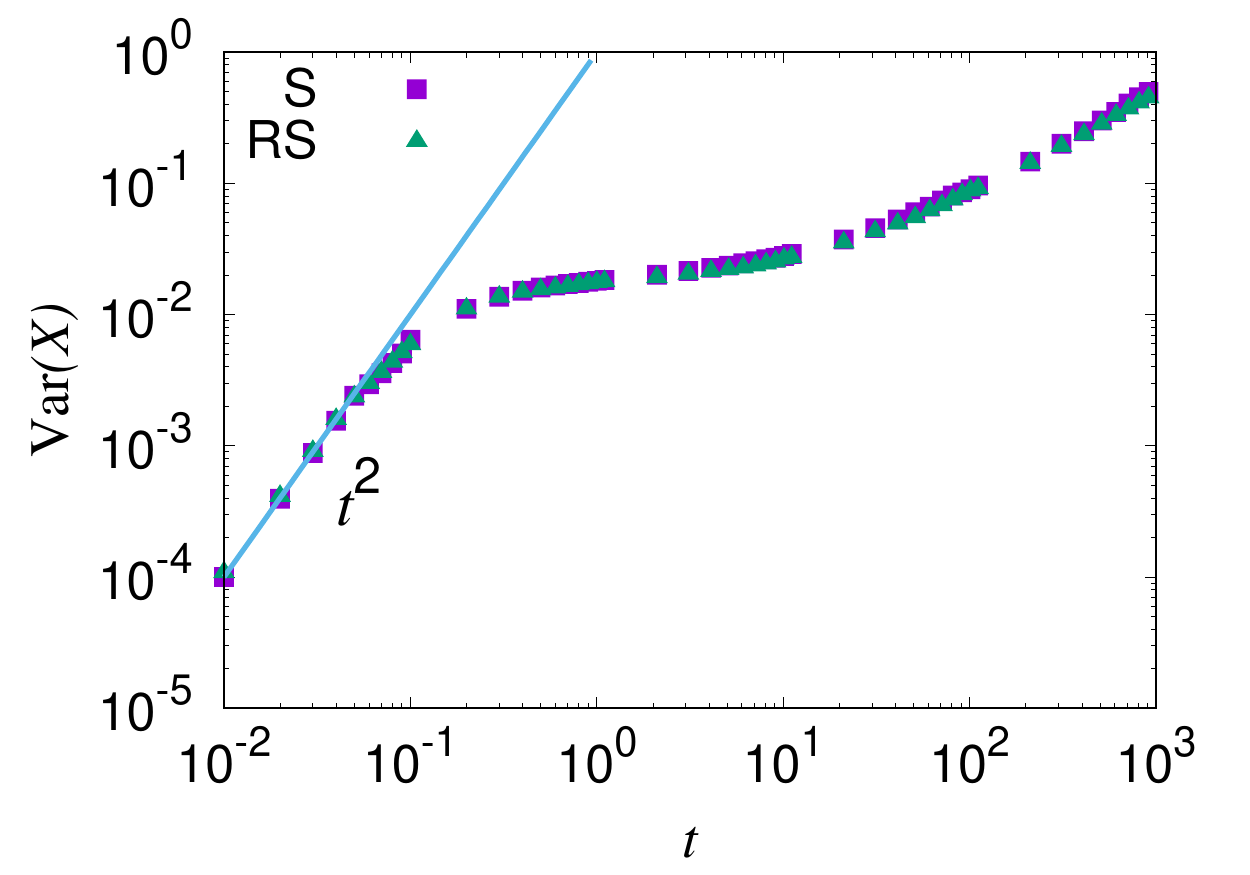}
\caption{\var \,  as a function of time, comparing S and RS updates for the {\em AAH} (Left) and {\em AAN} (Right) cases. The yellow line is a fit of $\sqrt{t}$ to the late-time data. $N=2000, \rho=10, \Delta t=0.001$. For {\em AAH}, S data is averaged over $1.5 \times 10^4$ realizations and RS data over $7 \times 10^3$ realizations; for {\em AAN} both data sets averaged over $7 \times 10^3$ realizations. For both panels $v=1=\gamma$. }\label{AAH-AAN-rseq}
\end{figure}

\begin{figure}
\centering
\includegraphics[width=7cm]{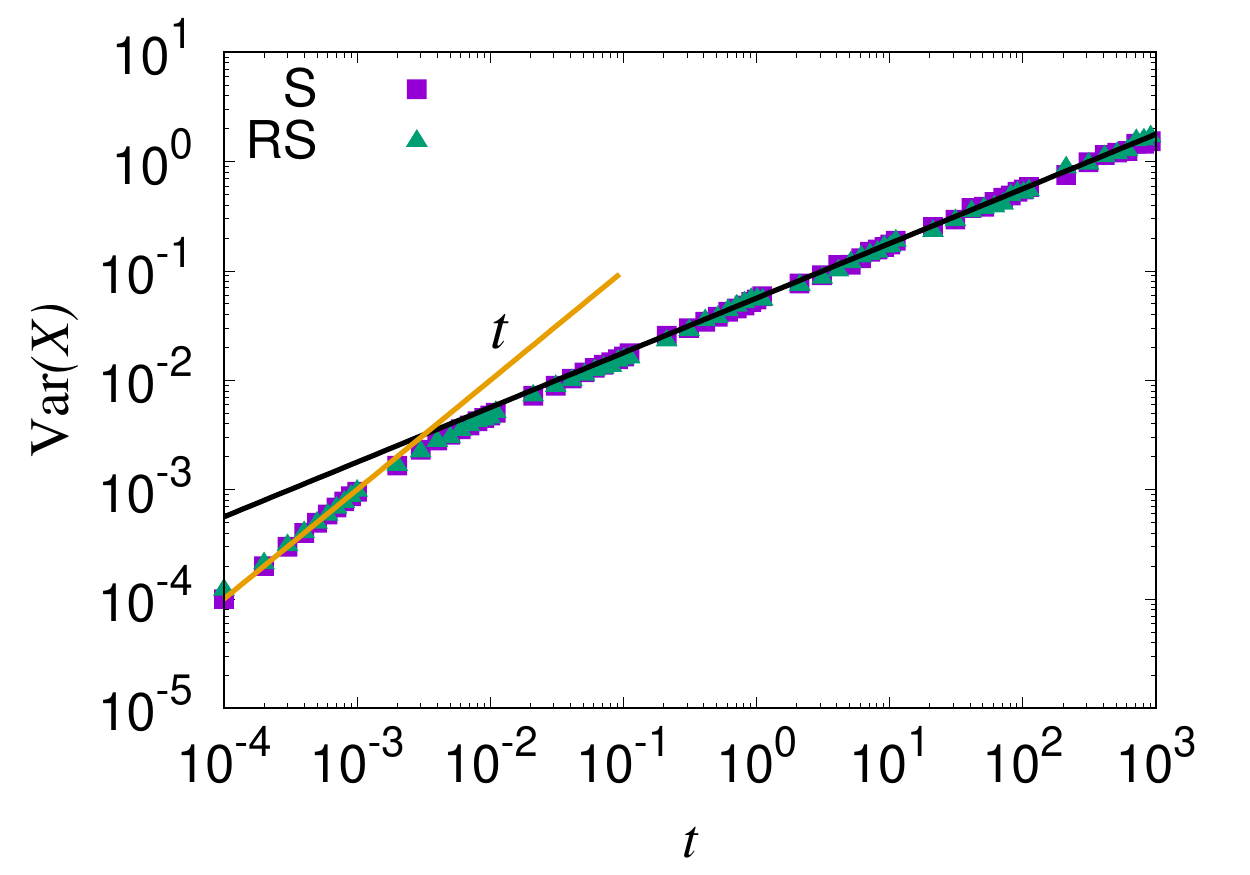}\includegraphics[width=7cm]{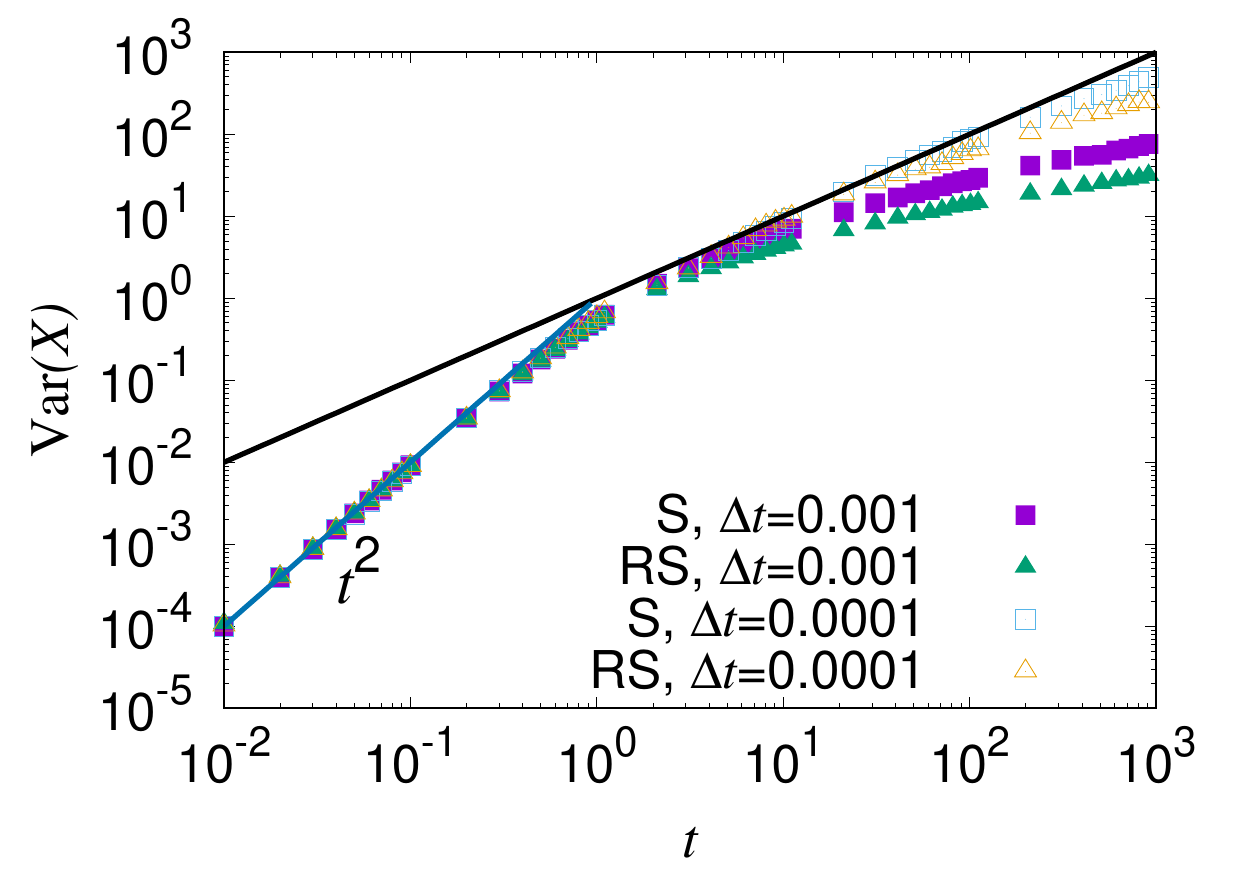}
\caption{ \var \,  as a function of time, comparing sequential (S) and random-sequential (RS) updates for: (Left) the {\em PPH} case (S data same as in Fig.~\ref{trajectory-hard-diff}(Right)); RS data over $400$ realizations . Solid black line is Eq.~\ref{Xvar-sadhugen}. $N=2000, \rho=10, \Delta t=0.0001, D=0.5$. (Right) the {\em PAH} case. Solid black line is $2 D t$. $N=2000, \rho=10,v=1=\gamma, D=0.5$; various $\Delta t$ as given in legend. RS data for $\Delta t=0.001$ averaged over $3.5 \times 10^3$ realizations and $\Delta t=0.0001$ over $350$ realizations. S data is the same as in Fig.~\ref{trajectory-PAH} (Left).}\label{PPH-rseq}
\end{figure}

\section{Finite size effects: illustrative results}\label{very-late-time-nonint}
{Here we investigate the finite-size effect on the late-time tracer dynamics stemming from the finite number of environment particles.}
Fig.~\ref{nonint-latetime-N501} (Left) illustrates the finite-size-induced restoration of linear scaling for \var \, {\em vs.} $t$ at very late times for a system with $N=500$ particles in the {\em PAN, AAN} and {\em APN} cases, as discussed in Secs.~\ref{FSE} and \ref{TVFSE}. {In Fig.~\ref{nonint-latetime-N501}(Left), we see that finite size effects are visible in the turn up of the {\em PPN} data which then lie above the exactly known result $\varm = \xi_{\textrm{PPH}}\sqrt{t}$ (solid line).  The latter holds only for $N\to\infty$ and this mismatch is exactly as expected from the arguments given in Sec.~\ref{FSE}. The same arguments apply equally to the {\em APN} and {\em AAN} curves which therefore also turn up cross the solid line (but not the actual {\em PPN} data).}

In Fig.~\ref{nonint-latetime-N501} (Right), we systematically vary $N$ and focus on the {\em AAN} scenario with $\rho=10$. For small system sizes ($N=100,500$), we observe that \var \, leaves the plateau and becomes superdiffusive before showing diffusive behaviour. It is notable that similar evolution for mean-squared displacements with a caging regime followed by superdiffusive and diffusive regimes has been reported in passive colloidal systems under shear~\cite{colloid1} or for crowded systems involving active Brownian particles~\cite{active-superdiff}. {The apparently `superdiffusive' catch-up region of very high positive slope ($d \ln \varm /d\ln t \gg1$), as far as we can tell, is not physically significant: plotting instead \var \, against $\sqrt{t}$ on linear axes will show a simple crossover from the constant plateau to a linear late time regime.  We do not attach direct physical significance to the superdiffusive regime for the problems studied here.}

\begin{figure}
\centering
\includegraphics[width=7cm]{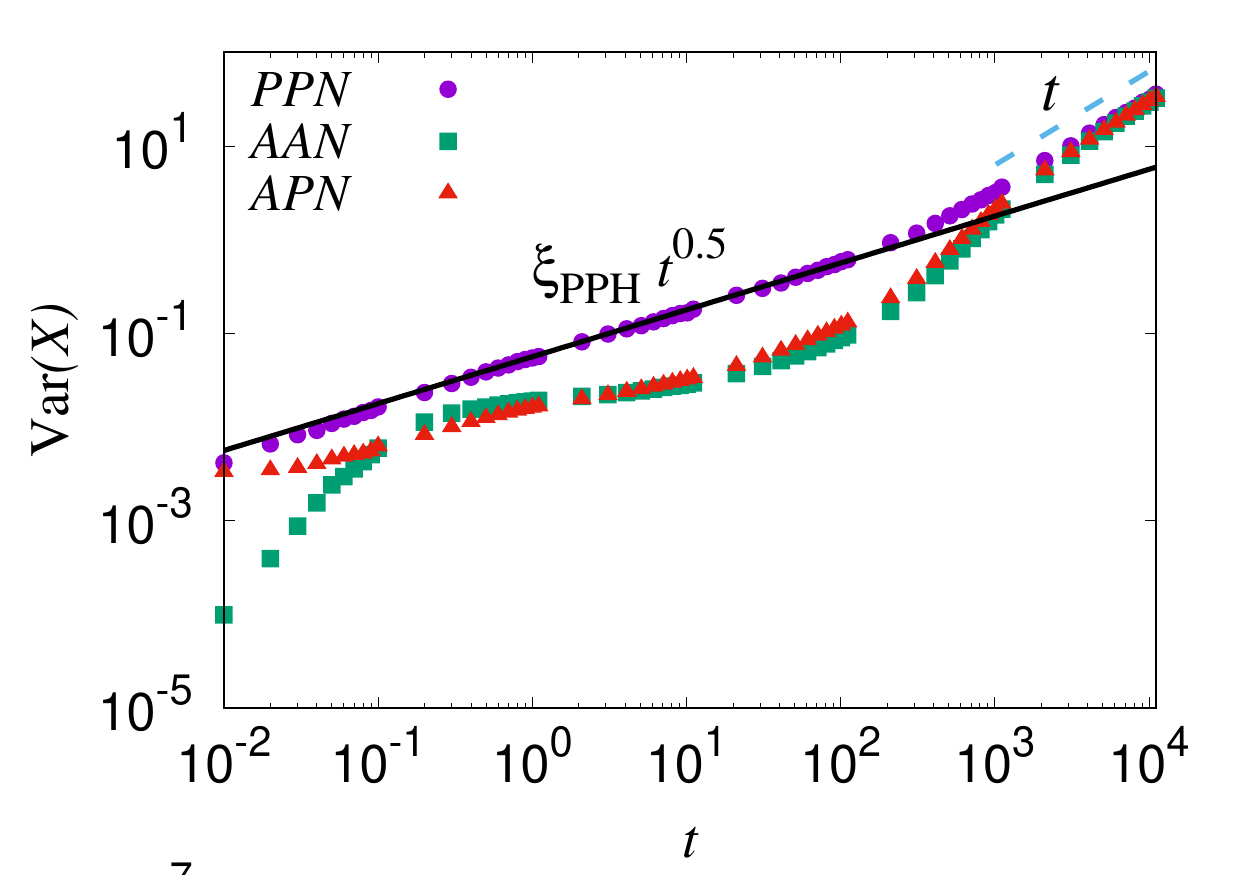}\includegraphics[width=7cm]{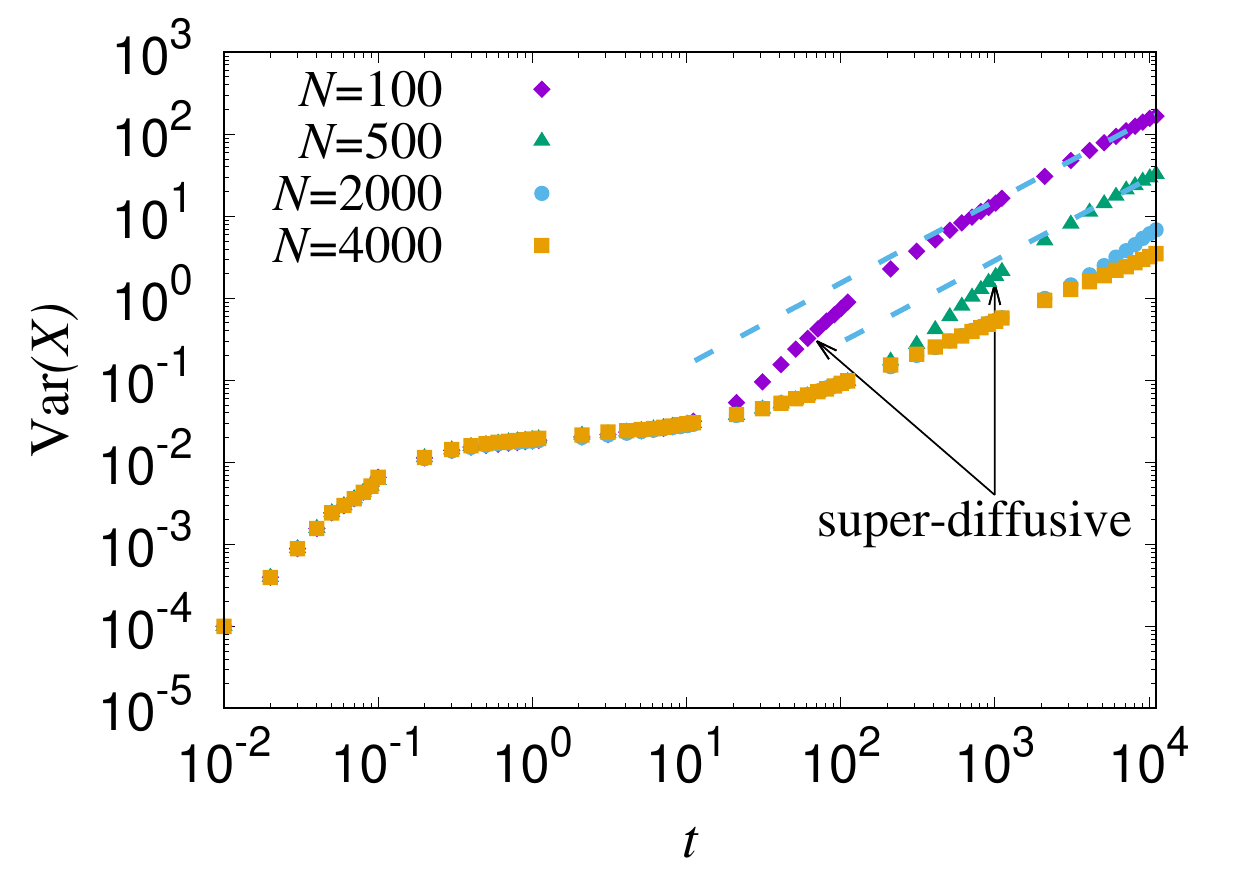}
\caption{ (Left)  \var \,for various tracers in different non-interacting environments for a high density, $\rho=10,N=500,\Delta t=0.001$. Finite size effects cause a crossover back to linear scaling of \var \, (dashed line). Black solid line is Eq.~\ref{Xvar-sadhugen}. (Right) \var \, against time in the {\em AAN} case for different $N$, with $\rho=10$ and $\Delta t=0.001$. Dashed lines show linear scaling. All data sets averaged over a minimum of $7 \times 10^3$ realizations.}\label{nonint-latetime-N501}
\end{figure}

\section{Role of delaying the start of observations}\label{t0}

\begin{figure}
\centering
\includegraphics[width=7cm]{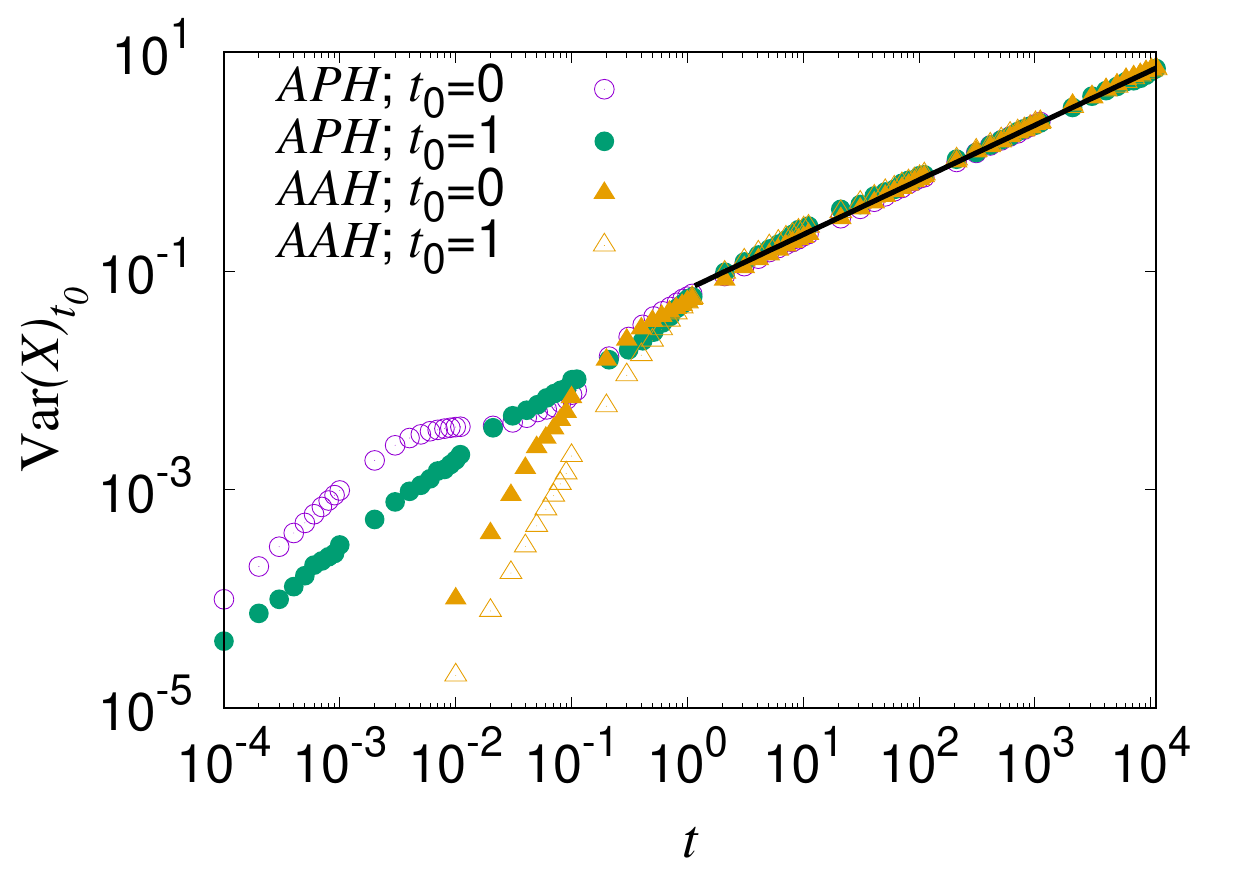}\includegraphics[width=7cm]{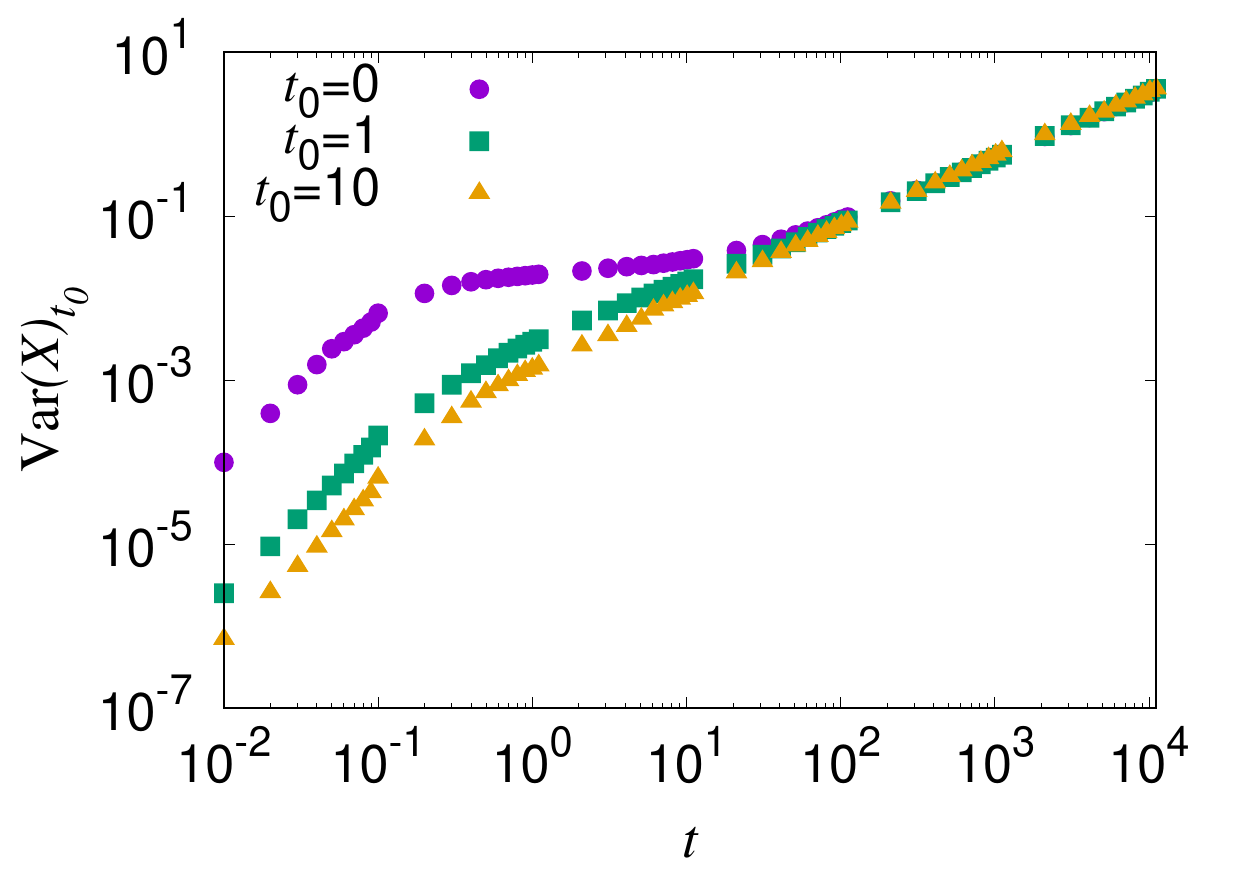}
\caption{ \var \, against time for different system ages $t_0$ when observations begin at $t=0$. (Left) {\em APH} and {\em AAH} results. $N=2000, \rho=10,v=1=\gamma, D=0.5$. The black solid line is $\xi_{\textrm{AAH}} \sqrt{t}$, with $\xi_{\textrm{AAH}}$ obtained via fitting the {\em AAH} data for $t_0=0$. All data averaged over at least $10^{4}$ realizations or more. (Right) The same for {\em AAN} with $N=4000, \rho=10$. For all {\em AAH} and {\em AAN} cases, $\Delta t=0.01$. For {\em APH} data, $\Delta t=0.0001$ for $t \leq 1$ and $\Delta t=0.01$ for $t>1$ (overlap of data sets checked in the range $1 \leq t \leq 10$).}\label{t0-fig}
\end{figure}

All results in the main text were obtained with the time interval between preparation of the equispaced initial state and the start of observations to be zero. {We now introduce a {\em lag-time} $t_0$ between the preparation of the initial equispaced state and the onset of observations for a given initial density $\rho =10$. The tracer's displacement $X$ is then measured from its value at $t=t_0$. The variance of $X$ is then given by }
\begin{equation}
\textrm{Var}(X)_{t_0}  = \langle [X(t_0+t)-X(t_0)]^2\rangle - \langle [X(t_0+t)-X(t_0)]\rangle^2\, \,. 
\end{equation}
We show results for {\em AAH} and {\em APH} (Fig.~\ref{t0-fig} (Left)) and for {\em AAN} (Fig.~\ref{t0-fig} (Right)). 
We observe that the late-time results remain independent of the choice of $t_0$. However, the early-time effects, such as caging, do depend strongly on $t_0$ and indeed disappear if $t_0$ is large enough. {This simply means that the caging in these $1d$ systems is transient, and to capture it properly, observations need be taken preferably right from the start of the experiment. From Fig.~\ref{t0-fig} we can see that the tracers become more and more constricted as $t_0 \uparrow \infty$, which effectively means that the respective tracers remain stuck against their neighbours. Therefore this limit can be viewed as one in which caging remains present but the cage-size becomes zero.}


\end{document}